\documentclass[aps,prb,reprint,superscriptaddress,floatfix]{revtex4-1}

\usepackage{appendix}
\usepackage{booktabs}
\usepackage{natbib}
\usepackage{graphicx}
\usepackage[T1]{fontenc}
\usepackage{nicefrac}
\usepackage{siunitx}
\usepackage{dcolumn}
\usepackage[para,online,flushleft]{threeparttable}
\usepackage[version=4]{mhchem}
\usepackage{url}
\usepackage{todonotes}
\begin{document}
	
\title{A Comprehensive Study of $g$-Factors, Elastic, Structural and Electronic 
Properties of III-V Semiconductors using Hybrid-Density Functional Theory}
	
\author{Carlos M. O. Bastos}
\affiliation{S\~ao Carlos Institute of Physics, University of S\~ao Paulo, PO Box 
369, 13560-970, S\~ao Carlos, SP, Brazil.}
	
\author{Fernando P. Sabino}
\affiliation{S\~ao Carlos Institute of Physics, University of S\~ao Paulo, PO Box 
369, 13560-970, S\~ao Carlos, SP, Brazil.}
	
\author{Guilherme M. Sipahi}
\affiliation{S\~ao Carlos Institute of Physics, University of S\~ao Paulo, PO Box 
369, 13560-970, S\~ao Carlos, SP, Brazil.}
	
\author{Juarez L. F. Da Silva}
\affiliation{S\~ao Carlos Institute of Chemistry, University of S\~ao Paulo, PO 
Box 780, 13560-970, S\~ao Carlos, SP, Brazil.}

\begin{abstract}
Despite the large number of theoretical III-V semiconductor studies reported every 
year, our atomistic understanding is still limited. The limitations of the theoretical 
approaches to yield accurate structural and electronic properties on an equal 
footing, due to the unphysical self-interaction problem that affects mainly the 
band gap and spin-orbit splitting (SOC) in semiconductors and, in particular, III-V 
systems with similar magnitude of the band gap and SOC. In this work, we will report 
a consistent study of the structural and electronic properties of the III-V semiconductors 
employing the screening hybrid-DFT framework, fitting the $\alpha$ parameters for 
\num{12} different III-V compounds, namely, \ce{AlN}, \ce{AlP}, \ce{AlAs}, \ce{AlSb}, 
\ce{GaN}, \ce{GaP}, \ce{GaAs}, \ce{GaSb}, \ce{InN}, \ce{InP}, \ce{InAs}, \ce{InSb}, 
in order to minimize the deviation between the theoretical and experimental values 
of the band gap and SOC. Structural relaxation effects were also included. Except 
for \ce{AlP}, whose $\alpha =\num{ 0.127}$, we obtained $\alpha$ values spreading from 
\num{0.209} to \num{0.343}, deviating less than \num{0.1} from the universal value 
of \num{0.25}. Our results for the lattice parameter and elastic constants indicate 
that the fitting of the $\alpha$ does not affect those structural parameters when 
compared with the HSE06 functional, where $\alpha = \num{0.25}$. Our analysis of the 
band structure based on the $\textbf{k}{\cdot}\textbf{p}$ method shows that the 
effective masses are in agreement with the experimental values, which can be attributed 
to the simultaneous fitting of the band gap and SOC. Finally, we estimate the values 
of $g$-factors, extracted directly from band structure, which are close to experimental 
results indicating that the obtained band structure produced a realistic set of 
$\textbf{k}{\cdot}\textbf{p}$ parameters. 
\end{abstract}

\maketitle
	
\section{Introduction} \label{sec:Introduction}

Semiconductors have been playing a key role in the development of new technologies 
since the 50's, e.g., light-emitting diodes and lasers,\cite{Schubert_2006,Nakamura_1139_2015,Sun_596_2016} 
infrared detectors,\cite{Rogalski_2010} solar cells \cite{Nelson_2003} and more 
recently, spin-lasers,\cite{Basu_91119_2008} and etc. Those developments have been 
possible due to the large number of fundamental studies employing theoretical or 
experimental techniques along decades.\cite{Yu_1999,Enderlein_1997} which have 
contributed to the present understanding of the semiconductors electronic band-structure 
properties, punctual and extended defects, and structural control.\cite{Enderlein_1997,Madelung_2004} 
Beyond that, new fields have emerged along the years such as topological insulators\cite{Hasan_3045_2010} 
and Majorana fermions in nanowires,\cite{Stanescu_144522_2011,Mourik_1003_2012} 
which are expected to contribute to future technological applications. 

Among a wide range of semiconductor materials,\cite{Enderlein_1997,Madelung_2004,Yu_1999,Bushby_2013} 
the III-V \ce{\textit{AB}} semiconductors, where \ce{\textit{A}} = \ce{Al}, \ce{Ga}, 
\ce{In} and \ce{\textit{B}} = \ce{N}, \ce{P}, \ce{As}, \ce{Sb} (i.e., \num{12} 
compounds), occupy an important place due to their role in several technological 
developments.\cite{Nelson_2003,Schubert_2006,Basu_91119_2008,Rogalski_2010,Nakamura_1139_2015} 
Although there is an impressive number of papers published every year based on 
experimental and/or first-principles calculations,\cite{Gmitra_165202_2016,Sandoval_115008_2016,Polly_64_2016,Seghilani_38156_2016,FariaJunior_235204_2016,Zhao_71_2016,Nestoklon_305801_2016,Sun_596_2016}
our atomistic understanding is still limited, in particular, due to the limitations 
in theoretical approaches to describe structural and electronic properties on an 
equal footing. For example, first-principles calculations based on density functional 
theory (DFT) with local or semilocal exchange-correlation energy functionals suffer 
from the unphysical self-interaction problem,\cite{Perdew_5048_1981,Perdew_1884_1983,Jones_689_1989} 
which affects mainly the band gap and spin-orbit splitting (SOC) in semiconductors\cite{Perdew_1884_1983,Kim_205212_2010} 
and, in particular for few III-V \ce{\textit{AB}} semiconductors, where the SOC 
can have similar magnitude as the band gap, e.g., \ce{GaSb}, \ce{InP}, \ce{InAs} 
and \ce{InSb}.\cite{Adachi_1992,Madelung_2004} 

Along the years, the self-interaction problem has motivated the widespread use 
of approximations or alternative descriptions of the electronic states such as 
the GW \cite{Hedin_796_1965,Aryasetiawan_237_1998} or nonlocal hybrid density-functionals, 
e.g., PBE0,\cite{Perdew_9982_1996,Adamo_6158_1999} HSE06,\cite{Heyd_8207_2003,Heyd_1187_2004,Heyd_219906_2006} 
and B3LYP,\cite{Becke_5648_1993} and etc. In principle, both GW and hybrid-DFT 
can yield an improved description of the band structure compared with local or 
semilocal functionals, however, GW was designed to address electronic properties 
but not the structural properties.\cite{Hedin_796_1965,Aryasetiawan_237_1998} In 
contrast with GW, hybrid-DFT can describe both structural and electronic properties 
on an equal footing, however, the electronic properties such as the band gap depends 
strongly on the magnitude of the nonlocal Fock exchange, $\alpha$, that replaces 
part of the semilocal exchange term. Although an universal value for $\alpha$ was 
suggested, i.e., \SI{25}{\percent} in PBE0,\cite{Perdew_9982_1996,Heyd_8207_2003} 
it is not as universal as expected.\cite{Vines_781_2017} 

As the band increases almost linearly by increasing the magnitude of the nonlocal 
Fock exchange,\cite{Moses_084703_2011} several studies have proposed a fitting 
of the $\alpha$ parameter to improve the description of the band gap.\cite{Vines_781_2017,Bastos_105002_2016,Moses_084703_2011} 
However, those studies have employed, in most cases, atomic structures optimized 
with the local or semilocal functionals,\cite{Ramos_165210_2001,DalCorso_085135_2012,Paier_154709_2006} 
i.e., the differences in the structural parameters are neglected. This is a good 
approximation but it might fail in cases in which the fine details of the electronic 
structure depend strongly on the lattice parameters, e.g., III-V systems with large 
SOC.   

In this work, we propose to perform a consistent study of the structural and electronic 
properties of the III-V semiconductors employing the screening hybrid-DFT framework, 
where the $\alpha$ parameters for the different compounds are fitted by the minimization 
of the deviation between the theoretical and experimental values for the band gap 
and SOC. We also employed optimized lattice parameters based on screening hybrid-DFT, 
i.e., small differences in the lattice constant were taken into account for the 
electronic parameters. Based on this framework, we calculated the equilibrium lattice 
constant, elastic constants, bulk modulus, band structures, effective masses, and 
etc. Furthermore, we employed the $\mathbf{k{\cdot}p}$ method to perform a deep 
analysis of the band structures, from which were possible to extract the $\mathbf{k{\cdot}p}$ 
parameters and the electronic $g$-factors. 

Except for \ce{AlP}, we obtained $\alpha$ parameters spreading from \num{0.209} 
to \num{0.343}, i.e., close to the universal value of \num{0.25}. For \ce{AlP}, 
the calculated $\alpha$ value is \num{0.127}. Based on several analyses, we found 
with small exceptions, that the $\alpha$ values correlate well with the atomic 
radius of the cationic species and hence, further $\alpha$ values for different 
semiconductors could be extrapolated from this finding without additional calculations. 
Our results for the lattice and elastic constants indicate that the fitting of 
the $\alpha$ does not affect those structural parameters when compared with the 
HSE06 functional, where $\alpha = 0.25$. 

Our analysis of the band structure based on the $\mathbf{k{\cdot}p}$ method shows 
that the effective masses are in agreement with the experimental values,\cite{Madelung_2004,Vurgaftman_5815_2001} 
i.e, the fitting of $\alpha$ at the $\Gamma$ point improved the description of 
the band curvatures. We also determined the $g$-factors directly from the effective 
band structures finding values that are close to the experimental results.\cite{Adachi_1992,Madelung_2004} 
Finally, having both, $g$-factors and effective masses, in good agreement with 
the experimental results indicates that we have determined realistic sets of $\mathbf{k{\cdot}p}$ 
parameters.

\section{Theoretical Approach and Computational Details}\label{sec:TheoreticalApproach}

\subsection{Density Functional Theory}\label{subsection:DFT}

It has been known for decades that DFT \cite{Hohenberg_B864_1964,Kohn_A1133_1965} 
within local (local density approximation -- LDA) \cite{Perdew_13244_1992} or semilocal 
(generalized gradient approximation -- GGA) \cite{Perdew_6671_1992,Perdew_3865_1996} 
exchange-correlation energy functionals is unable to yield a correct description 
of the fundamental band gap even for the most simple systems,\cite{Perdew_1884_1983,Kim_205212_2010} 
which has been attributed mainly to the unphysical self-interaction problem.\cite{Perdew_5048_1981,Perdew_1884_1983,Jones_689_1989}
This limitation has motivated the widespread use of approximations or alternative 
descriptions of the electronic valence states such as the GW \cite{Hedin_796_1965,Aryasetiawan_237_1998} 
or nonlocal hybrid functional, e.g., PBE0,\cite{Ernzerhof_5029_1999,Adamo_6158_1999} 
HSE06,\cite{Heyd_8207_2003,Heyd_1187_2004,Heyd_219906_2006} and B3LYP.\cite{Becke_5648_1993} 
In principle, both GW and nonlocal hybrid functionals can yield an improved description 
of the band structure compared with LDA or GGA. However, in contrast with the GW 
framework, nonlocal hybrid functional can provide also a reliable description of 
the structural and energetics properties,\cite{Paier_154709_2006,DaSilva_045121_2007,GandugliaPirovano_026101_2009,Hinuma_155405_2014} 
which is a plus compared with GW.

In this work, we will employ the DFT framework within the GGA formulation proposed 
by Perdew--Burk--Ernzerhof\cite{Perdew_3865_1996} (PBE) and the hybrid functional 
proposed by Heyd--Scuzeria--Ernzerhof\cite{Heyd_8207_2003,Heyd_1187_2004,Heyd_219906_2006} 
(HSE) in which the magnitude of the nonlocal Fock exchange replaces part of the 
PBE exchange. As will be described bellow, we will fit the magnitude of the nonlocal 
Fock exchange based on the experimental results of the fundamental band gap and 
the spin-orbit splitting, while using the same screening parameter derived for 
the HSE06 functional.\cite{Krukau_224106_2006,Heyd_219906_2006} 

To describe the electronic states, we employed the scalar-relativistic approximation\cite{Koelling_3107_1977,Takeda_43_1978} 
in which relativistic corrections are considered for the core-states, while the 
SOC is not considered for the valence states, and hence, the spin-orbit splitting 
for the III-V semiconductors cannot be described. For example, for \ce{InSb} the 
SOC splitting at the $\Gamma$-point and valence band maximum (VBM), has similar 
magnitude as the fundamental band gap,\cite{Madelung_1982,Kim_205212_2010} and 
hence, it plays a crucial role for the characterization of the band structure parameters, 
e.g., effective mass, $g$-factor, etc. Thus, to improve the description of the 
band structure properties, we employed the addition of the SOC for the valence 
states via the second-variational approach.\cite{Koelling_3107_1977}

To solve the Kohn--Sham equations, we employed the projected augmented wave (PAW) 
method,\cite{Blochl_17953_1994} as implemented in the Vienna \textit{ab initio} 
simulation package (VASP, version $5.4.1$),\cite{Kresse_13115_1993,Kresse_11169_1996} 
employing the PAW projectors provided with the package.\cite{Kresse_1758_1999} 
To describe the electronic states, the Kohn--Sham orbitals are expanded in plane 
waves employing a finite cutoff energy, which depends on the calculated properties. 
This is necessary since several properties, e.g., stress tensor and elastic constants 
converge slowly as the number of plane waves increases. 

From a large number of PBE and HSE convergence tests, we demonstrated that well 
converged total energies, band structures, and densities of states can be obtained 
by using a cutoff energy that is \num{1.125} times the recommended maximum cutoff 
energy (ENMAX$_{\text{max}}$). Stress tensor and elastic constants calculations, 
however, require at least $1.50\times\text{ENMAX}_{\text{max}}$. A further increase 
in the cutoff energy can improve the results slightly, and for the particular case 
of the PBE calculations, we increased the multiplication factor from \num{1.50} 
to \num{2.0} (stress tensor) and to \num{2.5} (elastic constants) aiming to provide 
reference data that can be used for further comparisons. For example, for \ce{AlN}, 
we employed \SI{473}{\electronvolt} for total energy and band structure; \SI{631}{\electronvolt} 
(HSE) and \SI{841}{\electronvolt} (PBE) for stress tensor; and \SI{631}{\electronvolt} 
(HSE) and \SI{1052}{\electronvolt} (PBE) for the elastic constants calculations. 
For the Brillouin zone integration, we employed a Monkhorst--Pack \textbf{k}-mesh 
of $10{\times}10{\times}10$, while the same \textbf{k}-point density was employed 
for the remaining III-V semiconductors. All those parameters are provided in the 
Supplementary Material. 

To obtain the equilibrium volume, we minimized the stress tensor, which was performed 
by several consecutive optimizations of the equilibrium volume to ensure that the 
optimized equilibrium volume is consistent with the initial set up of the basis 
size. To calculate the elastic properties, we considered the combination of two 
schemes, namely, $(i)$ rigid lattice parameters obtained from the stress-tensor 
optimization,\cite{Page_104104_2002} and $(ii)$ ionic volume relaxation from the 
inversion of the ionic Hessian matrix and internal stress tensor,\cite{Wu_035105_2005} 
as implemented in VASP. For those calculations, we employed atomic steps of \SI{0.010}{\angstrom}, 
which are slightly smaller than the recommended value by VASP, e.g., \SI{0.15}{\angstrom}.
  
\subsection{Hybrid HSE Functional: Fitting of the $\alpha$ Value}\label{subsection:HSE}

The hybrid PBE0 functional is composed by the PBE correlation energy and a fraction 
of \SI{25}{\percent} of the PBE exchange is replaced by the nonlocal Fock exchange 
of the Hartree--Fock (HF) method, i.e., $E_{\text{XC}}^{\text{PBE0}} = E_{\text{c}}^{\text{PBE}} 
+ {\alpha}E_{\text{x}}^{\text{HF}} + (1 - {\alpha})E_{\text{x}}^{\text{PBE}}$, 
where $\alpha = \num{0.25}$.\cite{Ernzerhof_5029_1999,Adamo_6158_1999} The hybrid 
HSE functional\cite{Heyd_8207_2003,Heyd_1187_2004,Heyd_219906_2006} is derived 
from PBE0 by using a screening function to split the semilocal PBE exchange and 
nonlocal Fock term into two parts, namely, a short- (SR) and long-range (LR) exchange 
contributions, in which the nonlocal Fock LR contribution cancel with part of the 
PBE LR exchange. Thus, the hybrid HSE functional is given by the following equation, 
\begin{equation} 
\begin{split}
E_{\text{XC}}^{\text{HSE}} = & E_{\text{c}}^{\text{PBE}} + E_\text{x}^{\text{PBE,LR}}(\omega) 
+ \alpha E_{\text{x}}^{\text{HF, SR}}(\omega) \\ & + (1-\alpha)E_\text{x}^{\text{PBE,SR}}(\omega)~,
\end{split}
\label{eq:HSE_equation}
\end{equation}
where the new parameter, $\omega$, measures the intensity of the screening, and 
hence, the extension of the nonlocal Fock interactions. For example, if $\omega$ 
is null, the SR contribution is equivalent to the full Fock operator and the LR 
contribution will become zero, while for ${\omega}{\rightarrow}{\infty}$, the range 
of SR terms decrease, recovering asymptotically the PBE functional. As defined 
in the PBE0 functional, the parameter $\alpha$ controls the amount of PBE exchange 
replaced by the nonlocal Fock exchange, and hence, in principle, it can ranges 
from \num{0} to \num{1}. For the hybrid HSE06 functional, $\alpha = 0.25$ and $\omega 
= \SI{0.206}{\angstrom^{-1}}$, which were obtained from the adiabatic perturbation 
theory and fitted from a large number of systems, respectively.\cite{Heyd_219906_2006,Krukau_224106_2006} 

Although, the hybrid HSE06 functional yield better results than the LDA and GGA 
functionals, HSE06 does not yield the experimental band gaps in most of the cases,\cite{Heyd_174101_2005,Kim_035203_2009,Kim_205212_2010} 
and hence, improved results can be obtained by fitting the $\omega$ or $\alpha$ 
parameters. Recently, Vi\~nes \textit{et al.} \cite{Vines_781_2017} suggested that 
a large number of combination of $\omega$ and $\alpha$ values can yield the band 
gap of oxides, however, it is important to mention that the fitting of $\omega$ 
can affect drastically the small contribution of the LR nonlocal Fock terms, which 
has the potential to decrease the stability of the electron density convergence. 
In contrast with $\omega$, the fitting of $\alpha$ affects mainly the SR nonlocal 
Fock contribution, which plays a crucial role in the physical properties.\cite{Perdew_2801_2017,Atalla_035140_2016,Sai_226403_2011,Yang_204111_2012} 

Therefore, to improve the description of the experimental fundamental band gap, 
$E_{\text{gap}}$, and the spin-orbit (SO) splitting energies, ${\Delta}_{\text{so}}$, 
we fixed the $\omega$ parameter to the same value used in HSE06, and fitted the 
$\alpha$ parameter to reproduce the experimental $E_{\text{gap}}$ and ${\Delta}_{\text{so}}$ 
results. The fitting was performed using the linear dependence of $E_{\text{gap}}$ 
and $\Delta_{\text{so}}$ as a function of $\alpha$, which is well known in the 
literature.\cite{Walsh_256401_2008,Komsa_075207_2011,Colleoni_495801_2016} Due 
to the nearly perfect linear dependence, the angular coefficient (slope), can be
 calculated using two points, namely, using the $E_{\text{gap}}$ or $E_{\Delta_{\text{so}}}$ 
values calculated with the PBE $(\alpha = 0)$ and HSE06 $(\alpha = 0.25)$ functionals. 
Thus, the angular coefficient for the fitting of the band gap is given by the following 
relation,  
\begin{equation}
m_{\text{gap}} = \frac{E_{\text{gap}}^{\text{HSE06}} - E_{\text{gap}}^{\text{PBE}}}{0.25}~,
\end{equation}
while $m_{\Delta_{\text{so}}}$ is obtained by replacing the band gap energies by 
the spin-orbit splittings. Thus, the $\alpha$ value that yields the experimental 
band gap, $E^{\text{exp}}_{\text{gap}}$, or the experimental spin-orbit splitting, 
$E^{\text{exp}}_{\Delta_{\text{so}}}$, can be obtained from the following equation, 
\begin{equation}
\alpha_{\text{gap}} = \frac{E^{\text{exp}}_{\text{gap}} - E_{\text{gap}}^{\text{PBE}}}{m_{\text{gap}}}~.
\label{eq:linAlpha}
\end{equation}

Consequently, from this scheme, we obtained two values for $\alpha$, namely, an 
optimized value for $\alpha_{\text{gap}}$ ($\alpha_{\Delta_\text{so}}$) that yields 
the experimental band gap (spin-orbit splitting). To obtain an unique $\alpha$ 
for each III-V system, we performed a minimization of the standard deviation between 
the experimental and the extrapolated $E^{\text{exp}}_{\text{gap}}$ and $E^{\text{exp}}_{\Delta_{\text{so}}}$ 
parameters obtained from the slope. Further technical details are reported in the 
Supplementary Material. Thus, from now on, the fitted hybrid HSE functional will 
be noted HSE$_{\alpha}$, where $\alpha$ is different for each semiconductor. Finally, 
in order to compare our results with literature and among different functionals, 
we adopted, as a measure, the normalized-root-mean-square deviation (NRMSD).\cite{Todeschini_2009}

\section{Results}\label{sec:Results}

\subsection{Magnitude of the Non-Local Fock Exchange}

\begin{figure}
\centering
\includegraphics[width=0.90\linewidth]{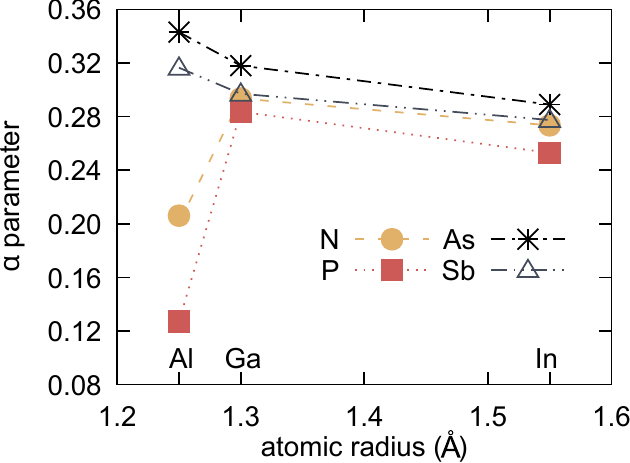}
\caption{Optimal magnitude of the nonlocal Fock exchange, $\alpha$, in the hybrid 
HSE$_{\alpha}$ functional obtained from the fitting of the fundamental band gap 
and the spin-orbit coupling versus the cation atomic radii obtained from Ref. \onlinecite{Slater_3199_1964}. 
The atomic radii are given in \si{\angstrom} and $\alpha$ is dimensionless.}
\label{fig:alpha}
\end{figure}

Except for \ce{AlP}, the optimal $\alpha$ values that minimize the relative errors 
for the undamental band gap and SOC splitting are in between \num{0.209} and \num{0.343}, 
i.e., close to the universal value of \num{0.25}.\cite{Perdew_9982_1996} For \ce{AlP}, 
the value if \num{0.127}. Therefore, we can conclude that an unique $\alpha$ value 
is unable to yield the fundamental band gap and SOC splitting for a wide range 
of compounds, as well as further physical properties, which have also been reported 
in the few previous studies.\cite{Walsh_256401_2008} Although the value of $\alpha$ 
has been obtained from a solid theoretical framework, few studies have tried to 
obtain a correlation between the magnitude of $\alpha$ and a particular physical 
property,\cite{Vines_781_2017} which can help in several applications. Thus, with 
the aim to identify the most important physical parameters that play the major 
role in the magnitude of $\alpha$, we performed several analyses (also in the Supplementary 
Material). Among all analyses, we found a good correlation between the magnitude 
of $\alpha$ versus the cationic radius,\cite{Slater_3199_1964} which is shown in 
Fig. \ref{fig:alpha}. Our results indicate that the value of the optimized $\alpha$ 
decreases almost linearly as a function of the atomic cationic radius, except for 
the cases of \ce{AlN} and  \ce{AlP}. 

\subsection{Equilibrium Lattice Parameter}\label{subsection:lattice_constants}

The most stable crystalline phase of the III-V semiconductors is the zinc-blende 
structure,\cite{Kittel_2004} which has a face-centered cubic (fcc) lattice and 
$\text{F}\bar{4}3\text{m}$ space group. The exception for this rule are the III-nitrides 
(\ce{AlN}, \ce{GaN}, \ce{InN}) that prefer to crystallize in the hexagonal wurtzite 
structure, belonging to the $\text{P}6_3\text{mc}$ space group,\cite{Morkoc_2009} 
but through the use of experimental techniques such as molecular beam epitaxy can 
also be grown as a zinc-blende structure.\cite{Yu_1999,Morkoc_2009} Thus, to rationalize 
our understanding, all the III-V semiconductors were studied in the zinc-blende 
structure, which contains two fourfold atoms with tetrahedral local symmetry. The 
equilibrium lattice parameter, $a_0$, was calculated using several approximations, 
namely, PBE, PBE+SOC, HSE06, and HSE$_{\alpha}$, and the results are summarized 
in Table \ref{tab:lattice_constants} along with the experimental results.

\begin{table}
\caption{Equilibrium lattice parameters $a_0$ (in \si{\angstrom}) for all the III-V 
semiconductors, calculated with the PBE, PBE with spin-orbit coupling for the valence 
states (PBE+SOC), HSE06, and HSE$_{\alpha}$ functionals. For the hybrid HSE$_{\alpha}$ 
functional, the adjusted $\alpha$ parameter is indicated within parentheses. The 
NRMSD indicates the normalized percentage deviation between theoretical and experimental 
parameters for the full series.}
\label{tab:lattice_constants}
\begin{threeparttable}
\addtolength{\tabcolsep}{1pt}  
\bgroup
\def\arraystretch{1.3}
\begin{tabular}{lccccl} \hline \hline
	& PBE & PBE+SOC & HSE06 & HSE$_{\alpha}$ ($\alpha$) & Exp. \\ \hline
	\ce{AlN}  & \num{4.399} &\num{4.399} &\num{4.361} &\num{4.367} (\num{0.219}) &\num{4.38}\tnote{a} \\ 
	\ce{AlP}  & \num{5.505} &\num{5.505} &\num{5.471} &\num{5.487} (\num{0.127}) &\num{5.46 }\tnote{b} \\ 
	\ce{AlAs} & \num{5.731} &\num{5.731} &\num{5.678} &\num{5.660} (\num{0.343}) &\num{5.661}\tnote{b} \\ 
	\ce{AlSb} & \num{6.213} &\num{6.215} &\num{6.160} &\num{6.146} (\num{0.318}) &\num{6.135}\tnote{b} \\ 
	\ce{GaN}  & \num{4.545} &\num{4.545} &\num{4.492} &\num{4.483} (\num{0.293}) &\num{4.52 }\tnote{a} \\ 
	\ce{GaP}  & \num{5.499} &\num{5.499} &\num{5.456} &\num{5.449} (\num{0.283}) &\num{5.45 }\tnote{b} \\ 
	\ce{GaAs} & \num{5.742} &\num{5.738} &\num{5.669} &\num{5.652} (\num{0.318}) &\num{5.653}\tnote{b} \\ 
	\ce{GaSb} & \num{6.203} &\num{6.203} &\num{6.124} &\num{6.124} (\num{0.297}) &\num{6.095}\tnote{b} \\ 
	\ce{InN}  & \num{5.042} &\num{5.039} &\num{4.976} &\num{4.956} (\num{0.274}) &\num{4.97 }\tnote{c} \\ 
	\ce{InP}  & \num{5.946} &\num{5.942} &\num{5.886} &\num{5.885} (\num{0.253}) &\num{5.868}\tnote{b} \\ 
	\ce{InAs} & \num{6.174} &\num{6.172} &\num{6.090} &\num{6.078} (\num{0.289}) &\num{6.058}\tnote{b} \\ 
	\ce{InSb} & \num{6.619} &\num{6.618} &\num{6.526} &\num{6.549} (\num{0.277}) &\num{6.479}\tnote{b} \\ \hline
	NRMSD     & \num{1.49}  &\num{1.47}  &\num{0.40}  &\num{0.49}                &    -       \\ \hline \hline
\end{tabular} 
\egroup
\addtolength{\tabcolsep}{2pt}  
\begin{tablenotes}
\footnotesize
\item[a] from Ref. \onlinecite{Pearton_2000},
\item[b] from Ref. \onlinecite{Madelung_2004},
\item[c] from Ref. \onlinecite{Rossler_2011}
\end{tablenotes}
\end{threeparttable}
\end{table}

In agreement with previous DFT-PBE calculations,\cite{Anua_105015_2013,Heyd_174101_2005,DalCorso_085135_2012} 
we obtained equilibrium PBE lattice constants that overestimate experimental results, 
with the largest deviation smaller than \SI{2.2}{\percent} for \ce{InSb}. The addition 
of the SOC for the valence states reduces the lattice constant only in the third 
decimal place, and hence, the improvement over the PBE compared with the experimental 
results is almost negligible and can be evaluated by the NRMSD show in Table \ref{tab:lattice_constants}. 
For example, the NRMSD is \SI{1.49}{\percent} for PBE and \SI{1.47}{\percent} for 
PBE+SOC. Therefore, the SOC does not affect the equilibrium lattice constants in 
contrast with the electronic properties, where it plays an essential role (see 
below). 

In order to reduce the computational cost, which increases substantially for HSE06+SOC, 
the HSE06 and HSE$_{\alpha}$ equilibrium lattice constants were calculated using 
stress tensor without the addition of the SOC for the valence states. The HSE06 
and HSE$_{\alpha}$ functionals yield $a_0$ parameters closer to the experimental 
results, and hence, with smaller relative errors compared with the PBE results. 
These small errors were expected since we provided an improved description of the 
exchange energy by the nonlocal Fock term. The differences between the HSE06 and 
HSE$_{\alpha}$ results are very small, i.e., the NRMSD changes from \SI{0.4}{\percent} 
(HSE06) to \SI{0.49}{\percent} (HSE$_{\alpha}$), which is an important result since 
it shows that the lattice parameters were only slightly affected by the improvement 
of the description of the fundamental band gap and spin-orbit splitting at the 
$\Gamma$-point. The HSE06 results are in excellent agreement with previous hybrid 
HSE06 results,\cite{Kim_035203_2009} e.g., indium composites have differences 
smaller than \SI{0.3}{\percent}, \SI{0.4}{\percent}, \SI{0.5}{\percent} for \ce{InP}, 
\ce{InAs} and \ce{InSb}, respectively. 

\begin{figure}
\centering
\includegraphics[width=0.90\linewidth]{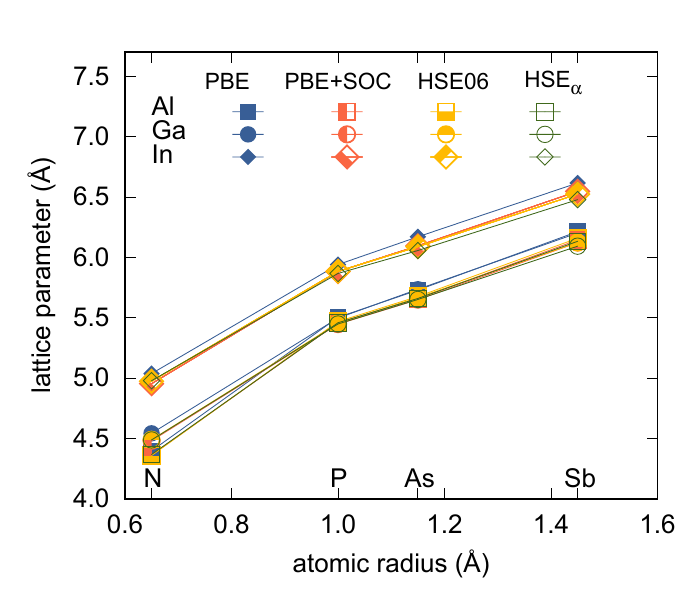}
\caption{Lattice parameters versus the atomic radius of the anion species. The 
lattice parameters were obtained using the following exchange-correlation functionals: 
PBE, PBE+SOC, HSE06, HSE$_\alpha$. The atomic radii were extracted from Ref. \onlinecite{Slater_3199_1964}.}
\label{fig:lat_rad}
\end{figure}

The lattice parameter have a slight linear dependence with the atomic radius of 
the material compound elements (occupation of the electronic shells), i.e., if 
the anionic atom size increases, the lattice parameter also increases, as showed 
in Fig. \ref{fig:lat_rad}. This effect occurs due to the electrostatic repulsion 
between atoms, i.e., the bond length depends on the atom size, changing the lattice 
parameter. As \ce{Al} and \ce{Ga} have similar atomic radii, with \SI{1.25}{\angstrom} 
and \SI{1.30}{\angstrom}, respectively,\cite{Slater_3199_1964} the lattice parameters 
for the \ce{Al}-V compounds are closer to the \ce{Ga}-V values, as showed in the 
figure. The \ce{In} has a greater atomic radius, \SI{1.55}{\angstrom},\cite{Slater_3199_1964} 
resulting in a larger lattice parameter for the \ce{In}-V compounds. In order to 
investigate the linearity break shown for the III-\ce{N} compounds, we evaluated 
the effective Bader charge, as shown in the table \ref{tab:EffectiveBader} in appendix 
\ref{Appendix:BaderCharge}. The III-\ce{N} compounds have a large charge transfer, 
which suggests that the break in the linearity behavior is due to the high ionicity 
combined with the smaller atomic radius of the \ce{N} when when compared with the 
other cations of the III-A column of the Periodic Table. This hypothesis is also 
supported by the fact that the III-\ce{N} compounds showed the highest elastic 
constants values in the series, as will be discussed in the next section, and the 
larger observed bond strengths.\cite{Nakamura_2000}

\subsection{Elastic constants}\label{subsection:elastic_constants}

\begin{table*}

\caption{Elastic constants calculated with PBE and $\alpha$-optimized HSE (HSE$_{\alpha}$), 
functionals. The bulk moduli, $B_0$, were obtained from the elastic constants results, 
namely, $B_0 = (C_{11}+2 C_{12})/3$, and are compared with the experimental values 
obtained from elastic constants (Exp.$^{C_{ij}}$) and direct measures (Exp.$^{B_0}$). 
All constants are given in \si{\giga\pascal}.}
\begin{threeparttable}
\addtolength{\tabcolsep}{2pt}  
\bgroup
\def\arraystretch{1.2}
\begin{tabular}{lcccccccccccccccc} \hline\hline
	& & $C_{11}$ &  &  && $C_{12}$ &  && & $C_{44}$ & & && &$B_0$ & \\ \cline{2-4}\cline{6-8}\cline{10-12} \cline{14-17}
	& PBE & HSE$_{\alpha}$ & Exp. && PBE & HSE$_{\alpha}$ & Exp. && PBE & HSE$_{\alpha}$ & Exp. && PBE & HSE$_{\alpha}$ & Exp.$^{C_{ij}}$ & Exp.$^{B_0}$ \\ \hline
\ce{AlN}           & \num{283.1} & \num{308.1} & -                   &&  \num{149.8} &  \num{161.6}  & -                  && \num{179.2} & \num{197.2} &-                   && \num{194.2} & \num{210.5} & -  		   & -  \\
\ce{AlP}           & \num{125.8} & \num{133.6} & \num{141}\tnote{a}  &&  \num{ 61.3} &  \num{ 65.1}  & \num{ 62}\tnote{a} && \num{ 60.8} & \num{ 64.1} & \num{ 70}\tnote{a} && \num{ 82.8} & \num{ 87.9} & \num{ 88.3} & -  \\
\ce{AlAs}          & \num{103.9} & \num{122.1} & \num{122}\tnote{a}  &&  \num{ 49.1} &  \num{ 58.5}  & \num{ 57}\tnote{a} && \num{ 51.1} & \num{ 59.0} & \num{ 60}\tnote{a} && \num{ 67.4} & \num{ 79.7} & \num{ 78.7} & 74\tnote{c}   \\
\ce{AlSb}          & \num{77.0 } & \num{ 93.7} & \num{ 88}\tnote{a}  &&  \num{ 35.8} &  \num{ 43.2}  & \num{ 40}\tnote{a} && \num{ 36.8} & \num{ 47.5} & \num{ 43}\tnote{a} && \num{ 49.5} & \num{ 60.0} & \num{ 56.0} & 55.1 \tnote{d} \\
\ce{GaN}           & \num{253.3} & \num{290.0} & -                   &&  \num{125.2} &  \num{149.0}  & -                  && \num{146.4} & \num{173.9} & -                  && \num{167.9} & \num{196.0} & - 		   & -  \\   
\ce{GaP}           & \num{124.6} & \num{150.0} & \num{140}\tnote{a}  &&  \num{ 56.0} &  \num{ 64.3}  & \num{ 62}\tnote{a} && \num{ 65.2} & \num{ 78.4} & \num{ 70}\tnote{a} && \num{ 78.9} & \num{ 92.9} & \num{ 88.0} & -  \\
\ce{GaAs}          & \num{98.1 } & \num{122.0} & \num{119}\tnote{a}  &&  \num{ 42.1} &  \num{ 49.4}  & \num{ 53}\tnote{a} && \num{ 50.8} & \num{ 65.3} & \num{ 60}\tnote{a} && \num{ 60.8} & \num{ 73.6} & \num{ 75.0} & - \\
\ce{GaSb}          & \num{74.6 } & \num{90.1 } & \num{ 88}\tnote{a}  &&  \num{ 32.0} &  \num{ 35.7}  & \num{ 40}\tnote{a} && \num{ 35.9} & \num{49.0 } & \num{ 43}\tnote{a} && \num{ 46.2} & \num{ 53.8} & \num{ 56.0} & -  \\
\ce{InN}           & \num{159.3} & \num{188.7} &-				     &&  \num{102.0} &  \num{125.8}  & -                  && \num{ 78.9} & \num{93.3 } & -					&& \num{121.1} & \num{146.8} & -		   & -  \\
\ce{InP}           & \num{87.4 } & \num{105.9} & \num{101}\tnote{b}  &&  \num{ 45.9} &  \num{ 56.4}  & \num{ 56}\tnote{b} && \num{ 41.9} & \num{49.3 } & \num{ 46}\tnote{b} && \num{ 59.7} & \num{ 72.9} & \num{ 71.0} & -  \\
\ce{InAs}          & \num{70.9 } & \num{ 91.5} & \num{ 83}\tnote{a}  &&  \num{ 37.8} &  \num{ 48.8}  & \num{ 45}\tnote{a} && \num{ 33.1} & \num{42.4 } & \num{ 40}\tnote{a} && \num{ 48.8} & \num{ 62.7} & \num{ 57.6} & 58\tnote{e}  \\
\ce{InSb}          & \num{55.3 } & \num{ 74.1} & \num{ 69}\tnote{a}  &&  \num{ 29.1} &  \num{ 34.8}  & \num{ 37}\tnote{a} && \num{ 25.4} & \num{39.0 } & \num{ 31}\tnote{a} && \num{ 37.8} & \num{ 47.9} & \num{ 47.7} & -  \\ \hline
NRMSD ( \%)        & \num{14.3 } & \num{4.9 }  &  -                  &&  \num{ 15.0} &  \num{  5.1}  &   -                && \num{13.8}  & \num{9.2 } & -                 &&  \num{14.4}\tnote{$\dagger$}  & \num{3.7}\tnote{$\dagger$}   &  -          & - \\  \hline \hline
\end{tabular} 
\egroup
\addtolength{\tabcolsep}{7pt}  
\begin{tablenotes}
\footnotesize
\item[a] Ultrasound Ref. \onlinecite{Martienssen_2005},
\item[b] Ultrasonic-wave transit times Ref. \onlinecite{Martienssen_2005},
\item[c] X-ray diffraction data from Ref. \onlinecite{Madelung_2004},
\item[c] Energy dispersive X-ray from Ref. \onlinecite{Madelung_2004}
\item[d] Ultrasound from Ref. \onlinecite{Madelung_2004}
\item[$\dagger$] Comparison with Exp.$^{C_{ij}}$
\end{tablenotes}
\end{threeparttable}
\label{tab:ElasticConstants}
\end{table*}

The cubic zinc-blende crystal structure has the symmetry defined by the space group 
$\text{F}\bar{4}3\text{m}$ that is associated to the point group $T_d$.\cite{Yu_1999} 
The symmetry analysis show that it possess only three non-equivalent elastic constants: 
$C_{11}$, $C_{12}$ and $C_{44}$. $C_{11}$ represents the modulus for the axial 
compression, i.e., the stress in one direction induces a strain in the same direction. 
In contrast, $C_{12}$ represents the stress that induces a strain in the perpendicular 
directions and $C_{44}$, the shear modulus, represents the strain across the faces 
induced by the stress in a direction parallel to it. PBE, HSE$_{\alpha}$ and the 
respective experimental results of $C_{11}$, $C_{12}$ and $C_{44}$ are shown in 
Table \ref{tab:ElasticConstants}. We also present the bulk modulus, $B_0$, which 
were calculated from the expression $B_0 = (C_{11} + 2C_{12})/3$. 

\begin{figure}
\centering
\includegraphics[width=0.90\linewidth]{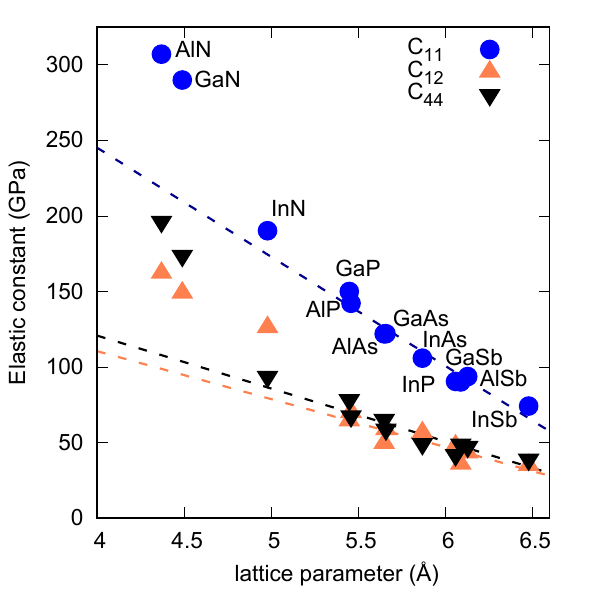}
\caption{Dependence of elastic constants on the equilibrium lattice parameter. 
Elastic constants and lattice parameters have been obtained using $\text{HSE}_{\alpha}$ 
functional.}
\label{fig:elasticConstants}
\end{figure}

Unrelated to the exchange and correlation functionals, the elastic constants in 
all directions decrease as the ionic radius increase. For the PBE functional, e.g., 
$C_{11}$ decreases from \num{283.1} in \ce{AlN} to \num{77} in \ce{AlSb}. The ionic 
bond character is responsible for the increase on the hardness of the material, 
and as shown in the table \ref{tab:EffectiveBader} in appendix \ref{Appendix:BaderCharge}, 
the ionicity (related with the Bader charge) decreases as the anion radius decreases. 
Therefore, it is expected that the elastic constant decreases as the lattice parameter 
(associated with the cation and anion radii) of the crystal structure increases. 
In fact, our results indicate a slight linear dependence with the lattice parameter, 
as showed in Fig. \ref{fig:elasticConstants}, where the dashed line shows a linear 
fitting using the all materials, nitrides excluded. This behavior was reported 
in the literature\cite{Adachi_1992,Keyes_3371_1962} and was traditionally used 
to estimate the elastic constants\cite{Willardson_1975,Adachi_R1_1985} by the 
extrapolation of the data.

In contrast with the lattice parameter overestimation by HSE functionals, the PBE 
functional underestimates the elastic constants in all the directions, which is 
consistent with the literature.\cite{Lopuszynski_045202_2007} On the other hand, 
the HSE$_{\alpha}$ results show better agreement with the experimental results, 
presenting NRMSDs of \SI{4.9}{\percent} and \SI{5.1}{\percent} for $C_{11}$ and $C_{12}$, 
respectively. The inclusion of nonlocal effects in the Fock exchange in HSE$_{\alpha}$ 
suggests an increasing of the bond hardness, consequently increasing the elastic 
constants when compared to the PBE functional. For $C_{44}$, our results when compared 
with the experimental data, show deviations for the HSE$_{\alpha}$ functional similar 
to the PBE ones, presenting NMRSDs of \SI{13.8}{\percent} and \SI{9.2}{\percent}, 
respectively. Similar $C_{44}$ values have been found by Caro \textit{et al.}\cite{Caro_014117_2012} 
using the HSE06 for nitrides. Nonetheless, PBE underestimates the experimental 
values of $C_{11}$ and $C_{12}$, while HSE$_{\alpha}$ overestimates them. Since 
the bulk modulus, $B_0$, in a cubic system has dependence only in the $C_{11}$ 
and $C_{12}$ elastic constant directions, the PBE functional also underestimates 
the $B_0$ values, while the HSE$_{\alpha}$ functional yields better results. This 
can be observed by the NRMSD which is \SI{3.7}{\percent} and \SI{14.4}{\percent} 
for HSE$_{\alpha}$ and PBE, respectively. 

\subsection{Band Structures}\label{subsection:BandStructure}

\begin{figure}
\centering
\includegraphics[width=0.90\linewidth]{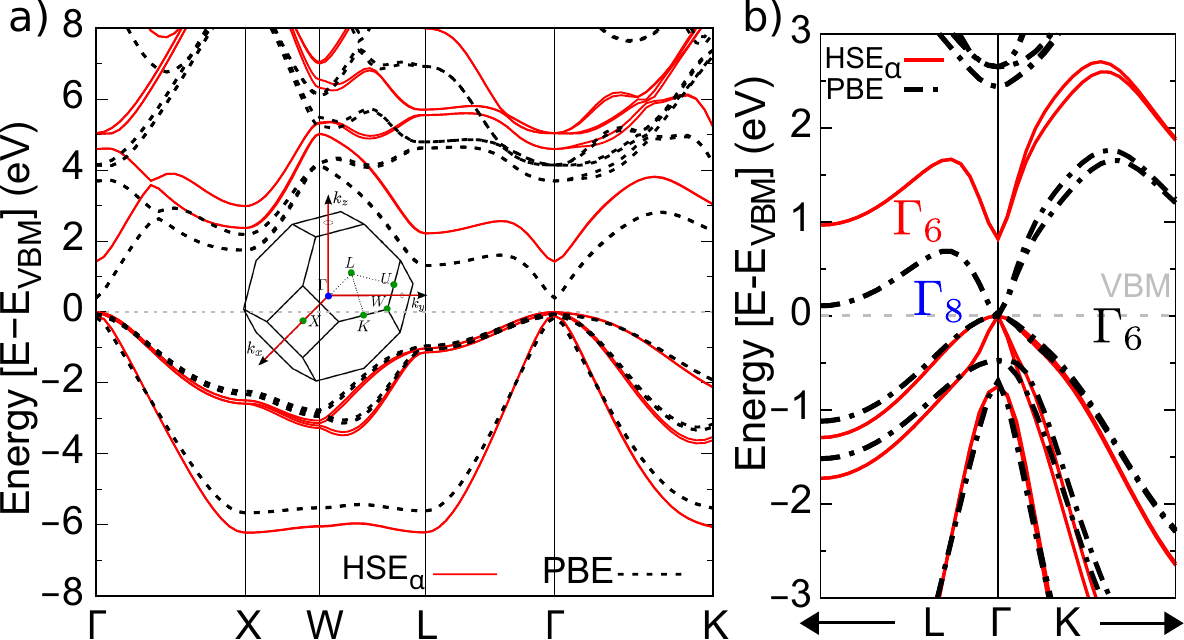}
\caption{\ce{InP} and \ce{GaSb} band structures evaluated with PBE (dashed line) 
and $\text{HSE}_{\alpha}$ (solid line) including spin-orbit coupling. a) band structure 
for \ce{InP}. The first Brillouin zone of zinc-blende phase is shown. b) \ce{GaSb} 
band structure close to $\Gamma$ point: PBE presents a negative band gap and the 
$\text{HSE}_{\alpha}$ shows a positive band gap. The band symmetries are indicated.}
\label{fig:PBEbandStructureandZBstructure}
\end{figure}

As discussed previously,\cite{Perdew_1884_1983,Kim_205212_2010} the PBE functional 
strongly underestimates the band gap energy. In the specific case of the small 
band gap III-V materials, leading to results presenting a null band gap or even 
the inversion of the ordering of $\Gamma_6$ and $\Gamma_8$ states, as showed in 
Fig. \ref{fig:PBEbandStructureandZBstructure}b. This result is completely inconsistent 
with the experimental data.\cite{Madelung_2004,Vurgaftman_5815_2001} To get rid 
of this problem, we employed the HSE06 and $\text{HSE}_{\alpha}$ exchange and correlation 
functions. The differences on the results using the different functionals may be 
clarified by analyzing the results as shown in Figure \ref{fig:PBEbandStructureandZBstructure}. 
The HSE06 and HSE$_{\alpha}$ gaps are closer to the experimental values than PBE 
predictions, e.g., \ce{InP} shows an increase of the value of \SI{68}{\percent} 
from PBE to HSE$_{\alpha}$. An even more dramatic example is the wrong predictions 
of negative band gaps for \ce{InSb}, \ce{GaSb}, \ce{InAs} and \ce{InN} made using 
the PBE functional. HSE06 or HSE$_{\alpha}$ functionals show the correct trend.

Although both HSE06 or HSE$_{\alpha}$ functionals predict the correct trend, the 
tuning of $\alpha$ provides a much better agreement for the band gap value, changing 
the deviations from the experimental results from \SI{0.8}{\percent} (\ce{InP}) 
up to \SI{42.1}{\percent} (\ce{InSb}) when using HSE06 to \SI{0.3}{\percent} (\ce{AlN}) 
and \SI{10.5}{\percent} (\ce{InN}) with HSE$_{\alpha}$. The accuracy in the description 
of $\Delta_{\text{so}}$ is also improved when using the hybrid functionals instead 
of PBE, as shown in Table \ref{Table:gapsAndspinOrbit}. The band structures for 
all the other materials are presented in the Supplementary Material. 

\begin{table*}
\caption{Band gap, $E_\text{gap}=\Gamma^{c}_{6}-\Gamma^{v}_{8}$, and spin-orbit 
splitting, $\Delta_\text{so}=\Gamma^{v}_{8}-\Gamma^{v}_{7}$, energies using the 
different functionals: PBE, PBE+SOC, HSE06, HSE06+SOC and $\text{HSE}_{\alpha}$+SOC. 
The contribution of nonlocal exchange in the HSE, $\alpha$, adjusted to obtain 
the experimental values of gap energy and spin-orbit splitting is also shown. The 
energies are given in \si{\electronvolt}.}
\addtolength{\tabcolsep}{1pt}  
\bgroup
\def\arraystretch{1.1}
\begin{threeparttable}
\begin{tabular}{lccrccrccccccccccc}
	\hline \hline
	&&&PBE &&\multicolumn{2}{c}{PBE+SOC} & & \multicolumn{1}{c}{HSE06}& & \multicolumn{2}{c}{HSE06+SOC} && 	\multicolumn{2}{c}{$\text{HSE}_{\alpha}$+SOC} & &\multicolumn{2}{c}{literature} \\
	\cline{4-4}  \cline{6-7} \cline{9-9} \cline{11-12} \cline{14-15} \cline{17-18} 	&  $\alpha$  & & $E_{\text{gap}}$  & & $E_{\text{gap}}$ & 	$\Delta_{\text{so}}$ & & $E_{\text{gap}}$ & & $E_{\text{gap}}$ & 	$\Delta_{\text{so}}$&& $E_{\text{gap}}$ & $\Delta_{\text{so}}$&& 	$E_{\text{gap}}$ & $\Delta_{\text{so}}$\\
	\hline
	\ce{AlN}  & \num{0.219} && \num{4.003}  && \num{3.997} & \num{0.019} && \num{5.609}  && \num{5.601} & \num{0.021} 	&& \num{5.383} & \num{0.022} && \num{5.40 }\tnote{f} &\num{0.019}\tnote{f} \\
	\ce{AlP}  & \num{0.127} && \num{3.090}  && \num{3.070} & \num{0.059} && \num{4.164}  && \num{4.153} & \num{0.064} 	&& \num{3.611} & \num{0.061} && \num{3.62 }\tnote{c} &\num{0.06 }\tnote{b} \\
	\ce{AlAs} & \num{0.343} && \num{1.757}  && \num{1.662} & \num{0.290} && \num{2.819}  && \num{2.732} & \num{0.316} 	&& \num{3.157} & \num{0.324} && \num{3.13 }\tnote{c} &\num{0.3  }\tnote{c} \\
	\ce{AlSb} & \num{0.318} && \num{1.314}  && \num{1.105} & \num{0.652} && \num{2.286}  && \num{2.111} & \num{0.691} 	&& \num{2.397} & \num{0.700} && \num{2.38 }\tnote{c} &\num{0.673}\tnote{c} \\
	\ce{GaN}  & \num{0.293} && \num{1.564}  && \num{1.560} & \num{0.012} && \num{3.043}  && \num{3.042} & \num{0.021} 	&& \num{3.312} & \num{0.022} && \num{3.30 }\tnote{d} &\num{0.017}\tnote{f} \\
	\ce{GaP}  & \num{0.283} && \num{1.603}  && \num{1.576} & \num{0.082} && \num{2.748}  && \num{2.739} & \num{0.092} 	&& \num{2.915} & \num{0.093} && \num{2.895}\tnote{c} &\num{0.08 }\tnote{f} \\
	\ce{GaAs} & \num{0.318} && \num{0.166}  && \num{0.072} & \num{0.325} && \num{1.297}  && \num{1.210} & \num{0.358} 	&& \num{1.471} & \num{0.365} && \num{1.519}\tnote{c} &\num{0.341}\tnote{c} \\
	\ce{GaSb} & \num{0.297} && \num{-0.259} &&\num{-0.477} & \num{0.694} && \num{0.782}  && \num{0.614} & \num{0.743} 	&& \num{0.819} & \num{0.751} && \num{0.82 }\tnote{c} &\num{0.756}\tnote{c} \\
	\ce{InN}  & \num{0.274} && \num{-0.504} &&\num{-0.497} & \num{0.002} && \num{0.507}  && \num{0.530} & \num{0.016} 	&& \num{0.674} & \num{0.017} && \num{0.61 }\tnote{g} &\num{0.005}\tnote{f} \\
	\ce{InP}  & \num{0.253} && \num{ 0.468} &&\num{ 0.452} & \num{0.095} && \num{1.402}  && \num{1.408} & \num{0.111} 	&& \num{1.422} & \num{0.111} && \num{1.42 }\tnote{c} &\num{0.108}\tnote{c} \\
	\ce{InAs} & \num{0.289} && \num{-0.525} &&\num{-0.626} & \num{0.335} && \num{0.372}  && \num{0.301} & \num{0.373} 	&& \num{0.373} & \num{0.385} && \num{0.418}\tnote{c} &\num{0.38 }\tnote{c} \\
	\ce{InSb} & \num{0.277} && \num{-0.558} &&\num{-0.782} & \num{0.716} && \num{0.335}  && \num{0.136} & \num{0.778} 	&& \num{0.232} & \num{0.783} && \num{0.235}\tnote{c} &\num{0.81 }\tnote{c} \\
	\hline
	\multicolumn{2}{l}{NRMSD(\si{\percent})} 		&&\num{55.7}   	&&\num{59.1} & \num{12.0} && \num{9.1} && \num{10.9} &\num{4.8} 	&&\num{1.3} &\num{5.4} &&  &  \\
	\hline \hline
\end{tabular} 
\begin{tablenotes}
\item[a] Exp. from ref. \onlinecite{Thompson_3331_2001},
\item[b] Theory GW from ref. \onlinecite{Chantis_086405_2006},
\item[c] Exp. from ref. \onlinecite{Madelung_2004},
\item[d] Exp. from ref. \onlinecite{Monemar_676_1974},
\item[e] Exp. from ref.  \onlinecite{Walukiewicz_119_2004},
\item[f] Theory from ref. \onlinecite{Vurgaftman_5815_2001}.
\item[g] Exp. from ref. \onlinecite{Buss_225701_2015}.
\end{tablenotes}
\end{threeparttable}
\egroup
\addtolength{\tabcolsep}{1pt} 
\label{Table:gapsAndspinOrbit}
\end{table*}

\begin{figure}
\centering
\includegraphics[width=0.90\linewidth]{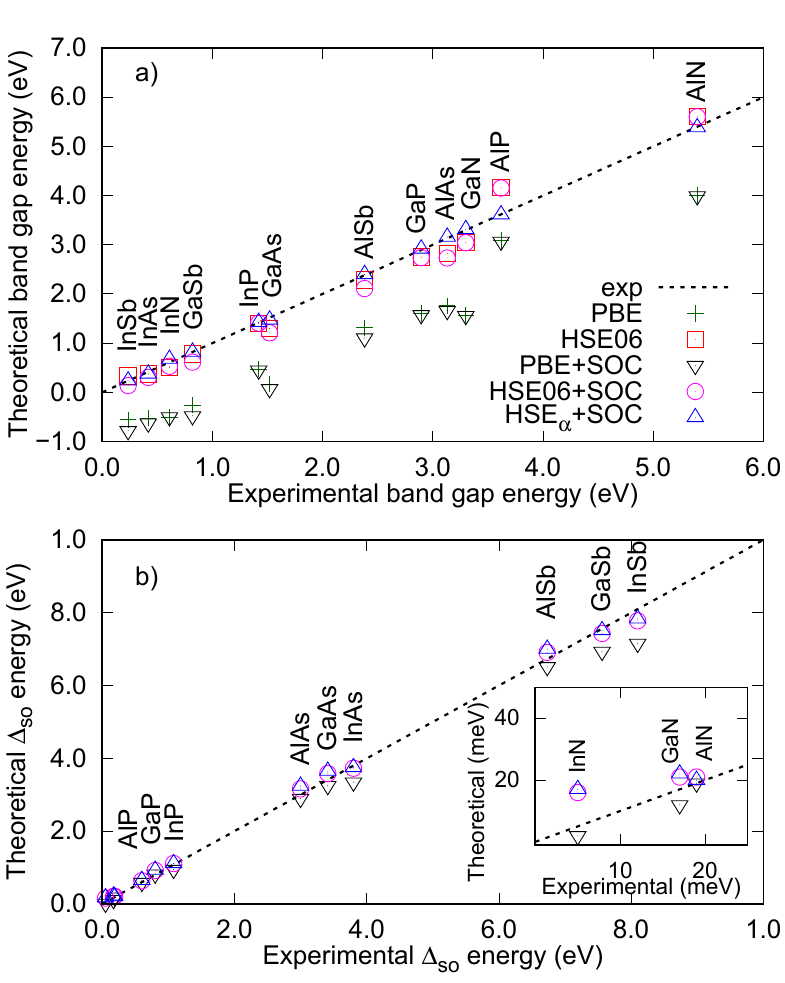}
\caption{Energies of the \num{12} compounds determined with the different functionals 
with and without the inclusion of the spin-orbit coupling: a) band gaps b) spin-orbit 
splittings. In both figures, the dashed lines show the literature values.}
\label{fig:gapAndSO}
\end{figure}

Regardless of the gap adjustment, i.e., using the HSE06 calculations, one can notice 
a monotonic relation between the anionic radius and the band gap energies, e.g., 
in \ce{Al} compounds, we observe that $E_{\text{gap}}^{\ce{AlN}} > E_{\text{gap}}^{\ce{AlP}} 
> E_{\text{gap}}^{\ce{AlAs}} > E_{\text{gap}}^{\ce{AlSb}}$. This trend is also 
valid for the \ce{As} compounds, as shown in Fig. \ref{fig:gapAndSO}. It fails, 
however, for the \ce{In} compounds, in which the calculated \ce{InN} band gap is 
smaller than the trend suggests. Carrier \textit{et al.}\cite{Carrier_33707_2005} 
suggested, when this same rule was violated for wurtzite compounds, that this was 
due to the high electronegativity of \ce{N} and the smaller band gap deformation 
potentials. In our understanding, the same explanation should be applied to the 
zinc-blende case.  

\begin{figure*}
\centering
\includegraphics[width=0.90\textwidth]{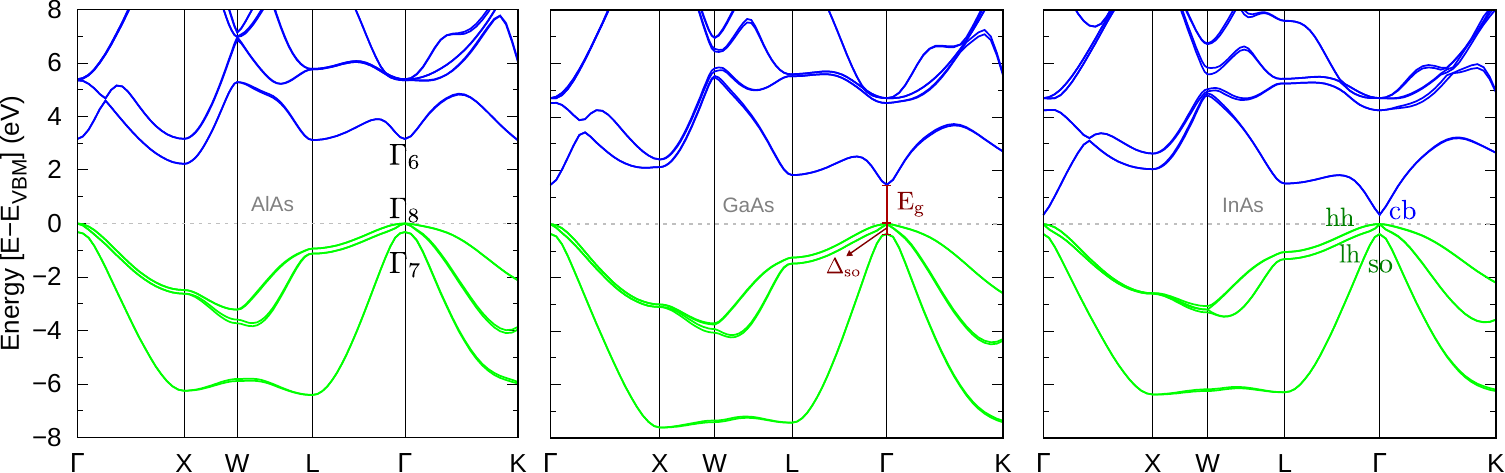}
\caption{Band structures obtained using $\text{HSE}_{\alpha}$+SOC along high symmetry 
lines for \ce{AlAs}, \ce{GaAs} and \ce{InAs}. The blue lines indicate conduction 
bands, while the green lines, the valence bands. The band structures for the other 
III-V materials are shown in the Supplementary Materials.}
\label{fig:bandStructure}
\end{figure*}

\subsection{$\mathbf{k{\cdot}p}$ Parameters}\label{subsection:k.pParameters}

Despite the fact that band gaps and $\Delta_{\text{so}}$ are  close to the experimental 
values, there is no guarantee that the calculated band structures are in agreement 
with the experimental results. To perform this analysis, we calculated the effective 
masses, using the $\mathbf{k{\cdot}p}$ approach. In the $\mathbf{k{\cdot}p}$ method, 
the interactions involving electrons and nuclei are described through an effective 
potential with the same periodicity of the lattice, allowing the utilization of 
the Bloch's theorem. To found an effective Hamiltonian, we used the perturbative 
technique proposed by L\"owdin,\cite{Lowdin_1396_1951} where the basis set is divided 
in two classes: A and B. States within the class A, are the states of interest 
and are described exactly, while states from the class B are taken into account 
perturbativelly through the interactions with the states of class A. Class A states 
are chosen according to the energy range at the point of the first Brillouin zone 
(FBZ) in aim, defining the effective Hamiltonian.

In this work, we used the $6{\times}6$ zinc-blende effective $\mathbf{k{\cdot}p}$ 
Hamiltonians proposed by Luttinger-Kohn \cite{Luttinger_869_1955} (LK6) and extended 
by Kane to an $8{\times}8$ model.\cite{Kane_75_1966} In the LK6 model, in the vicinity 
of the $\Gamma$ point, the class A is composed by the three topmost valence bands, 
i.e., HH, LH and SO, and the matrix elements are determined using perturbation 
theory up to the second order. In the Kane model, the same bands and perturbative 
order were used but class A also includes the first conduction band (CB). The use 
the symmetry properties of zinc-blende crystals and some algebraic manipulation\cite{Willatzen_2009,Enderlein_1997} 
shows that the Kane Hamiltonian depend on \num{5} different effective mass parameters 
$\tilde{\gamma}_1$, $\tilde{\gamma}_2$, $\tilde{\gamma}_3$, $\tilde{\text{e}}$ 
and $P$, plus the gap and $\Delta_{\text{so}}$, while the LK6 depends only on \num{3} 
parameters, $\gamma_1$, ${\gamma}_2$ and ${\gamma}_3$, plus the $\Delta_{\text{so}}$.\footnote{We 
denoted the $6{\times}6$ parameters, the so called Luttinger-Kohn parameters without 
tilde. In opposition, the $8\times8$ ones are referred as Kane parameters.} As 
the $\mathbf{k{\cdot}p}$ method is semi-empirical, all effective mass parameters 
are obtained, with little algebraic manipulations, from the direct measurements 
of the effective masses of the carriers in the materials, except for the $P$ parameter.

Distinctly from the other effective mass parameters, $P$ can not be obtained by 
direct measures, but must be extracted from the interband (CB-VB) interaction energy 
$E_P$. An accurate measure of $E_P$ is hard to obtain due to the inherent difficulties 
associated with the decoupling of the CB-VB interaction to the interaction of them 
with the remote bands. $E_P$ values have been estimated indirectly by experimental 
techniques such as electron-spin-resonance, through interband matrix elements.\cite{Weisbuch_816_1977,Chadi_4466_1976} 
and from measures of the $g$-factors, which have small influence from the remote 
bands and yield more accurate values.\cite{Vurgaftman_5815_2001} Due to the difficulties 
involved in the measuring the $g$-factor in III-V semiconductors, the traditional 
procedure is to obtain the $P$ parameter from the effective mass parameters using 
the $6{\times}6$ $\mathbf{k{\cdot}p}$ Hamiltonian. When the $P$ is determined, the 
$8{\times}8$ parameters can be evaluated using the relations showed in appendix 
\ref{Appendix:LuttingerParameters}.

We chose in this work an alternative method to determine the effective mass and 
$P$ parameters from band structures evaluated by DFT calculations. We fitted the 
HSE$_{\alpha}$ band structure using the secular equation of the $8{\times}8$ Hamiltonian 
proposed in the Ref. \onlinecite{Bastos_105002_2016}, determining simultaneously 
all the parameters. All points have the same weight and the same distance for all 
materials. Using different percentages of the FBZ around $\Gamma$ point, we determined  
different parameter sets and the choice of the final set of parameters was done 
by root mean square deviation (RMSD) analysis,\cite{Bastos_105002_2016} using the 
euclidean distance between the band structures from HSE$_{\alpha}$ and the effective 
Hamiltonian $\mathbf{k{\cdot}p}$ with the adjusted parameters. Technical details 
about the fitting are available in the Supplementary Material.

\begin{figure*}
\centering
\includegraphics[width=0.90\linewidth]{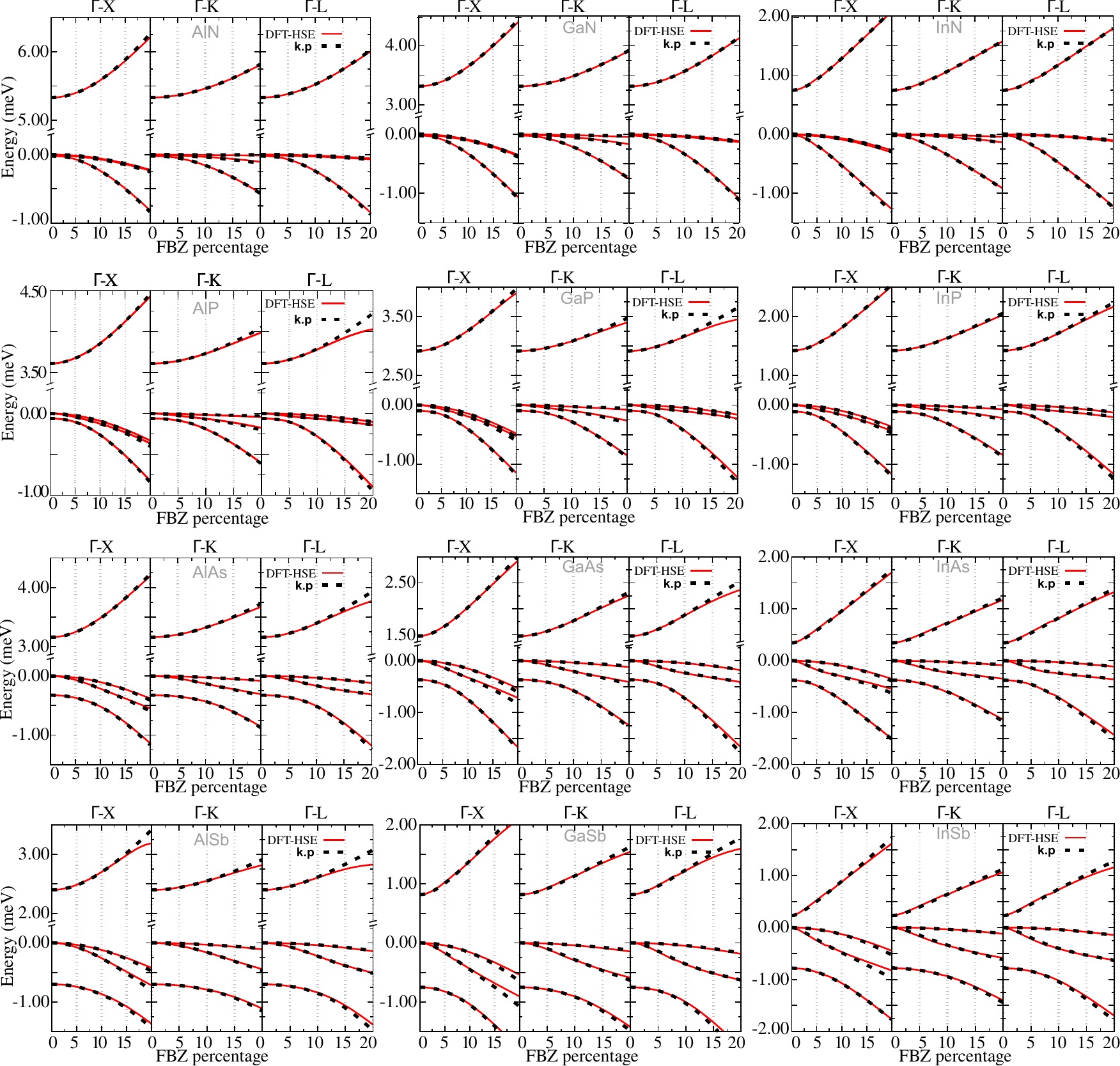}
\caption{HSE$_{\alpha}$ and $\mathbf{k{\cdot}p}$ band structure comparisons. The 
optimal parameters were obtained from fitting using \SI{10}{\percent} of the FBZ. 
Three high symmetry directions are shown: $\Gamma$-X, $\Gamma$-K and $\Gamma$-L.}
\label{fig:bandStructureFitting}
\end{figure*}

As the difference between $\mathbf{k{\cdot}p}$ and HSE$_{\alpha}$ increases considerably 
for large FBZ percentages, we recommend the values from the fitting for \SI{10}{\percent} 
as showed in Table IV. This choice is a compromise between describing a reasonable 
percentage of the band and obtaining a small deviation of HSE$_{\alpha}$ band structure. 
A general feature of the band structures is the non-parabolicity in the region 
between \num{6} and \SI{8}{\percent} of the FBZ, better seen on CB and SO bands, 
as showed in Fig. \ref{fig:bandStructureFitting}. Another non-parabolicity also 
arises near \SI{15}{\percent}. Depending on the magnitude of this second non-parabolicity 
in an specific direction, the deviations of the values become more or less important 
at \SI{20}{\percent} of the FBZ.

\begin{table*}
\caption{Kane and Luttinger--Kohn parameters obtained through the fitting of the 
Kane $\mathbf{k{\cdot}p}$ Hamiltonian in the band structure obtained using $\text{HSE}_{\alpha}$+SOC 
close to $\Gamma$ point. A range of Luttinger-Kohn parameters found in the compilation 
of the literature in references \onlinecite{Madelung_1982,Vurgaftman_5815_2001,Vurgaftman_3675_2003} 
is given for comparison. The $\tilde{\gamma}_s$, ${\gamma}_s$, \~e and e values 
are in $\hbar^2/2m_0$ units, while $\text{E}_{\text{p}}$ are in \si{\electronvolt}.}
\addtolength{\tabcolsep}{.8pt}  
\bgroup
\def\arraystretch{1.2}
\begin{tabular}{lccccccccccccccccc}
	\hline \hline
	&\multicolumn{5}{c}{Kane }
	&  \multicolumn{4}{c}{Luttinger-Kohn } 
	&&\multicolumn{6}{c}{range in literature for Luttinger-Kohn parameters}\\ 
	\cline{2-6}\cline{8-11}\cline{13-17}
	& $\tilde{\gamma}_1$ & $\tilde{\gamma}_2$ & $\tilde{\gamma}_3$ & 
	$\tilde{\text{e}}$ &$E_{P}$ &&
	$\gamma_1$ & $\gamma_2$ & 	$\gamma_3$ & e   && 
	$\gamma_1$ & $\gamma_2$ & $\gamma_3$ & e & $E_{P}$ \\
	\hline
	\ce{AlN}      &  \num{0.36}  &  \num{-0.20}  &  \num{ 0.06}  &  \num{-0.10 } & \num{18.8}  && \num{ 1.52}  & \num{ 0.38}  & \num{ 0.64} &  \num{ 3.38}    &&  	\num{1.92 }            & \num{0.47}               &   \num{0.85}             & \num{ 3.03}-\num{4.0  }  & \num{27.1} \\
	\ce{AlP}      &  \num{0.81}  &  \num{-0.48 } &  \num{0.09 } &  \num{-1.01 } & \num{22.1}  && \num{ 2.85}  & \num{ 0.54}  & \num{ 1.11} &  \num{ 5.07}    &&  	\num{3.35 }-\num{3.47} & \num{0.06}-\num{0.71}    &   \num{1.15}-\num{1.19}  & \num{ 4.55} 			    & \num{17.7} \\
	\ce{AlAs}     &  \num{0.95}  &  \num{-0.61 } &  \num{0.06 } &  \num{-1.26 } & \num{27.2}  && \num{ 3.79}  & \num{ 0.92}  & \num{ 1.50} &  \num{ 6.99}    &&  	\num{3.25 }-\num{4.04} & \num{0.65}-\num{0.9 }    &   \num{1.21}-\num{1.38}  & \num{ 6.67}              & \num{21.1} \\
	\ce{AlSb}     &  \num{0.89}  &  \num{-1.05 } &  \num{-0.11 }&  \num{-3.66 } & \num{27.9}  && \num{ 5.02}  & \num{ 1.15}  & \num{ 1.95} &  \num{ 7.88}    &&  	\num{4.15 }-\num{5.18} & \num{1.01}-\num{1.19}    &   \num{1.71}-\num{1.81}  & \num{ 3.03}-\num{8.33 }  & \num{18.7} \\
	\ce{GaN}      &  \num{0.68}  &  \num{-0.28 } &  \num{0.08 } &  \num{-0.27 } & \num{16.4}  && \num{ 2.39}  & \num{ 0.60}  & \num{ 0.94} &  \num{ 4.87}    &&  	\num{2.67 }-\num{3.07} & \num{0.75}-\num{0.86}    &   \num{1.1 }-\num{1.16}  & \num{ 6.67}              & \num{25  } \\
	\ce{GaP}      &  \num{1.38}  &  \num{-0.62 } &  \num{0.16 } &  \num{-1.67 } & \num{25.2}  && \num{ 4.20}  & \num{ 0.87}  & \num{ 1.58} &  \num{ 6.68}    &&  	\num{4.05 }-\num{4.2 } & \num{0.49}-\num{0.98}    &   \num{1.25}-\num{1.95}  & \num{ 7.69}-\num{10.81}  & \num{31.4} \\
	\ce{GaAs}     &  \num{1.37}  &  \num{-0.81 } &  \num{0.10 } &  \num{-2.02 } & \num{25.9}  && \num{ 7.10}  & \num{ 2.15}  & \num{ 2.99} &  \num{14.05}    &&  	\num{6.8  }-\num{7.8 } & \num{2.02}-\num{2.50}    &   \num{1.0 }-\num{2.43}  & \num{14.93}-\num{15.43}  & \num{25.9}-\num{27.6}  \\
	\ce{GaSb}     &  \num{1.74}  &  \num{-1.15 } &  \num{0.15 } &  \num{-3.23 } & \num{24.8}  && \num{11.78}  & \num{ 3.87}  & \num{ 5.19} &  \num{22.04}    &&  	\num{11.08}-\num{13.4} & \num{4.03}-\num{4.7 }    &   \num{5.26}-\num{5.74}  & \num{24.27}-\num{25.64}  & \num{23.7}-\num{25.1}  \\
	\ce{InN}      &  \num{0.65}  &  \num{-0.25 } &  \num{0.05 } &  \num{-0.16 } & \num{11.1}  && \num{ 6.13}  & \num{ 2.49}  & \num{ 2.79} &  \num{16.15}    &&  	\num{3.72 }            & \num{1.26}               &   \num{1.63}             & \num{ 8.33}-\num{14.29}  & \num{17.2}-\num{21.1}  \\
	\ce{InP}      &  \num{1.23}  &  \num{-0.54 } &  \num{0.14 } &  \num{-1.18 } & \num{18.3}  && \num{ 5.33}  & \num{ 1.58}  & \num{ 2.20} &  \num{10.86}    &&  	\num{4.95 }-\num{6.28} & \num{0.94}-\num{2.08}    &   \num{1.62}-\num{2.08}  & \num{12.38}-\num{14.71}  & \num{18.1}-\num{19.6}  \\
	\ce{InAs}     &  \num{1.29}  &  \num{-0.77 } &  \num{0.10 } &  \num{-1.38 } & \num{18.9}  && \num{18.20}  & \num{ 7.69}  & \num{ 8.55} &  \num{40.75}    &&  	\num{19.67}-\num{20.5} & \num{8.30}-\num{8.50}    &   \num{9.10}-\num{9.17}  & \num{37.74}-\num{45.66}  & \num{21.5}-\num{21.9}  \\
	\ce{InSb}     &  \num{1.68}  &  \num{-1.04 } &  \num{0.12 } &  \num{-2.09 } & \num{20.4}  && \num{31.05}  & \num{13.65}  & \num{14.80} &  \num{63.37}    &&  	\num{34.5 }-\num{37.1} & \num{14.5}-\num{16.5}    &   \num{15.7}-\num{17.7}  & \num{68.49}-\num{84.75}  & \num{23.1}-\num{23.5}  \\
	\hline \hline
\end{tabular} 
\egroup
\addtolength{\tabcolsep}{7pt}  
\label{tab:k.pParameters}
\end{table*}

In Table \ref{tab:k.pParameters}, we show the values for Kane ($\tilde{\gamma}_1$, 
$\tilde{\gamma}_2$, $\tilde{\gamma}_3$, \~e and $E_{P}$), and Luttinger--Kohn 
($\gamma_1$, $\gamma_2$, $\gamma_3$ and e) parameters, determined with the data 
using the range of up to \SI{10}{\percent} of the FBZ (and in any up to range from 
\SI{2}{} to \SI{20}{\percent} in the Supplementary Materials). The ranges of the 
values observed in literature are also given for comparison. Due to the small number 
of results founded in the literature, we included both, experimental and theoretical 
works, indicating the maximum and minimum values extracted from traditional sources 
such as Madelung \textit{et al.},\cite{Madelung_1982} Vurgaftman \textit{et al.}\cite{Vurgaftman_5815_2001,Vurgaftman_3675_2003} 
and Winkler.\cite{Winkler_2003} Our results are in agreement with the literature, 
i.e., the obtained effective mass parameters are inside the range of the most accepted 
values. In addition, the highest deviation comes from the nitrides. Since the most 
stable phase for the III-nitrides  is wurtzite and not zinc-blende, the scarce 
experimental data on it, prevents a more controlled comparison.

On the Kane models, the $P$ parameter, or its related energy $E_{P}$ is 
essential. This parameter represents the influence of the conduction band on the 
masses of the valence states and consequently the influence of the valence band 
on the conduction states. Our results for the $E_{P}$ parameter differ from 
the literature ones. The main reason for this difference is a divergence on the 
interpretation of the influence of the remote bands on the experimental measurements 
of the electron spin resonance as pointed out by Shantharama \textit{et al.}.\cite{Shantharama_4429_1984,Adams_401_1986} 
In their article they compare, e.g., Chadi \textit{et al.}\cite{Chadi_4466_1976} 
value for \ce{GaAs} of $E_{P} = \SI{29}{\electronvolt}$ with their 
estimation  based on an analysis of a $14{\times}14$ $\mathbf{k{\cdot}p}$ Hamiltonian 
of $E_{P} = \SI{25\pm0.5}{\electronvolt}$. The reason for the divergence 
is atributed to an overestimation of the influence of the remote bands. Our suggested 
value for this parameter is $ E_{P} = \SI{25.9}{\electronvolt}$. 
As $E_{P}$ is directly related to the Kane parameter $P$, its fitting 
is essential.

\begin{figure}
\centering
\includegraphics[width=0.90\linewidth]{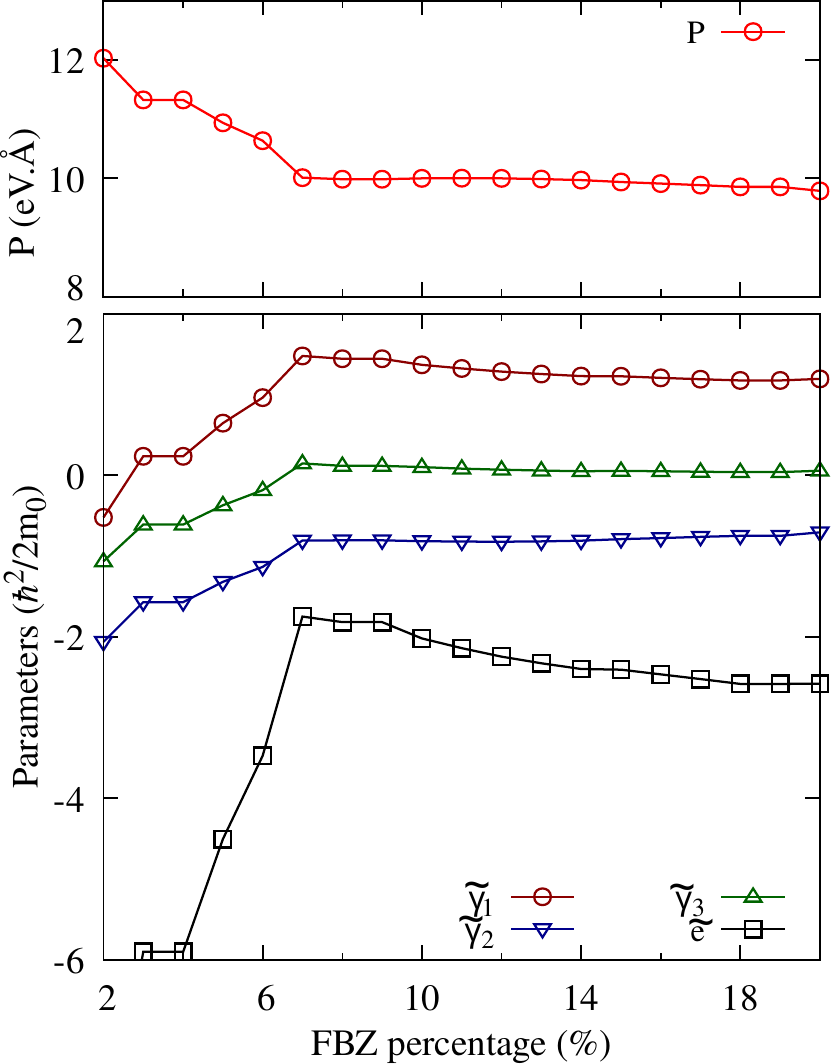}
\caption{\ce{GaAs} dependence of $P$ and Kane parameters with FBZ region used in 
the fitting process.}
\label{fig:kpParameters}
\end{figure}

As we have previously shown,\cite{Bastos_105002_2016} to correctly assign a value 
to $P$, it is necessary to include the first non-parabolicity in the range used 
for the fitting. If only the values below it are included, there is a fast variation 
of the values of $P$ depending on the range used. However, the fittings done with 
ranges including the non-parabolicity show a stabilization of the value. As an 
example, in Fig. \ref{fig:kpParameters}. we present a curve of the fitted parameter 
for \ce{GaAs}, showing the fast variation for ranges very close to the $\Gamma$-point 
and the stabilized values for ranges above \SI{8}{\percent}. The stabilization 
of our values suggests that our method improves the evaluation of the $P$ parameter, 
providing a way to distinguish the interactions of near and distant bands in the 
effective mass tensor. In the Supplementary Materials we provide the stabilization 
curves for the $P$, as well as the effective mass parameters, for all the materials.  

\subsection{Effective mass and $g$-factors}\label{subsection:EffectiveMass}

In order to verify the accuracy of our calculations, we compared the experimental 
effective masses with the ones obtained from our effective mass parameters (see 
relations in Appendix \ref{Appendix:LuttingerParameters}). Using the electronic 
$g$-factors, that are directly related to the spin splitting of the carrier bands, 
we have compared the measured values in the literature with our own values estimated 
from the $\mathbf{k{\cdot}p}$ parameters using the Roth's formula,\cite{Roth_90_1959}
\begin{equation}
\text{g}^*_{\text{c}} = 2-\frac{2\text{E}_\text{p} \Delta_{\text{so}}}{3
\text{E}_\text{gap}(\text{E}_\text{gap}+\Delta_{\text{so}})}\text{,}
\label{eq:g-factor}
\end{equation}
and the values for $E_P$, $E_{gap}$ and ${\Delta}_{\text{so}}$. This equation includes 
only the interaction between VB and CB, while the interactions between the other 
bands are neglected. 

Table \ref{table:EffectiveMass} shows the effective masses and $g$-factors obtained 
by the $\mathbf{k{\cdot}p}$ parameters calculation. SO and CB electronic effective 
masses are considered to be isotropic and HH and LH were evaluated along three 
different directions of the FBZ: [111], [110] and [100]. The CB $g$-factors have 
been estimated using equation \ref{eq:g-factor}. Tabulated values, extracted from 
Refs. \onlinecite{Madelung_1982,Vurgaftman_5815_2001,Vurgaftman_3675_2003,Martienssen_2005} 
are presented for comparison. The Supplementary Material provides tables for all 
calculated parameter sets.

\begin{table*}
\caption{Effective masses for light and the heavy hole along three directions ([100], 
[110] and [111]), isotropic masses from conduction band electrons and spin-orbit 
splitting holes and electron $g$-factors, obtained around the $\Gamma$ point, from 
the Luttinger-Kohn parameters. The ranges of values comprising the values found 
in the 
literature (experimental and theoretical results) are also shown for comparison.}
\def\arraystretch{1.2}
\addtolength{\tabcolsep}{3pt}  
\begin{threeparttable}
\begin{tabular}{ccccccccccccc}
	\hline \hline
	& & \multicolumn{2}{c}{{[}100{]}} &  & \multicolumn{2}{c}{{[}110{]}} &  & \multicolumn{2}{c}{{[}111{]}} &  &  & \tabularnewline
	\cline{3-4} \cline{6-7} \cline{9-10} 
	& & $\text{m}_{\text{HH}}^{*}$ & $\text{m}_{\text{LH}}^{*}$ &  & $\text{m}_{\text{HH}}^{*}$ & $\text{m}_{\text{LH}}^{*}$ &  & $\text{m}_{\text{HH}}^{*}$ & $\text{m}_{\text{LH}}^{*}$ & $\text{m}_{\text{so}}^{*}$ & $\text{m}_{\text{e}}^{*}$ & $\text{g}_{\text{c}}^{*}$\tabularnewline
	\hline 
	\ce{AlN} & this work &\num{1.320} & \num{0.437} &  & \num{4.123} & \num{0.357} &  & \num{2.858} & \num{0.371} & \num{0.659} & \num{0.296} & \num{1.99}\tabularnewline
	         & literature	 &  &  &  &  &  &  &  &  &  & \num{0.330}\tnote{a} & 
	\tabularnewline
	&  &  &  &  &  &  &  &  &  &  & \tabularnewline
	\ce{AlP} & this work& \num{0.563} & \num{0.255} &  & \num{1.599} & \num{0.197} &  & \num{1.176} & \num{0.206} & \num{0.355} & \num{0.197} & \num{1.93}\tabularnewline
	&literature & \num{0.513}\tnote{b} & \num{0.211}\tnote{b} &  &  &  &  & 
	\num{1.372}\tnote{b} & \num{0.145}\tnote{b} & \num{0.30}\tnote{a} &  & 
	\tabularnewline
	&  &  &  &  &  &  &  &  &  &  & \tabularnewline
	\ce{AlAs} & this work& \num{0.503} & \num{0.177} &  & \num{0.955} & \num{0.152} &  & \num{1.253} & \num{0.146} & \num{0.282} & \num{0.143} & \num{1.46}\tabularnewline
	&literature & \num{0.409}\tnote{c} & \num{0.153}\tnote{c} &  &  &  &  &	
	\num{1.022}\tnote{c} & \num{0.109}\tnote{c} & \num{0.28}\tnote{a} & 
	\num{0.15}\tnote{c} & \tabularnewline
	&  &  &  &  &  &  &  &  &  &  & \tabularnewline
	\ce{AlSb} & this work& \num{0.359} & \num{0.136} &  & \num{0.682} & \num{0.115} &  & \num{0.890} & \num{0.111} & \num{0.240} & \num{0.127} & \num{0.20}\tabularnewline
	&literature & \num{0.336}\tnote{c} & \num{0.123}\tnote{c} &  &  &  &  & 
	\num{0.872}\tnote{c} & \num{0.091}\tnote{c} &  & \num{0.14}\tnote{a} & 
	\tabularnewline
	&  &  &  &  &  &  &  &  &  &  & \tabularnewline
	\ce{GaN} & this work& \num{0.841} & \num{0.279} &  & \num{1.542} & \num{0.243} &  & \num{1.991} & \num{0.234} & \num{0.421} & \num{0.205} & \num{1.98}\tabularnewline
	&literature &  &  &  &  &  &  &  &  &  &  & 
	\num{1.95}\tnote{d}\tabularnewline
	&  &  &  &  &  &  &  &  &  &  & \tabularnewline
	\ce{GaP}& this work & \num{0.405} & \num{0.168} &  & \num{0.754} & \num{0.141} &  & \num{0.966} & \num{0.136} & \num{0.243} & \num{0.150} & \num{1.82}\tabularnewline
	&literature &  & &  &  &  &  & \num{0.670}\tnote{c} & \num{0.170}\tnote{c} & 
	\num{0.465}\tnote{e} & \num{0.13}\tnote{a} & \tabularnewline
	&  &  &  &  &  &  &  &  &  &  & \tabularnewline
	\ce{GaAs}& this work & \num{0.357} & \num{0.088} &  & \num{0.672} & \num{0.079} &  & \num{0.898} & \num{0.076} & \num{0.169} & \num{0.071} & \num{-0.34}\tabularnewline
	&literature & \num{0.34}\tnote{b} & \num{0.094}\tnote{b} &  &  &  &  & 
	\num{0.75}\tnote{b} & \num{0.082}\tnote{b}  &  & \num{0.0662}\tnote{b} & 
	\num{-0.44}\tnote{f}\tabularnewline
	&  &  &  &  &  &  &  &  &  &  & \tabularnewline
	\ce{GaSb}& this work & \num{0.247} & \num{0.051} &  & \num{0.500} & \num{0.046} &  & \num{0.710} & \num{0.045} & \num{0.144} & \num{0.045} & \num{-7.66}\tabularnewline
	&literature & \num{0.29}\tnote{c} &  &  & \num{0.36}\tnote{c} &  &  & 
	\num{0.40}\tnote{c} &  & \num{0.12}\tnote{a}	& \num{0.039}\tnote{b} & 
	\num{-9.1}\tnote{c}\tabularnewline
	&  &  &  &  &  &  &  &  &  &  & \tabularnewline
	\ce{InN}& this work & \num{0.874} & \num{0.090} &  & \num{1.851} & \num{0.085} &  & \num{1.460} & \num{0.086} & \num{0.167}  & \num{0.062} & \num{1.73}\tabularnewline
	&literature &  &  &  &  &  &  &  &  &  &  & \tabularnewline
	&  &  &  &  &  &  &  &  &  &  & \tabularnewline
	\ce{InP}& this work & \num{0.460} & \num{0.118} &  & \num{0.828} & \num{0.106} &  & \num{1.073} & \num{0.103} & \num{0.199} & \num{0.092} & \num{1.38}\tabularnewline
	&literature &  &  &  &  &  &  &  &  & \num{0.121}\tnote{b} & 
	\num{0.0808}\tnote{b} & 1.48\tnote{b}\tabularnewline
	&  &  &  &  &  &  &  &  &  &  & \tabularnewline
	\ce{InAs}& this work & \num{0.354} & \num{0.030} &  & \num{0.911} & \num{0.028} &  & \num{0.661} & \num{0.029} & \num{0.104} & \num{0.025} & \num{-15.18}\tabularnewline
	&literature & \num{0.35}\tnote{b} &  &  &  &  &  & \num{0.85}\tnote{b} &  & 
	\num{0.14}\tnote{a} & \num{0.0265}\tnote{b} & 
	\num{-15.3}\tnote{b}\tabularnewline
	&  &  &  &  &  &  &  &  &  &  & \tabularnewline
	\ce{InSb}& this work & \num{0.266} & \num{0.017} &  & \num{0.498} & \num{0.017} &  & \num{0.692} & \num{0.016} & \num{0.119} & \num{0.016} & \num{-43.30}\tabularnewline
	&literature & \num{0.32}\tnote{c} &  &  & \num{0.42}\tnote{c} &  &  & 
	\num{0.44}\tnote{c} &  & \num{0.11}\tnote{a}& \num{0.013} \tnote{b}& 	
	\num{-51.31}\tnote{b}\tabularnewline
	\hline \hline
\end{tabular}
\begin{tablenotes}
\item[a] Theory from ref. \onlinecite{Vurgaftman_5815_2001},
\item[b] Exp. from ref. \onlinecite{Madelung_2004},
\item[c] Exp. from ref. \onlinecite{Martienssen_2005},
\item[d] Exp. from ref. \onlinecite{Buss_225701_2015},
\item[e] Theory from ref. \onlinecite{Martienssen_2005},
\item[f] Exp. from ref. \onlinecite{Oestreich_2315_1995}.
\end{tablenotes}
\end{threeparttable}
\label{table:EffectiveMass}
\end{table*}

As we can distinguish the effects of the interactions from inner and outer bands, 
our $g$-factors show excellent agreement with the literature values, exception 
done to the materials with large spin-orbit coupling, such as antimonides and indium 
compounds. In these materials we have found large deviations from the reference 
values of HH and LH effective masses along the [110] and [111] directions. The 
lack of $k$ dependence on the spin-orbit coupling on the $\mathbf{k{\cdot}p}$ Hamiltonian 
used in our description may be responsible for such deviation. However, even for 
these materials, the LH and HH $g$-factors along the [100] direction present good 
agreement with the experimental values, since the specific symmetry of the zinc-blende 
systems prevent the splitting of the bands along that specific direction. Finally, 
CB and SO bands also present good agreement with the experimental values, since 
the spin split for them is small. 

\section{Summary and conclusion}\label{section:Summary}

We reported an extensive \textit{ab initio} study of electronic and structural 
properties of the III-V semiconductors (\num{12} systems) based on DFT within the 
PBE, PBE+SOC, HSE06, HSE$_{\alpha}$, and HSE$_{\alpha}$+SOC functionals. For the 
hybrid HSE$_{\alpha}$ functional, we fitted the magnitude of the nonlocal Fock 
exchange that replaces part of the PBE exchange based on the experimental results 
for the fundamental band gap and spin-orbit splitting energies. Except for the 
\ce{AlP} compound, whose $\alpha$ is \num{0.127}, our $\alpha$ parameters are in 
between \num{0.209} and \num{0.343}, deviating less than \num{0.1} from the universal 
value of \num{0.25} estimated by Perdew.\cite{Perdew_9982_1996}

Although the electronic properties were improved by the fitting, our results and 
analysis indicate clearly that HSE$_{\alpha}$ does not yield a significant improvement 
of the structural properties when compared with HSE06. In fact, it is an excellent 
result as it shows that is possible to improve the electronic properties without 
affecting the structural parameters by using fitted HSE$_{\alpha}$ functionals. 
This conclusion is valid, at least, for small changes in $\alpha$ near to the \num{0.25} 
value. Furthermore, based on several analysis, we found a correlation between the 
values of $\alpha$ with the cationic radius, namely, the optimized $\alpha$ value 
descreases almost linearly as a function of the atomic cationic radius, except 
for the case of \ce{AlN}. Therefore, our findings combined with previous results 
obtained by Vi\~nes \textit{et al.}\cite{Vines_781_2017} suggested that is possible 
to correlate the values of the ${\alpha}$ with the physical properties, and hence, 
it opens new possibilities in the study of much more complex materials. 

We found that the HSE$_{\alpha}$ overestimates the elastic constants, while PBE 
underestimates them. However, the magnitude of the relative error is smaller employing 
the HSE$_{\alpha}$ functional. We obtained from our results that the elastic constants 
decrease as the ionic radius increase, and hence, the elastic constants decrease 
by increasing the lattice parameter of the crystal structures. This behavior was 
reported  in the literature\cite{Adachi_1992,Keyes_3371_1962} and was traditionally 
used to estimate the elastic constants\cite{Willardson_1975,Adachi_R1_1985} by 
the  extrapolation of the data.

In order to provide a deeper understanding of the band structure curvatures, we 
used the DFT band structures to determine accurate $\mathbf{k{\cdot}p}$ parameters, 
and, from them, obtained the effective masses and the $g$-factors beyond the parabolic 
model. For the antimonides and indium compounds in specific directions, we observed 
large deviations of the $g$-factors from the experimental results indicating that 
the $8{\times}8$ $\mathbf{k{\cdot}p}$ Hamiltonian may still not be adequate for 
describing systems with small band gap or large spin-orbit splittings. The $k$-dependent 
spin-orbit term, responsible for the spin-orbit splitting outside of the $\Gamma$ 
point, is neglected in our model, resulting in the deviations observed. Finally, 
we tabulated the effective masses and $\mathbf{k{\cdot}p}$ parameters, presenting 
a full set of III-V parameters that may be used in realistic simulations of systems 
with higher complexity, such as nanowires and quantum dots, or devices based on 
these compounds.

\begin{appendices}

\appendix

\section{Luttinger--Kohn Parameters and Effective mass Relations}\label{Appendix:LuttingerParameters}

Since class A and B states differ among the $6\times6$ and $8\times8$ models, the 
definitions of the effective mass parameters differ from one model to the other.\cite{Enderlein_1997,Willatzen_2009} 
The relation among both model parameters, for zinc-blende structures, is given 
by the following expressions

\begin{equation*}
\begin{split}
\gamma_1=&\tilde{\gamma}_1+\frac{\text{E}_{\text{p}}}{3\text{E}_\text{gap}}, 
\hspace*{0.5cm}\gamma_2=\tilde{\gamma}_2+\frac{\text{E}_{\text{p}}}{6\text{E}_\text{gap}},\\
\gamma_3=&\tilde{\gamma}_3+\frac{\text{E}_{\text{p}}}{6\text{E}_\text{gap}},
\hspace*{0.65cm}
e=\tilde{e}+\frac{(\text{E}_{\text{gap}}+\frac{2}{3}\Delta_{\text{so}})\text{E}_{\text{p}}}{\text{E}_{\text{gap}}(\text{E}_{\text{gap}}+\Delta_{\text{so}})},\\
\text{E}_{\text{p}}=&\frac{2\text{m}_0}{\hbar^2}\text{P}^2.
\end{split}
\end{equation*}
The effective masses may be determined from the parameters using the following 
relations
\begin{equation*}
\begin{split}
\text{m}_{\text{lh}[100]}=&(\gamma_1+2\gamma_2)^{-1},\\
\text{m}_{\text{hh}[100]}=&(\gamma_1-2\gamma_2)^{-1}, \\
\text{m}_{\text{e}}=&e^{-1}, 
\end{split}
\quad
\begin{split}
\text{m}_{\text{lh}[110]}=&(\gamma_1+2\gamma_3)^{-1},\\
\text{m}_{\text{hh}[110]}=&(\gamma_1-2\gamma_3)^{-1},\\
\text{m}_{\text{lh}[111]}=&(\gamma_1+\sqrt{\gamma_2^2+3\gamma_3^2})^{-1},
\end{split}
\end{equation*}
\begin{equation*}
\begin{split}
\text{m}_{\text{hh}[111]}=&(\gamma_1-\sqrt{\gamma_2^2+3\gamma_3^2})^{-1},\\
\text{m}_{\text{so}}=& 
\left(\gamma_1-\frac{1}{3}\frac{\Delta_{\text{so}}\text{E}_{\text{p}}}{\text{E}_{\text{gap}}(\text{E}_{\text{gap}}+\Delta_{\text{so}})}\right)^{-1}.
\end{split}
\end{equation*}

\section{Effective Bader Charge} \label{Appendix:BaderCharge}

\begin{table}[h]
\caption{Effective Bader charge evaluated using the PBE functional for III-V semiconductors. 
All units are in \si{\coulomb}.}
\addtolength{\tabcolsep}{3pt}  
\bgroup
\def\arraystretch{1.2}%
\begin{tabular}{lcccc} \hline \hline
	        & \ce{N} & \ce{P} & \ce{As} & \ce{Sb}  \\ \hline
	\ce{Al} & \num{2.37} & \num{2.06} & \num{1.92} & \num{1.63} \\
	\ce{Ga} & \num{1.52} & \num{0.84} & \num{0.68} & \num{0.34} \\
	\ce{In} & \num{1.40} & \num{0.88} & \num{0.74} & \num{0.47} \\ \hline \hline
\end{tabular}
\egroup
\addtolength{\tabcolsep}{7pt}  
\label{tab:EffectiveBader}
\end{table}

\end{appendices}


%

\setcounter{section}{0}
\setcounter{equation}{0}
\setcounter{figure}{0}
\setcounter{table}{0}
\setcounter{page}{1}

\pagebreak
\widetext

\begin{center}
\textbf{\large Supplementary Material: A comprehensive study of g-factors, 
		elastic, structural and electronic properties of III-V semiconductors 
		using {H}ybrid-{D}ensity {F}unctional {T}heory}
\end{center}	

\onecolumngrid 
	
	\section{Computational details}
	
	\begin{table*}[h!]
		\centering
		\caption{HSE functional PAW VASP projectors used in this work, together with
			the number of valence electrons (ZVAL) and the 
			cut-off energies recommended by VASP, used in the minimization of the stress/elastic 
			constants tensor and in the determination of the total energy/band structures. }
		\label{}
		\begin{tabular}{lllccccccc}
			\hline
			\hline
			&& & & &  \multicolumn{5}{c}{cut-off energy (\si{\electronvolt})} \\
			\cline{6-10}
			&&\multicolumn{1}{c}{PAW} & ZVAL& valence & recommended && stress/elastic 
			constants&& total energy/band structure \\
			\hline
			Al && Al\_GW\_19Mar2012    & 3  & 3s$^\text{2}$ 3p$^\text{1}$                
			& 240.300 && 360.450 && 270.337 \\ 
			Ga && Ga\_d\_GW\_06Jul2010 & 3  & 4s$^\text{2}$ 4p$^\text{1}$                
			& 134.678 && 202.017 && 151.513 \\ 
			In && In\_d\_GW\_29May2007 & 13 & 4d$^\text{10}$ 5s$^\text{2}$ 5p$^\text{1}$ 
			& 278.624 && 417.936 && 313.452 \\ 
			N  && N\_GW\_10Apr2007     & 5  & 2s$^\text{2}$ 2p$^\text{3}$                
			& 420.902 && 631.353 && 473.515 \\ 
			P  && P\_GW\_19Mar2012     & 5  & 3s$^\text{2}$ 3p$^\text{3}$                
			& 255.040 && 382.560 && 286.920 \\ 
			As && As\_GW\_20Mar2012    & 5  & 4s$^\text{2}$ 4p$^\text{3}$                
			& 208.702 && 313.053 && 234.790 \\ 
			Sb && Sb\_d\_GW\_22Apr2009 & 15 & 4d$^\text{10}$ 5s$^\text{2}$ 5p$^\text{3}$ 
			& 172.069 && 258.103 && 193.578 \\
			\hline
			\hline
		\end{tabular} 
	\end{table*}
	
	\begin{table*}[h!]
		\centering
		\caption{PBE functional PAW VASP projectors used in this work, together with
			the number of valence electrons (ZVAL) and the 
			cut-off energies recommended by VASP, used in the minimization of the stress/elastic 
			constants tensor and in the determination of the total energy/band structures.}
		\begin{tabular}{lllccccccc}
			\hline
			\hline
			&& & & &  \multicolumn{5}{c}{cut-off energy (\si{\electronvolt})} \\
			\cline{6-10}
			&&\multicolumn{1}{c}{PAW} & ZVAL& valence & recommended && stress&elastic 
			constants& total energy/band structure \\
			\hline
			Al && Al\_GW\_19Mar2012    & 3  & 3s$^\text{2}$ 
			3p$^\text{1}$                
			& 240.300 && 480.600 &600.750 & 270.337 \\ 
			Ga && Ga\_d\_GW\_06Jul2010 & 3  & 4s$^\text{2}$ 
			4p$^\text{1}$                
			& 134.678 && 269.356 & 336.695& 151.513 \\ 
			In && In\_d\_GW\_29May2007 & 13 & 4d$^\text{10}$ 5s$^\text{2}$ 
			5p$^\text{1}$ 
			& 278.624 && 557.248 &696.560& 313.452 \\ 
			N  && N\_GW\_10Apr2007     & 5  & 2s$^\text{2}$ 
			2p$^\text{3}$                
			& 420.902 && 841.804 &1052.255& 473.515 \\ 
			P  && P\_GW\_19Mar2012     & 5  & 3s$^\text{2}$ 
			3p$^\text{3}$                
			& 255.040 && 510.080 &637.600& 286.920 \\ 
			As && As\_GW\_20Mar2012    & 5  & 4s$^\text{2}$ 
			4p$^\text{3}$                
			& 208.702 && 417.404 &521.755& 234.790 \\ 
			Sb && Sb\_d\_GW\_22Apr2009 & 15 & 4d$^\text{10}$ 5s$^\text{2}$ 
			5p$^\text{3}$ 
			& 172.069 && 344.138 &430.172& 193.578 \\
			\hline
			\hline
		\end{tabular} 
	\end{table*}
	
	\pagebreak
	
	\section{Dependence of properties with parameters}
	
	\subsection{Dependence of the band gap and $\Delta_{\text{so}}$ on the $\alpha$ 
		parameter}
	
	\begin{figure}[h!]
		\centering
		\includegraphics[width=0.8\linewidth]{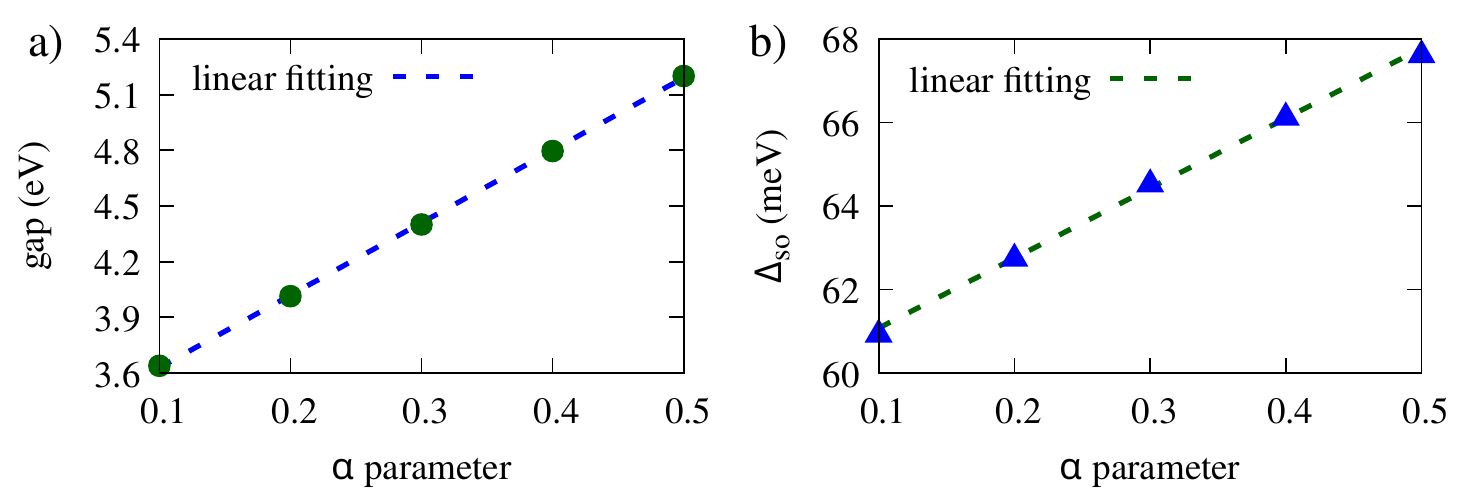}
		\caption{Linear dependence of the band gap (right) and 
			{$\Delta_{\text{so}}$} (left) on the $\alpha$ parameter for \ce{AlP}.}
		\label{fig:linearalphagap}
	\end{figure}

	\subsection{$\alpha$ parameters}
	
	\begin{figure}[h!]
		\centering
		\includegraphics[width=0.32\linewidth]{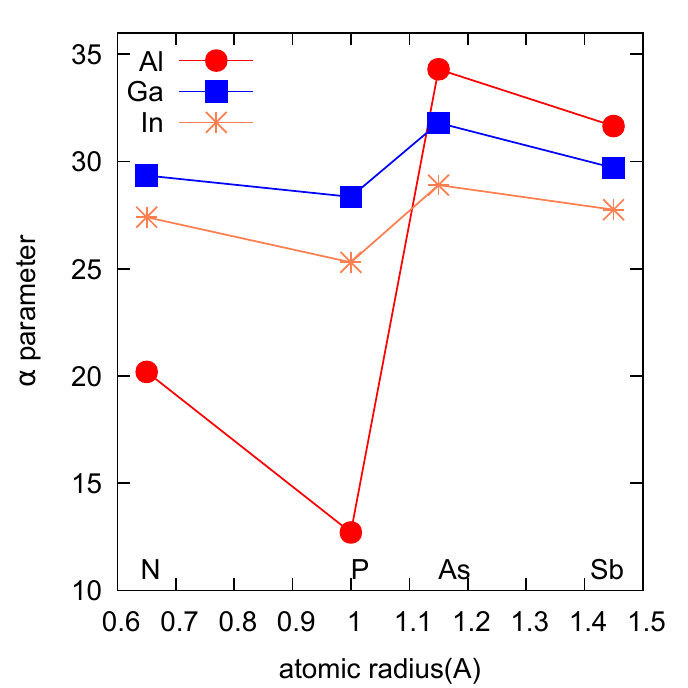}
		\includegraphics[width=0.32\linewidth]{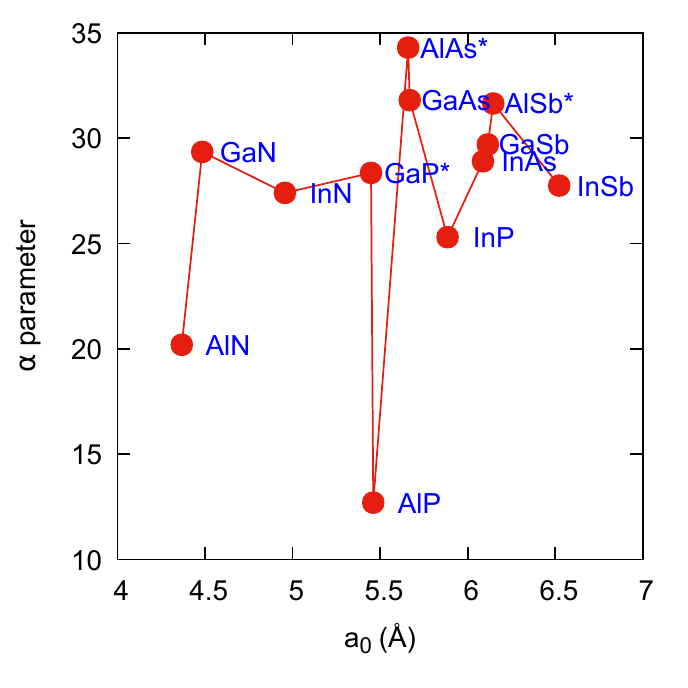}
		\caption{Cation atomic radius (left) and lattice parameter(right) dependence on the $\alpha$ parameter. 
		}
		\label{fig:alphavscatatom}
	\end{figure}
	
	\begin{figure}[h!]
		\centering
		\includegraphics[width=0.42\linewidth]{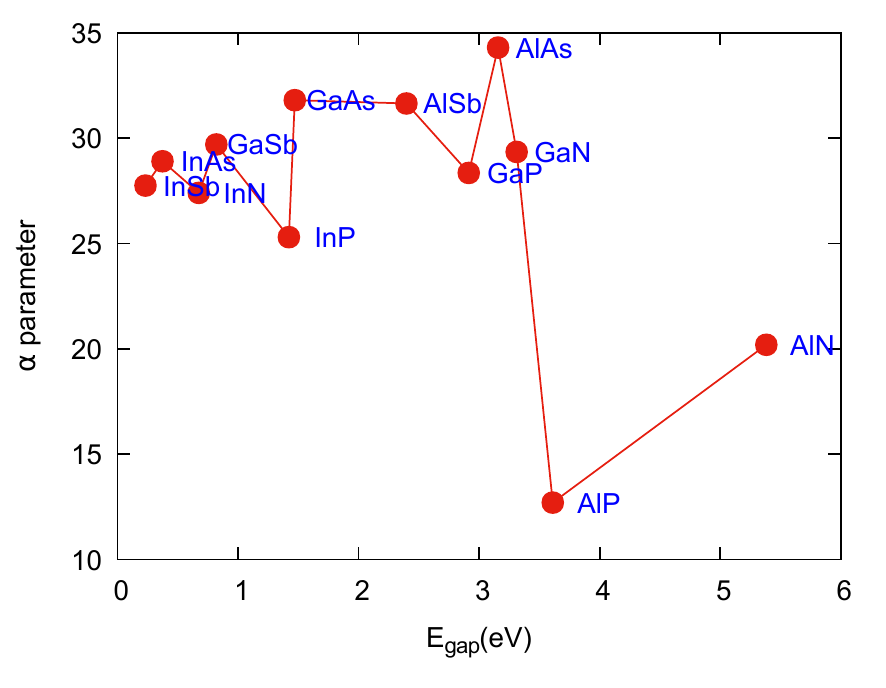}
		\includegraphics[width=0.32\linewidth]{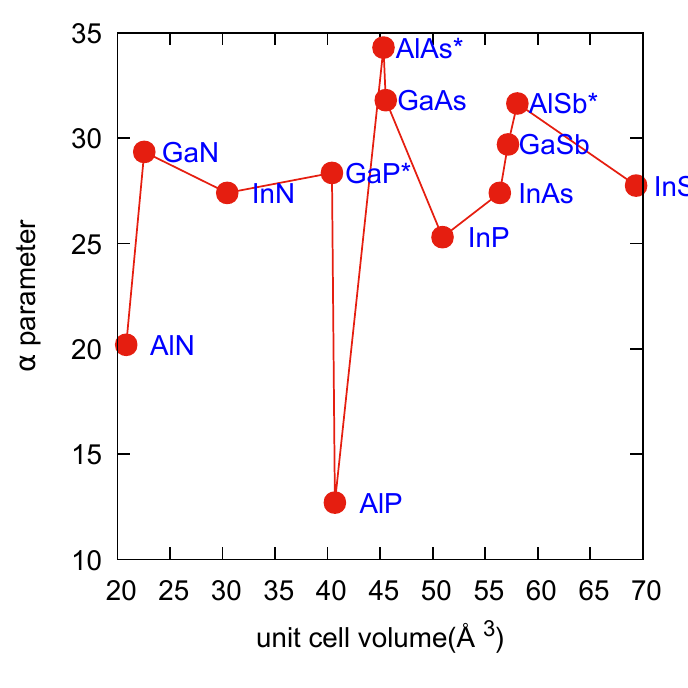}
		\caption{Energy gap (left) and equilibrium volume (right) dependence on the 
			$\alpha$ parameter. 
		}
		\label{fig:alphavsgap}
	\end{figure}
	
	\pagebreak
	\section{Band Structures}

	\begin{figure*}[h!]
		\centering
		\includegraphics[]{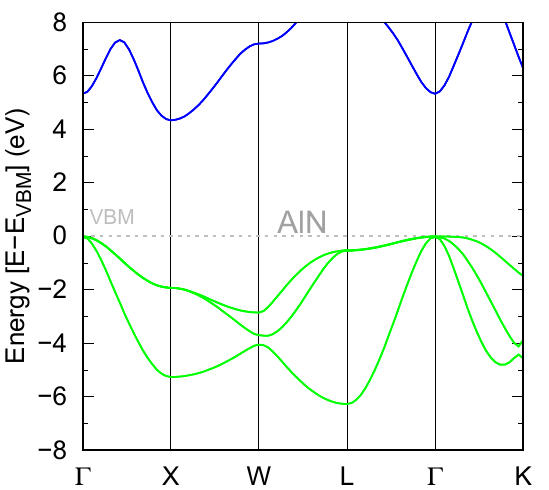}
		\includegraphics[]{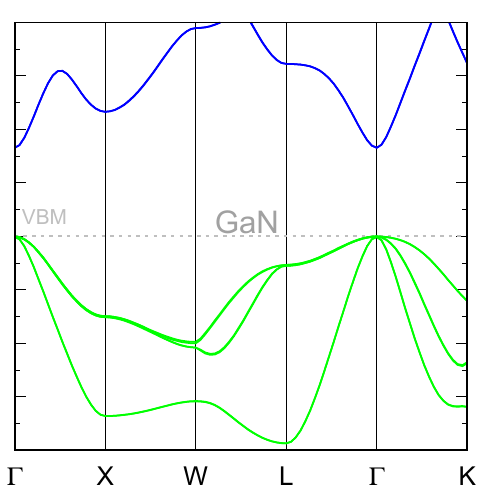}
		\includegraphics[]{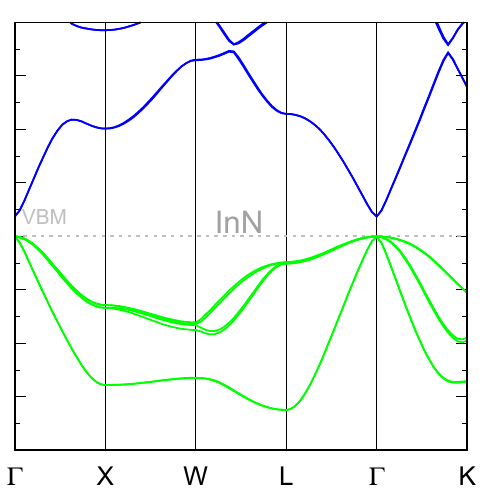}
		
		\includegraphics[]{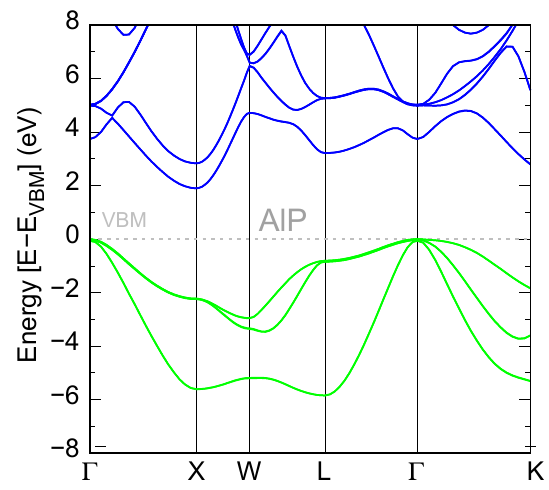}
		\includegraphics []{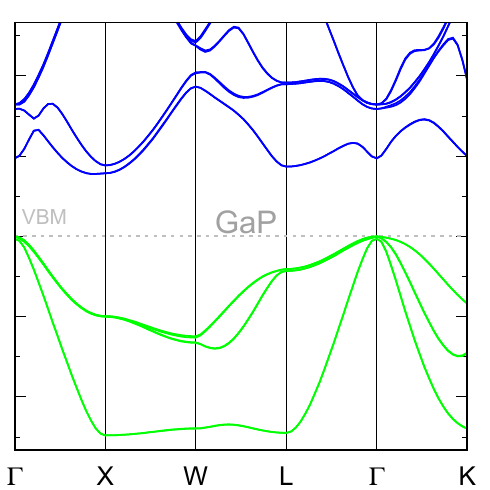}
		\includegraphics[]{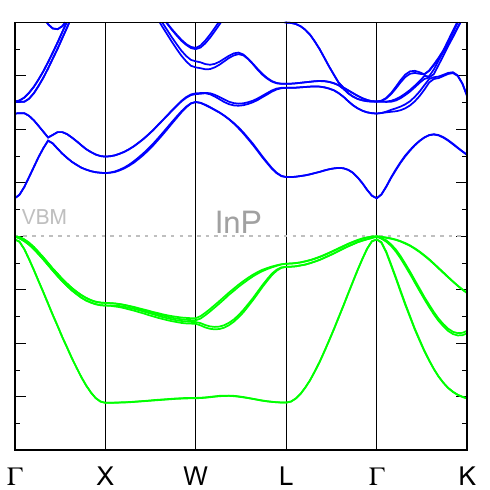}
		
		\includegraphics[]{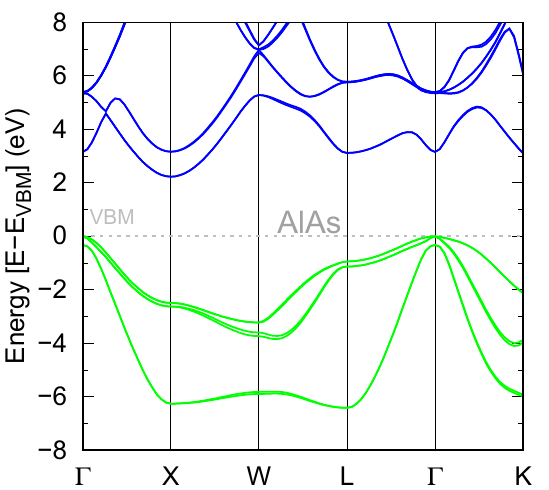}
		\includegraphics[]{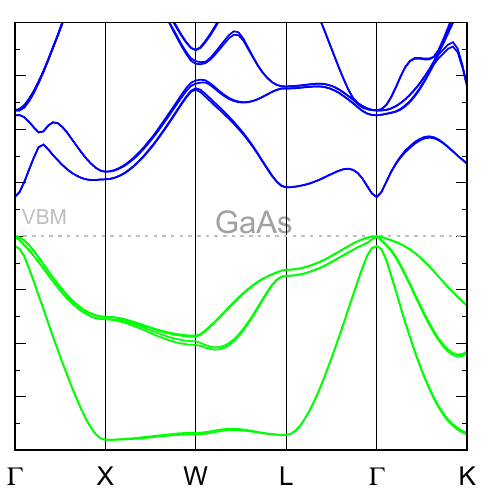}
		\includegraphics[]{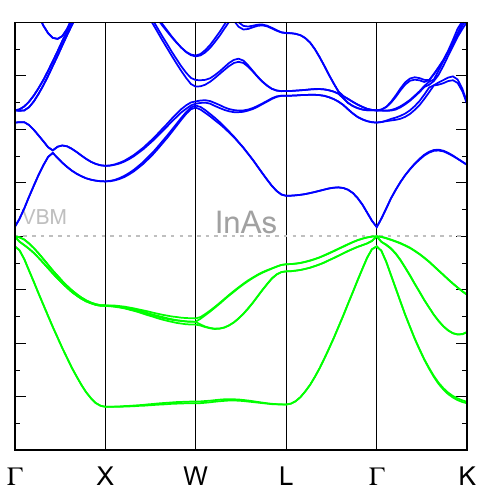}
		
		\includegraphics[]{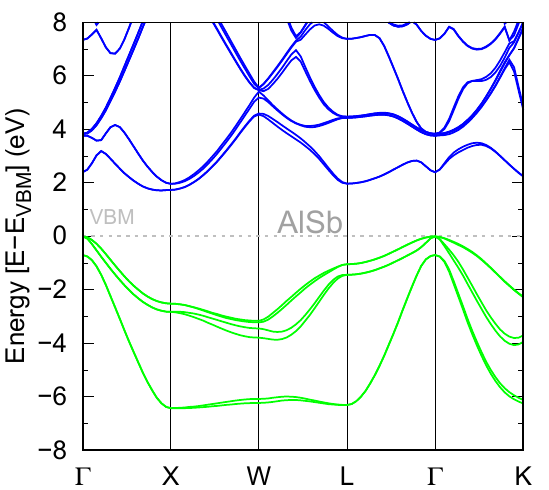}
		\includegraphics[]{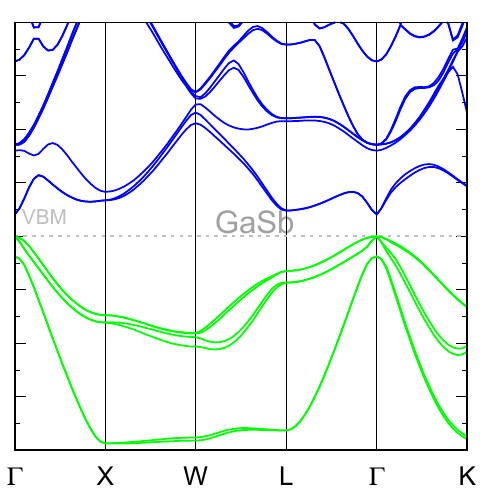}
		\includegraphics[]{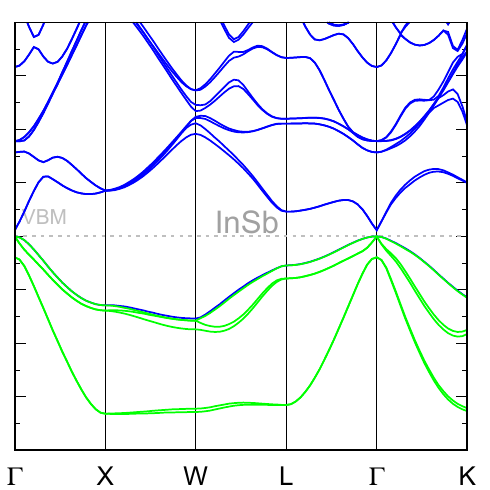}
		\caption{$\text{HSE}_{\alpha}$+SOC band structures obtained. Energy zero is set to the valence band maximum (VBM). 
			Blue and green lines indicate conduction and valence bands, respectively.
		}
		\label{fig:AlN_bs}
	\end{figure*}
	
	\pagebreak
	
	\section{$\mathbf{k \cdot p}$ parameters}
	\begin{figure*}[h!]
		\centering
		\includegraphics[]{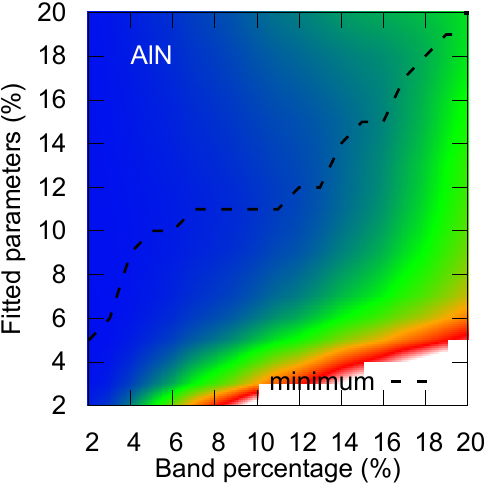}
		\includegraphics[]{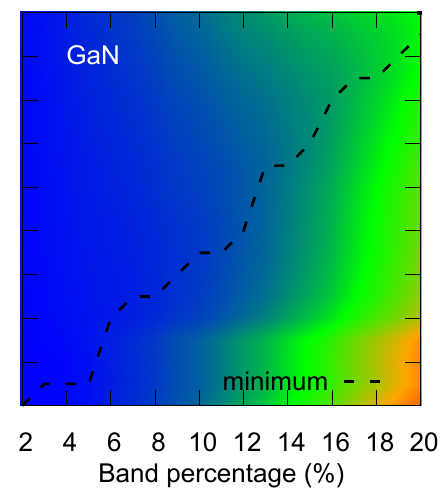}
		\includegraphics[]{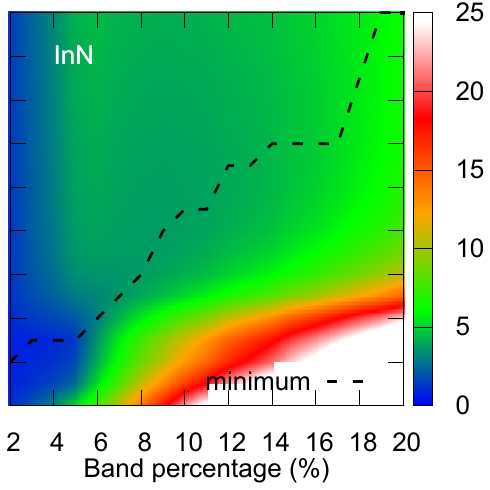}
		
		\includegraphics[]{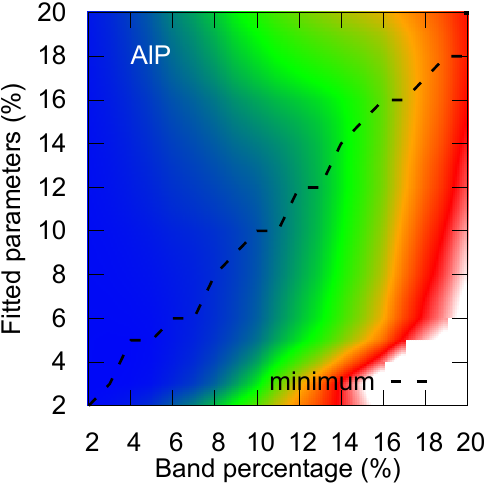}
		\includegraphics[]{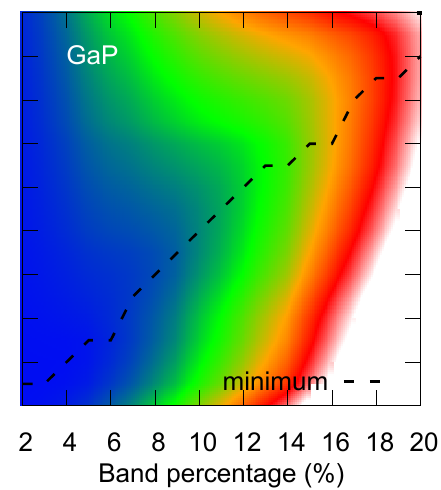}
		\includegraphics[]{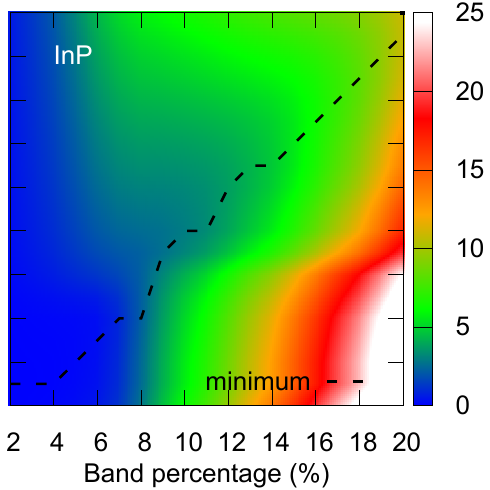}
		
		\includegraphics[]{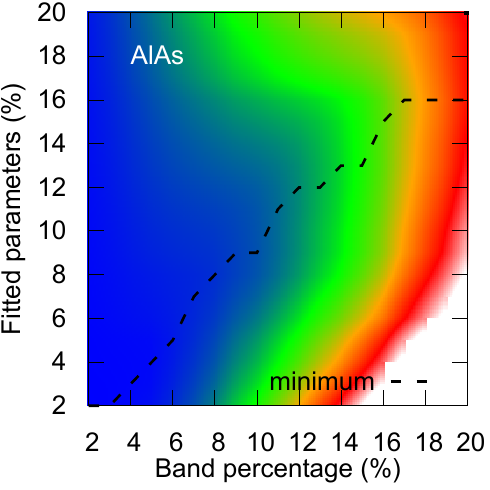}
		\includegraphics[]{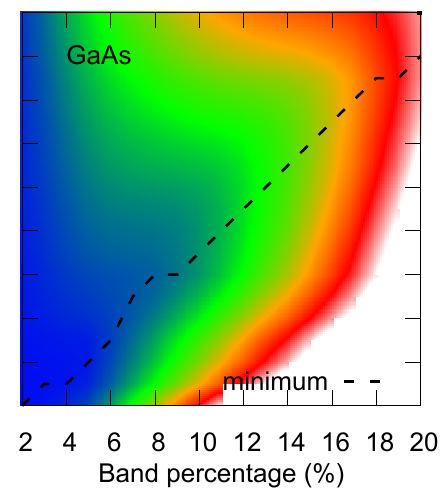}
		\includegraphics[]{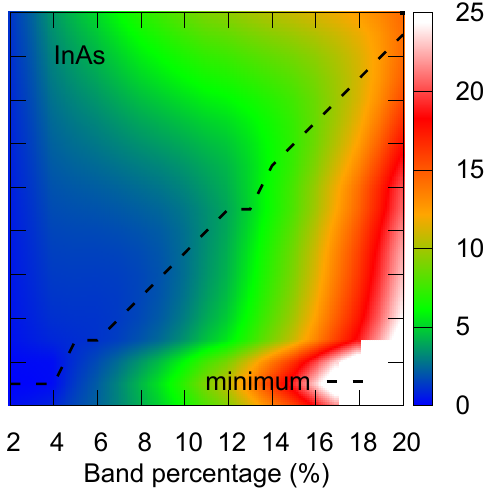}
		
		\includegraphics[]{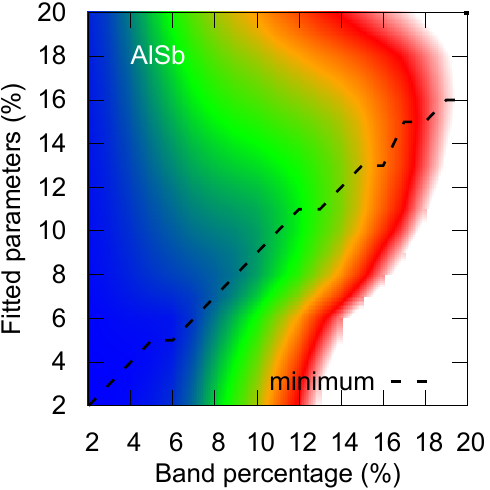}
		\includegraphics[]{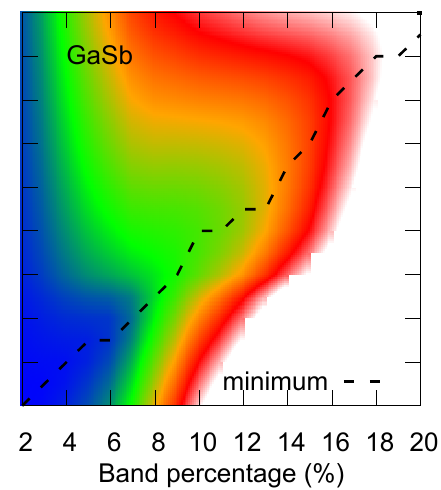}
		\includegraphics[]{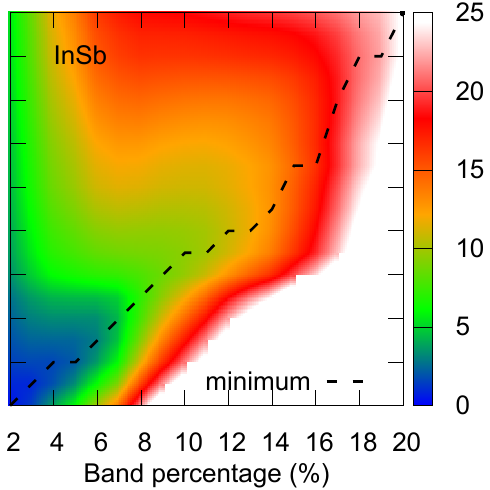}
		
		\caption{ Density map showing the agreement of the different adjusted parameter sets against the
			range around the $\Gamma$-point they are sampled, using the Root mean square deviation (RMSD) measure.
			The dashed line indicate the optimal sets parameter.}
	\end{figure*}
	
	\newpage
	\section{Kane parameters}
	
	\begin{figure*}[h!]
		\includegraphics[width=0.4\textwidth]{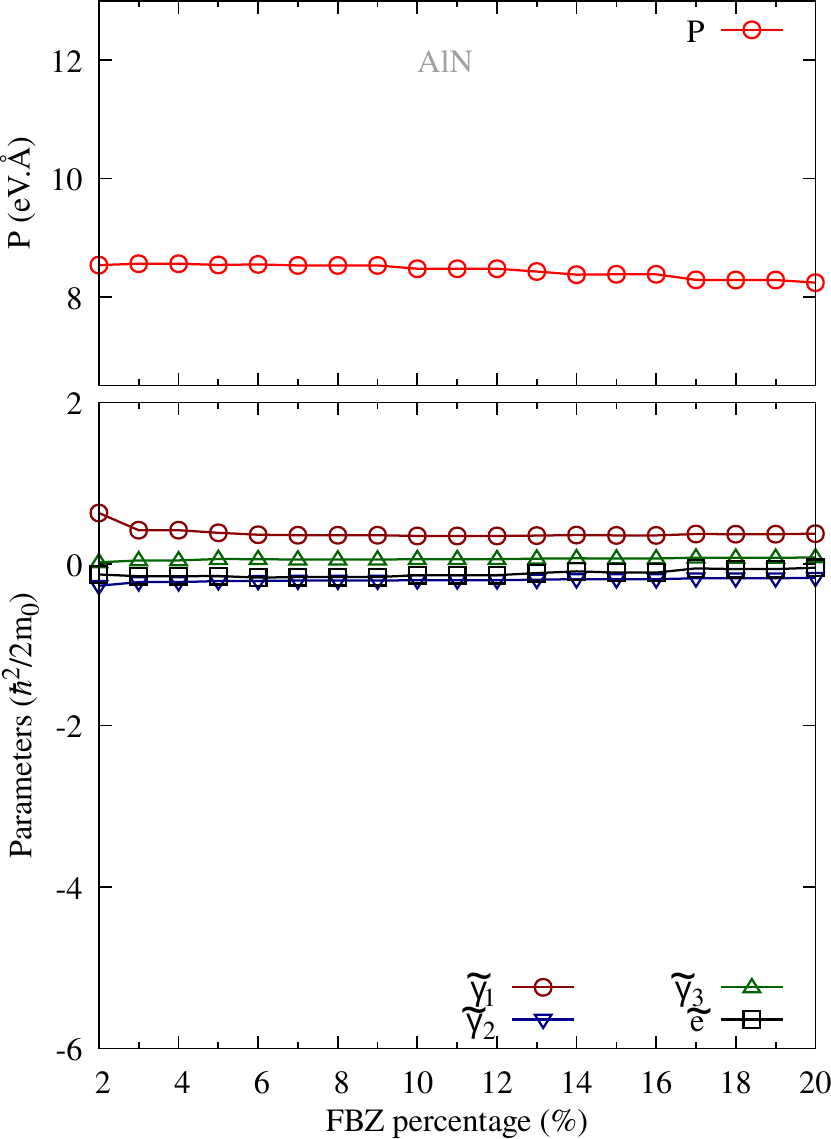}
		\includegraphics[width=0.4\textwidth]{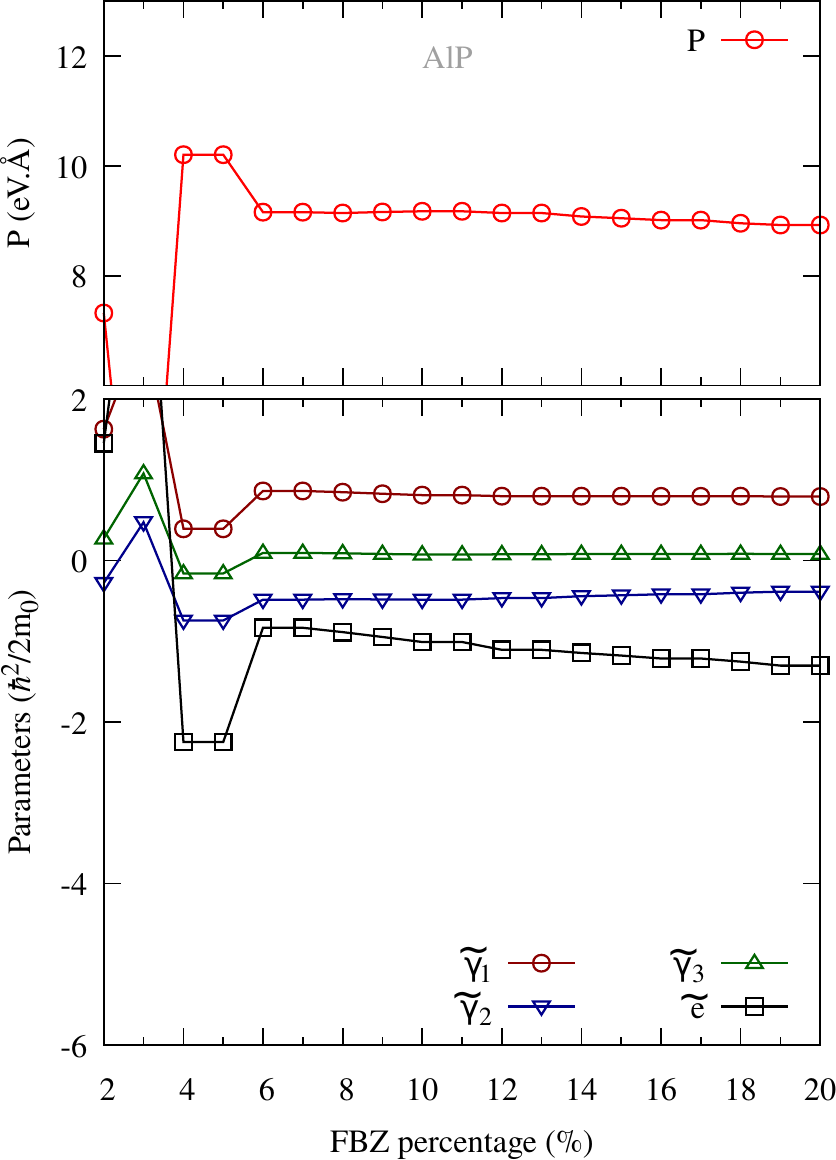}
		\includegraphics[width=0.4\textwidth]{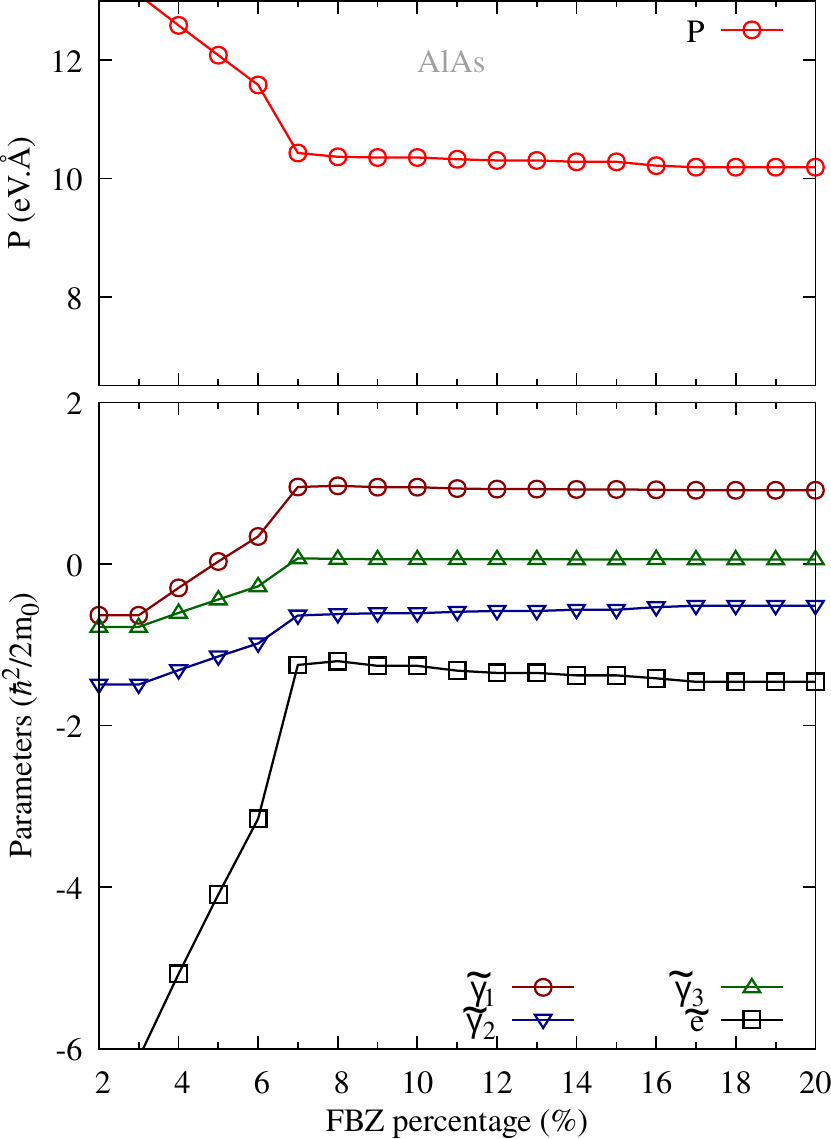}
		\includegraphics[width=0.4\textwidth]{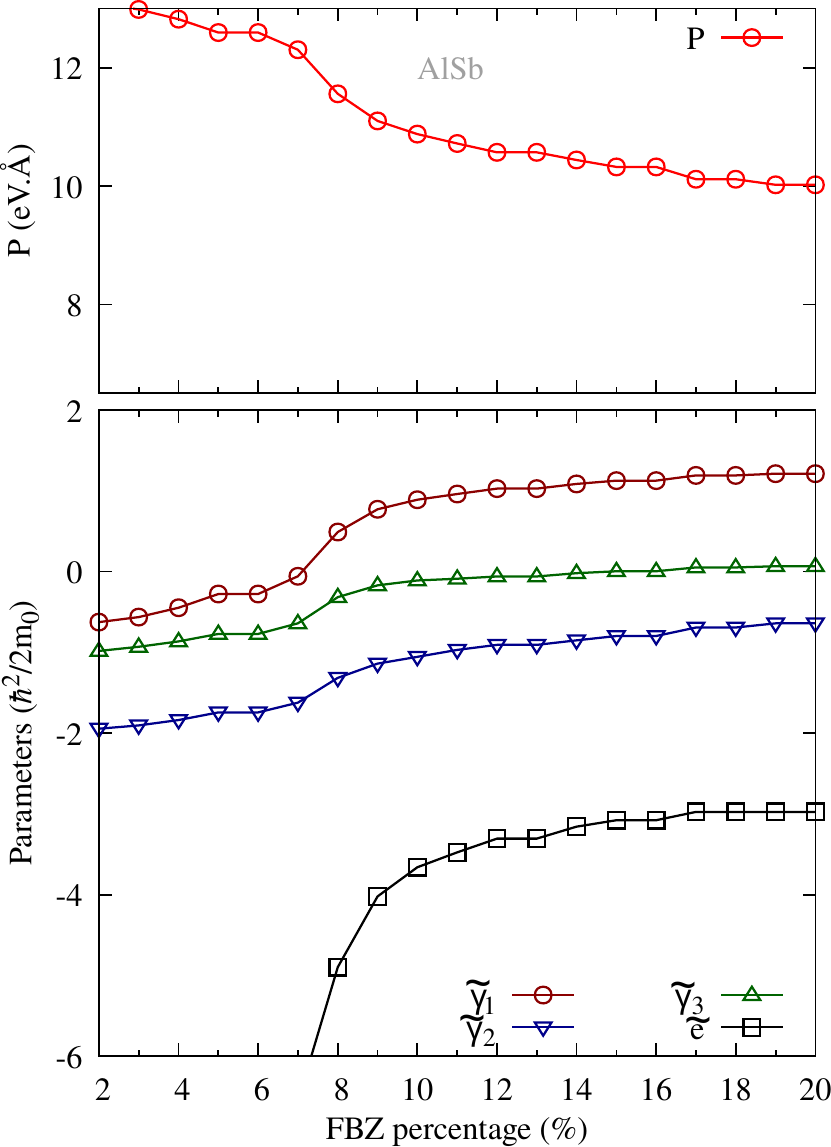}
		
		\caption{Recommended Kane parameters for each region near to $\Gamma$ point for aluminum-V compounds. In the x-axis is the percentage of the FBZ used in the fitting.}
	\end{figure*}

	\begin{figure*}[h!]
		\includegraphics[width=0.4\textwidth]{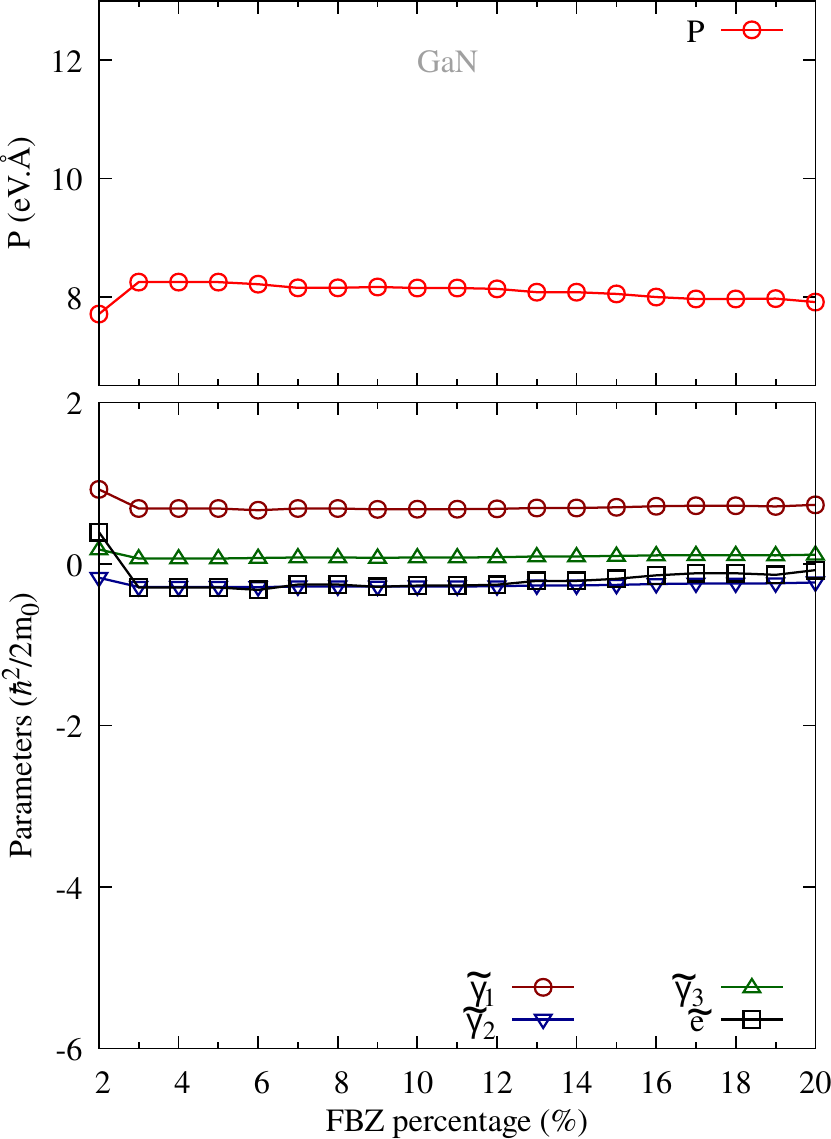}
		\includegraphics[width=0.4\textwidth]{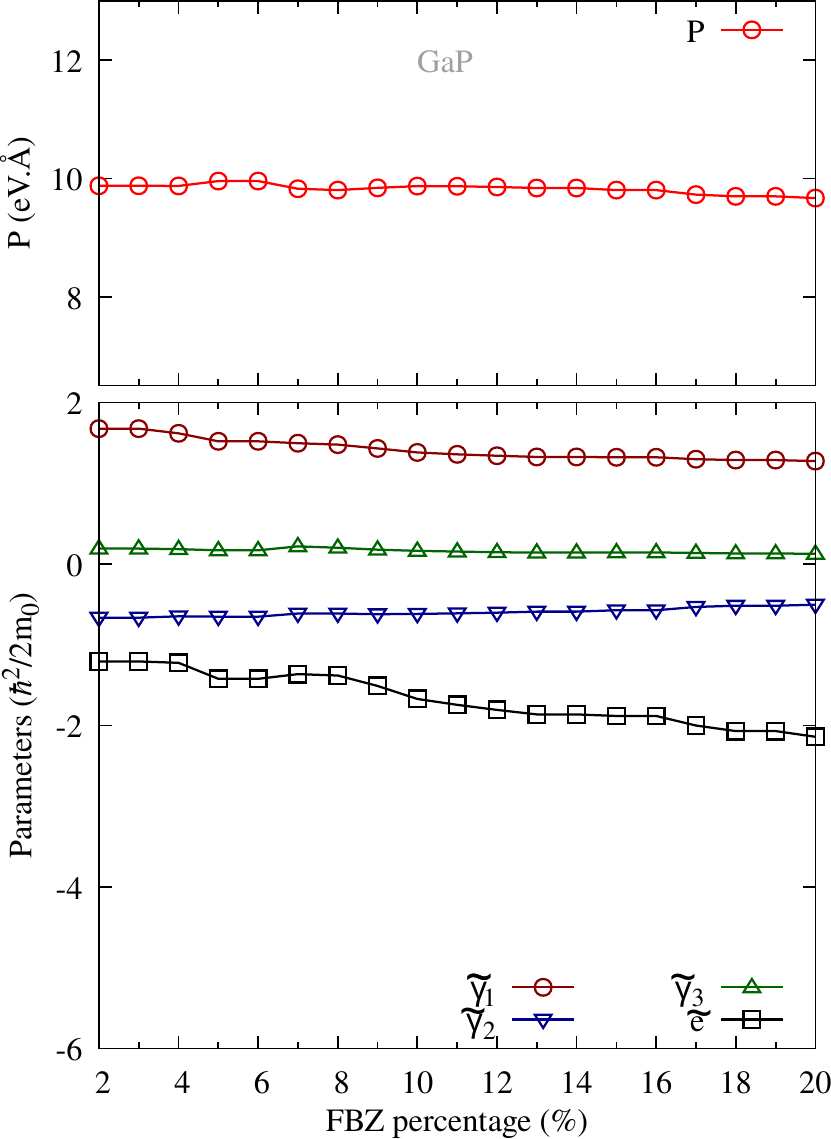}
		\includegraphics[width=0.4\textwidth]{GaAs_parameters}
		\includegraphics[width=0.4\textwidth]{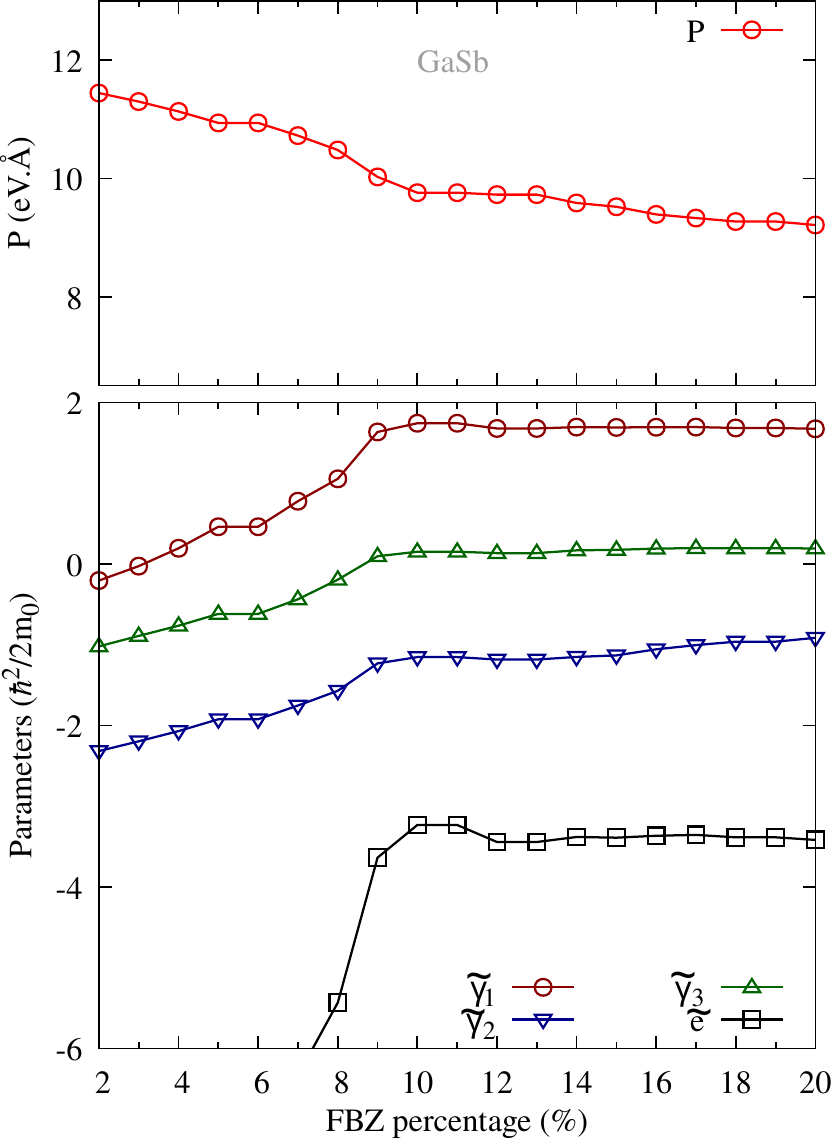}
		
		\caption{Recommended Kane parameters for each region near to $\Gamma$ point for gallium-V compounds. In the x-axis is the percentage of the FBZ used in the fitting.}
	\end{figure*}
	
	\begin{figure*}[h!]
		\includegraphics[width=0.4\textwidth]{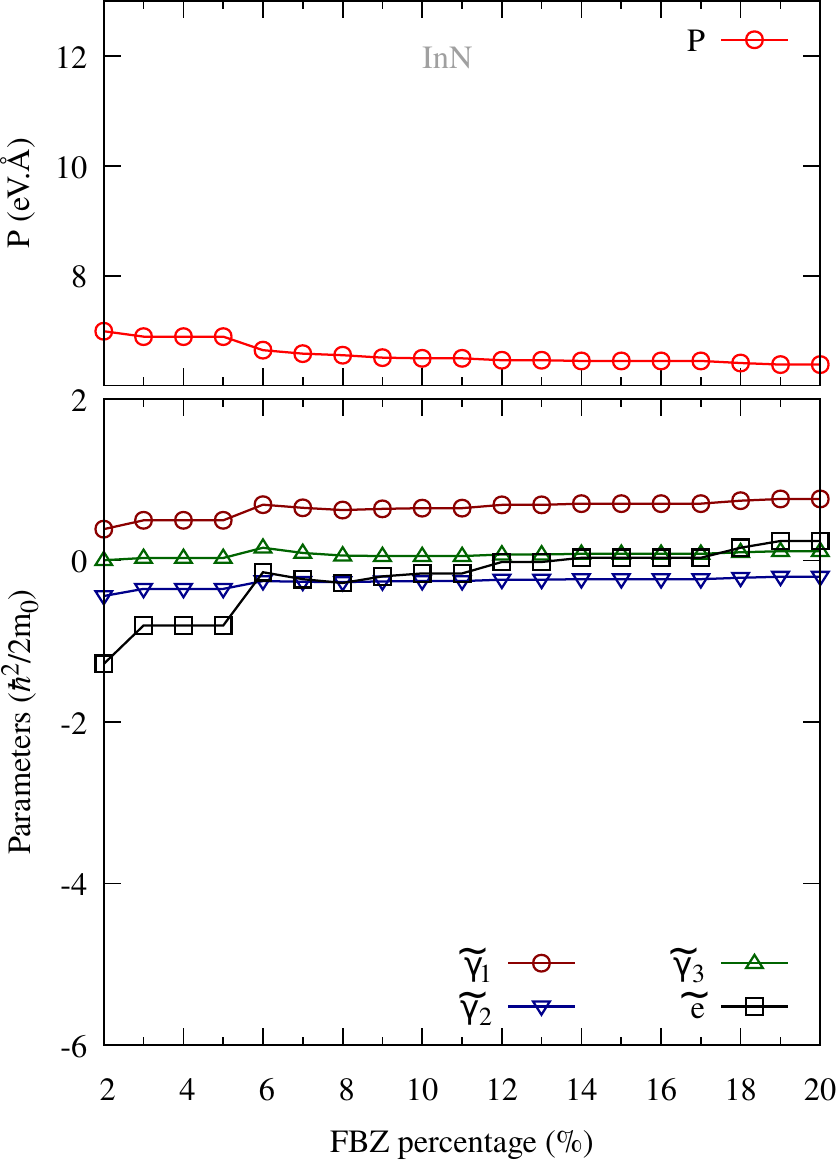}
		\includegraphics[width=0.4\textwidth]{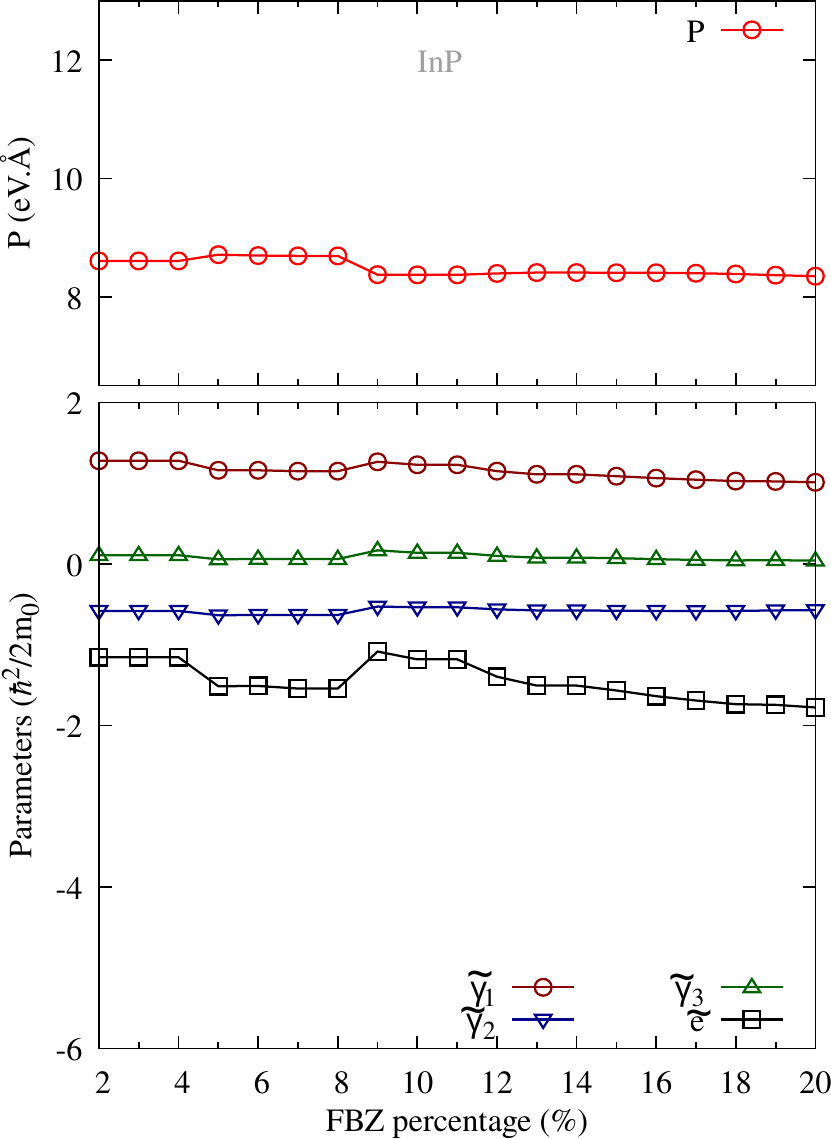}
		\includegraphics[width=0.4\textwidth]{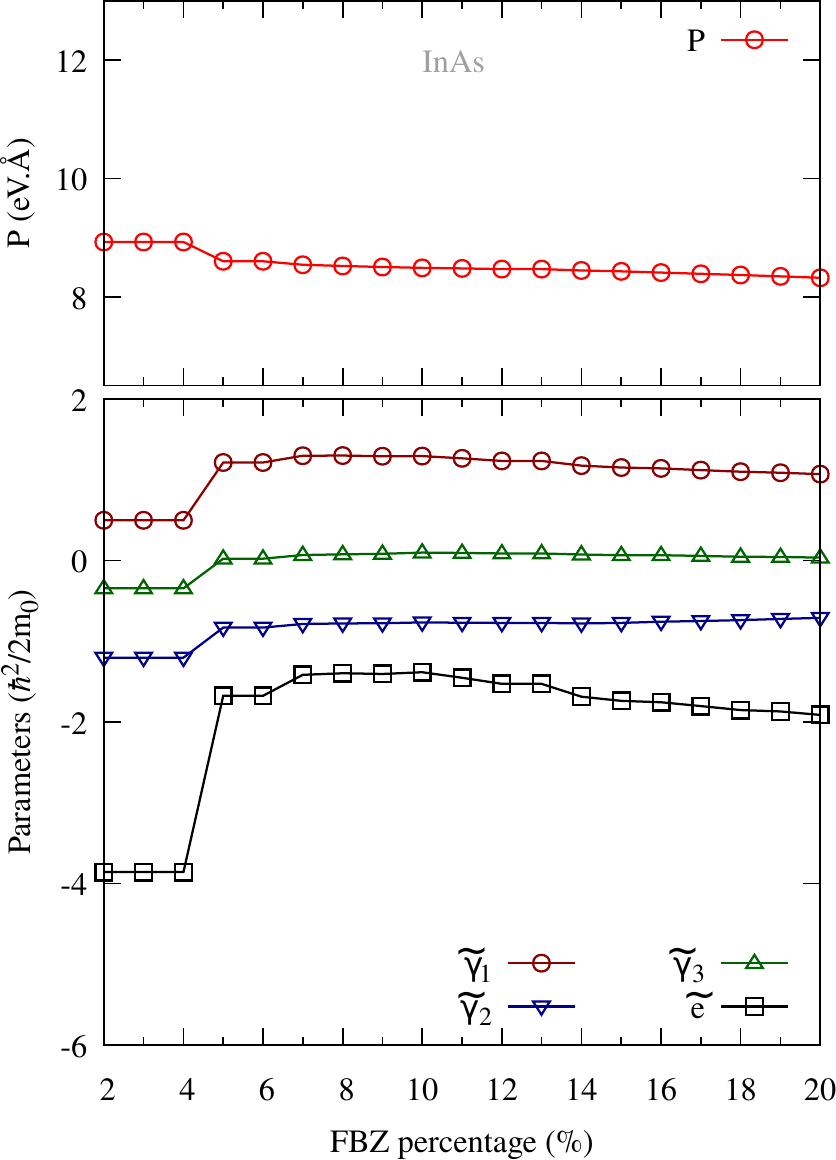}
		\includegraphics[width=0.4\textwidth]{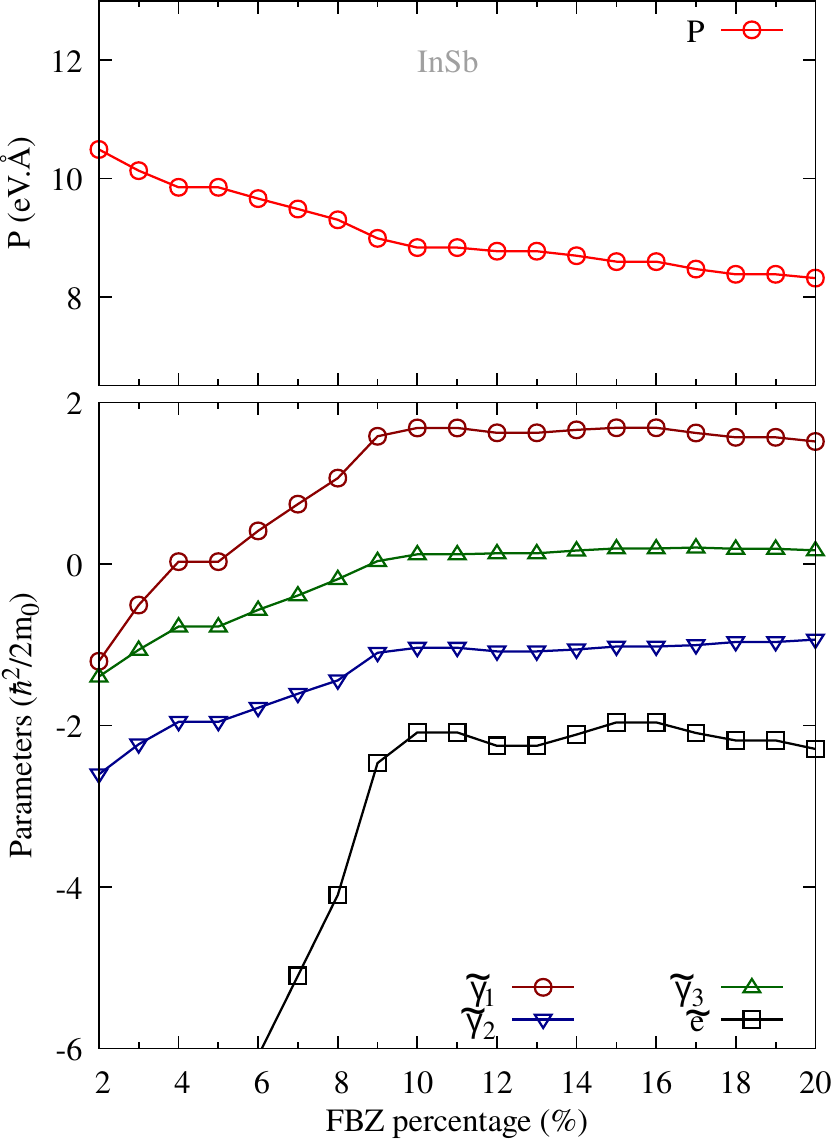}
		\caption{Recommended Kane parameters for each region near to $\Gamma$ point for Indium-V compounds. In the x-axis is the percentage of the FBZ used in the fitting.}
	\end{figure*}

	\section{Kane-Lutinger parameters}
	
	\begin{table*}[h!]
		
		\centering
		\caption{Adjusted Kane parameters and $\text{E}_{\text{p}}$ for \ce{AlN} and \ce{AlP} in function of the adjusted region. The RMSDs are also indicated.}
		\label{}
\begin{tabular}{SSSSSSSSSSSSSSS}
			
			\toprule
			\hline \\
			
			& \multicolumn{6}{c}{AlN} && \multicolumn{6}{c}{AlP} \\
			\cline{2-7} \cline{9-14} \\
			{{\si{\percent}}}& {RMSD} & {$\tilde{\gamma}_{1}$ }& {$\tilde{\gamma}_{2}$} & {$\tilde{\gamma}_{3}$} &  {{\~e}} & {${\text{P}}$} &&
			{ RMSD} & {$\tilde{\gamma}_{1}$} & {$\tilde{\gamma}_{2}$} & {$\tilde{\gamma}_{3}$} &  {{\~e}}  & {P } \\
			\midrule	  
			2 & 0.35 & 0.439 & -0.228 &  0.006 & -0.049 & 8.44 &&  0.19 & 1.627 & -0.278 &  0.272 &  1.453 &  7.32 \\
			3 & 0.37 & 0.399 & -0.221 &  0.022 & -0.101 & 8.49 &&  0.18 & 1.627 & -0.278 &  0.272 &  1.453 &  7.32 \\
			4 & 0.44 & 0.365 & -0.206 &  0.048 & -0.113 & 8.49 &&  0.20 & 0.392 & -0.742 & -0.161 & -2.248 & 10.20 \\
			5 & 0.52 & 0.358 & -0.204 &  0.051 & -0.119 & 8.49 &&  0.26 & 0.392 & -0.742 & -0.161 & -2.248 & 10.20 \\
\hline
			\hline
		\end{tabular}
	\end{table*}
	
	\begin{table*}[h!]
		\centering
		\caption{Adjusted Kane parameters and $\text{E}_{\text{p}}$ for \ce{AlAs} and \ce{AlSb} in function of the adjusted region. The RMSDs are also indicated.}
		\label{}
		\begin{tabular}{SSSSSSSSSSSSSSS}
			
			\toprule
			\hline \\
			
			& \multicolumn{6}{c}{AlAs} && \multicolumn{6}{c}{AlSb} \\
			\cline{2-7} \cline{9-14} \\
			{{\si{\percent}}}& {RMSD} & {$\tilde{\gamma}_{1}$ }& {$\tilde{\gamma}_{2}$} & {$\tilde{\gamma}_{3}$} &  {{\~e}} & {${\text{P}}$} &&
			{ RMSD} & {$\tilde{\gamma}_{1}$} & {$\tilde{\gamma}_{2}$} & {$\tilde{\gamma}_{3}$} &  {{\~e}}  & {P } \\
			\hline
			2 &  0.02 & -0.636 & -1.493 & -0.779 & -6.117 & 13.10   && 0.02  & -0.625 & -1.944 & -0.981 & -8.268 & 13.08 \\ 
			3 &  0.05 & -0.636 & -1.493 & -0.779 & -6.117 & 13.10   && 0.04  & -0.562 & -1.904 & -0.929 & -8.037 & 12.98 \\
			4 &  0.10 & -0.295 & -1.313 & -0.605 & -5.074 & 12.58   && 0.10  & -0.445 & -1.836 & -0.863 & -7.681 & 12.82 \\
			5 &  0.18 &  0.031 & -1.144 & -0.439 & -4.090 & 12.08   && 0.24  & -0.277 & -1.743 & -0.772 & -7.193 & 12.59 \\
\hline \hline
		\end{tabular}
	\end{table*}

	\begin{table*}[h!]
		\centering
		\caption{Adjusted Kane parameters and $\text{E}_{\text{p}}$ for \ce{GaN} and \ce{GaP} in function of the adjusted region. The RMSDs are also indicated.}
		\label{}
		\begin{tabular}{SSSSSSSSSSSSSSS}
			
			\toprule
			\hline \\
			
			& \multicolumn{6}{c}{GaN} && \multicolumn{6}{c}{GaP} \\
			\cline{2-7} \cline{9-14} \\
			{{\si{\percent}}}& {RMSD} & {$\tilde{\gamma}_{1}$ }& {$\tilde{\gamma}_{2}$} & {$\tilde{\gamma}_{3}$} &  {{\~e}} & {${\text{P}}$} &&
			{ RMSD} & {$\tilde{\gamma}_{1}$} & {$\tilde{\gamma}_{2}$} & {$\tilde{\gamma}_{3}$} &  {{\~e}}  & {P } \\
			\hline
			2 & 0.02 & 0.922 & -0.171 & 0.179 &  0.394 & 7.70 &&  0.30 & 1.677 & -0.665 & 0.190 & -1.206 & 9.87\\ 
			3 & 0.05 & 0.687 & -0.285 & 0.067 & -0.291 & 8.24 &&  0.31 & 1.677 & -0.665 & 0.190 & -1.206 & 9.87\\ 
			4 & 0.13 & 0.687 & -0.285 & 0.067 & -0.291 & 8.24 &&  0.32 & 1.616 & -0.648 & 0.183 & -1.224 & 9.87\\ 
			5 & 0.31 & 0.687 & -0.285 & 0.067 & -0.291 & 8.24 &&  0.47 & 1.518 & -0.653 & 0.169 & -1.421 & 9.95\\
\hline \hline
		\end{tabular}
	\end{table*}

	\begin{table*}[h!]
		\centering
		\caption{Adjusted Kane parameters and $\text{E}_{\text{p}}$ for \ce{GaAs} and \ce{GaSb} in function of the adjusted region. The RMSDs are also indicated.}
		\label{}
		\begin{tabular}{SSSSSSSSSSSSSSS}
			
			\toprule
			\hline \\
			
			& \multicolumn{6}{c}{GaAs} && \multicolumn{6}{c}{GaSb} \\
			\cline{2-7} \cline{9-14} \\
			{{\si{\percent}}}& {RMSD} & {$\tilde{\gamma}_{1}$ }& {$\tilde{\gamma}_{2}$} & {$\tilde{\gamma}_{3}$} &  {{\~e}} & {${\text{P}}$} &&
			{ RMSD} & {$\tilde{\gamma}_{1}$} & {$\tilde{\gamma}_{2}$} & {$\tilde{\gamma}_{3}$} &  {{\~e}}  & {P } \\
			\hline
			2 & 0.27  &-0.521 & -2.062 & -1.065 & -8.613  & 12.03 &&  0.04 &-0.205 & -2.315 & -1.020 & -9.431 & 11.44 \\
			3 & 0.34  & 0.236 & -1.569 & -0.609 & -5.899  & 11.32 &&  0.12 &-0.027 & -2.196 & -0.891 & -8.785 & 11.29 \\
			4 & 0.44  & 0.236 & -1.569 & -0.609 & -5.899  & 11.32 &&  0.32 & 0.195 & -2.069 & -0.764 & -8.086 & 11.13 \\
			5 & 0.57  & 0.645 & -1.318 & -0.371 & -4.504  & 10.93 &&  0.71 & 0.461 & -1.923 & -0.617 & -7.278 & 10.93 \\
\hline \hline
		\end{tabular}
	\end{table*}

	\begin{table*}[h!]
		\centering
		\caption{Adjusted Kane parameters and $\text{E}_{\text{p}}$ for \ce{InN} and \ce{InP} in function of the adjusted region. The RMSDs are also indicated.}
		\label{}
		\begin{tabular}{SSSSSSSSSSSSSSS}
			
			\toprule
			\hline \\
			
			& \multicolumn{6}{c}{InN} && \multicolumn{6}{c}{InP} \\
			\cline{2-7} \cline{9-14} \\
			{{\si{\percent}}}& {RMSD} & {$\tilde{\gamma}_{1}$ }& {$\tilde{\gamma}_{2}$} & {$\tilde{\gamma}_{3}$} &  {{\~e}} & {${\text{P}}$} &&
			{ RMSD} & {$\tilde{\gamma}_{1}$} & {$\tilde{\gamma}_{2}$} & {$\tilde{\gamma}_{3}$} &  {{\~e}}  & {P } \\
			\hline
			2 &  0.75 & 0.388 & -0.441 & 0.004 & -1.280 & 6.99 &&  0.03   & 1.277  & -0.581 & 0.110 & -1.156 & 8.60 \\
			3 &  0.86 & 0.498 & -0.352 & 0.031 & -0.805 & 6.89 &&  0.04   & 1.277  & -0.581 & 0.110 & -1.156 & 8.60 \\
			4 &  0.97 & 0.498 & -0.352 & 0.031 & -0.805 & 6.89 &&  0.07   & 1.277  & -0.581 & 0.110 & -1.156 & 8.60 \\
			5 &  1.27 & 0.498 & -0.352 & 0.031 & -0.805 & 6.89 &&  0.14   & 1.159  & -0.636 & 0.057 & -1.515 & 8.71 \\
\hline \hline
		\end{tabular}
	\end{table*}

	\begin{table*}[h!]
		\centering
		\caption{Adjusted Kane parameters and $\text{E}_{\text{p}}$ for \ce{InAs} and \ce{InSb} in function of the adjusted region. The RMSDs are also indicated.}
		\label{}
		\begin{tabular}{SSSSSSSSSSSSSSS}
			
			\toprule
			\hline \\
			
			& \multicolumn{6}{c}{InAs} && \multicolumn{6}{c}{InSb} \\
			\cline{2-7} \cline{9-14} \\
			{{\si{\percent}}}& {RMSD} & {$\tilde{\gamma}_{1}$ }& {$\tilde{\gamma}_{2}$} & {$\tilde{\gamma}_{3}$} &  {{\~e}} & {${\text{P}}$} &&
			{ RMSD} & {$\tilde{\gamma}_{1}$} & {$\tilde{\gamma}_{2}$} & {$\tilde{\gamma}_{3}$} &  {{\~e}}  & {P } \\
			\hline
			2 &  0.09 & 0.499 & -1.202 & -0.345 & -3.857 & 8.92 &&   0.29  & -1.204 & -2.604 & -1.393 & -10.959 & 10.48\\
			3 &  0.12 & 0.499 & -1.202 & -0.345 & -3.857 & 8.92 &&   0.90  & -0.508 & -2.233 & -1.062 &  -8.840 & 10.13\\
			4 &  0.25 & 0.499 & -1.202 & -0.345 & -3.857 & 8.92 &&   1.66  &  0.028 & -1.954 & -0.775 &  -7.220 &  9.85\\
			5 &  1.06 & 1.214 & -0.831 &  0.022 & -1.676 & 8.60 &&   2.10  &  0.028 & -1.954 & -0.775 &  -7.220 &  9.85\\
\hline \hline
		\end{tabular}
	\end{table*}

	\newpage
	
	\cleardoublepage
	\section{Lutinger parameters}
	
	\begin{table*}[h!]
		
		\centering
		\caption{Adjusted Luttinger-Kohn parameters and $\text{E}_{\text{p}}$ for \ce{AlN} and \ce{AlP} in function of the adjusted region. The RMSDs are also indicated.}
		\label{}
		\begin{tabular}{SSSSSSSSSSSSSSS}
			
			\toprule
			\hline \\
			
			& \multicolumn{6}{c}{AlN} && \multicolumn{6}{c}{AlP} \\
			\cline{2-7} \cline{9-14} \\
			{{\si{\percent}}}& {RMSD} & {$\gamma_{1}$ }& {$\gamma_{2}$} & {$\gamma_{3}$} &  {{e}} & {$\text{E}_{\text{P}}$} &&
			{RMSD} & {$\gamma_{1}$ }& {$\gamma_{2}$} & {$\gamma_{3}$} &  {{e}} & {$\text{E}_{\text{P}}$} \\
			\hline
			2 & 0.36 & 1.597 & 0.351 & 0.585 & 3.421 & 18.71 &&  0.19 &  2.926 & 0.372 & 0.921 & 5.327 & 14.07 \\
			3 & 0.38 & 1.572 & 0.366 & 0.608 & 3.410 & 18.93 &&  0.18 &  2.926 & 0.372 & 0.921 & 5.327 & 14.07 \\
			4 & 0.44 & 1.536 & 0.379 & 0.634 & 3.395 & 18.91 &&  0.20 &  2.912 & 0.518 & 1.100 & 5.272 & 27.31 \\
			5 & 0.52 & 1.529 & 0.382 & 0.636 & 3.388 & 18.91 &&  0.26 &  2.912 & 0.518 & 1.100 & 5.272 & 27.31 \\
\hline \hline
		\end{tabular}
	\end{table*}
	
	\begin{table*}[h!]
		\centering
		\caption{Adjusted Luttinger-Kohn parameters and $\text{E}_{\text{p}}$ for \ce{AlAs} and \ce{AlSb} in function of the adjusted region. The RMSDs are also indicated.}
		\label{}
		\begin{tabular}{SSSSSSSSSSSSSSS}
			
			\toprule
			\hline \\
			
			& \multicolumn{6}{c}{AlAs} && \multicolumn{6}{c}{AlSb} \\
			\cline{2-7} \cline{9-14} \\
			{{\si{\percent}}}& {RMSD} & {$\gamma_{1}$ }& {$\gamma_{2}$} & {$\gamma_{3}$} &  {{e}} & {$\text{E}_{\text{P}}$} &&
			{RMSD} & {$\gamma_{1}$ }& {$\gamma_{2}$} & {$\gamma_{3}$} &  {{e}} & {$\text{E}_{\text{P}}$} \\
			\hline
			2 &  0.30 & 24.311 & 10.358 & 11.224 & 52.550 & 27.0 &&  0.02 & 5.627 & 1.181 & 2.145 & 9.073 & 44.9 \\
			3 &  0.73 & 23.074 &  9.947 & 10.808 & 51.087 & 24.1 &&  0.04 & 5.592 & 1.173 & 2.148 & 9.035 & 44.2 \\
			4 &  1.28 & 23.074 &  9.947 & 10.808 & 51.087 & 24.1 &&  0.10 & 5.554 & 1.163 & 2.137 & 8.960 & 43.1 \\
			5 &  1.84 & 22.304 &  9.631 & 10.487 & 49.867 & 22.5 &&  0.24 & 5.513 & 1.152 & 2.123 & 8.866 & 41.6 \\
\hline \hline
		\end{tabular}
	\end{table*}

	\begin{table*}[h!]
		\centering
		\caption{Adjusted Luttinger-Kohn parameters and $\text{E}_{\text{p}}$ for \ce{GaN} and \ce{GaP} in function of the adjusted region. The RMSDs are also indicated.}
		\label{}
		\begin{tabular}{SSSSSSSSSSSSSSS}
			
			\toprule
			\hline \\
			
			& \multicolumn{6}{c}{GaN} && \multicolumn{6}{c}{GaP} \\
			\cline{2-7} \cline{9-14} \\
			{{\si{\percent}}}& {RMSD} & {$\gamma_{1}$ }& {$\gamma_{2}$} & {$\gamma_{3}$} &  {{e}} & {$\text{E}_{\text{P}}$} &&
			{RMSD} & {$\gamma_{1}$ }& {$\gamma_{2}$} & {$\gamma_{3}$} &  {{e}} & {$\text{E}_{\text{P}}$} \\
			\hline
			2 & 0.02 & 2.491 & 0.614 & 0.963 & 5.090 & 15.5 &&  0.30 &  4.604 & 0.799 & 1.654  & 7.483 & 25.5 \\
			3 & 0.05 & 2.484 & 0.614 & 0.965 & 5.089 & 17.8 &&  0.31 &  4.604 & 0.799 & 1.654  & 7.483 & 25.5 \\
			4 & 0.13 & 2.484 & 0.614 & 0.965 & 5.089 & 17.8 &&  0.32 &  4.542 & 0.814 & 1.646  & 7.460 & 25.5 \\
			5 & 0.31 & 2.484 & 0.614 & 0.965 & 5.089 & 17.8 &&  0.47 &  4.493 & 0.834 & 1.656  & 7.408 & 26.0 \\
\hline \hline
		\end{tabular}
	\end{table*}

	\begin{table*}[h!]
		\centering
		\caption{Adjusted Luttinger-Kohn parameters and $\text{E}_{\text{p}}$ for \ce{GaAs} and \ce{GaSb} in function of the adjusted region. The RMSDs are also indicated.}
		\label{}
		\begin{tabular}{SSSSSSSSSSSSSSS}
			
			\toprule
			\hline \\
			
			& \multicolumn{6}{c}{GaAs} && \multicolumn{6}{c}{GaSb} \\
			\cline{2-7} \cline{9-14} \\
			{{\si{\percent}}}& {RMSD} & {$\gamma_{1}$ }& {$\gamma_{2}$} & {$\gamma_{3}$} &  {{e}} & {$\text{E}_{\text{P}}$} &&
			{RMSD} & {$\gamma_{1}$ }& {$\gamma_{2}$} & {$\gamma_{3}$} &  {{e}} & {$\text{E}_{\text{P}}$} \\
			\hline
			2 &  0.27 & 8.088 & 2.242 & 3.240 & 15.497 & 37.9 &&   0.04 & 13.779 & 4.677 & 5.972 & 25.834 & 34.3 \\
			3 &  0.34 & 7.866 & 2.246 & 3.207 & 15.470 & 33.6 &&   0.12 & 13.607 & 4.621 & 5.926 & 25.597 & 33.5 \\
			4 &  0.44 & 7.866 & 2.246 & 3.207 & 15.470 & 33.6 &&   0.32 & 13.427 & 4.547 & 5.852 & 25.283 & 32.5 \\
			5 &  0.57 & 7.761 & 2.240 & 3.186 & 15.424 & 31.3 &&   0.71 & 13.238 & 4.466 & 5.771 & 24.943 & 31.4 \\
\hline \hline
		\end{tabular}
	\end{table*}

	\begin{table*}[h!]
		\centering
		\caption{Adjusted Luttinger-Kohn parameters and $\text{E}_{\text{p}}$ for \ce{InN} and \ce{InP} in function of the adjusted region. The RMSDs are also indicated.}
		\label{}
		\begin{tabular}{SSSSSSSSSSSSSSS}
			
			\toprule
			\hline \\
			
			& \multicolumn{6}{c}{InN} && \multicolumn{6}{c}{InP} \\
			\cline{2-7} \cline{9-14} \\
			{{\si{\percent}}}& {RMSD} & {$\gamma_{1}$ }& {$\gamma_{2}$} & {$\gamma_{3}$} &  {{e}} & {$\text{E}_{\text{P}}$} &&
			{RMSD} & {$\gamma_{1}$ }& {$\gamma_{2}$} & {$\gamma_{3}$} &  {{e}} & {$\text{E}_{\text{P}}$} \\
			\hline
			2 & 0.75 & 6.729 & 2.730 & 3.174 & 17.583 & 12.81 &&  0.03 & 5.833 & 1.697 & 2.387 & 12.182 & 19.4 \\
			3 & 0.86 & 6.662 & 2.730 & 3.113 & 17.536 & 12.46 &&  0.04 & 5.833 & 1.697 & 2.387 & 12.182 & 19.4 \\
			4 & 0.97 & 6.662 & 2.730 & 3.113 & 17.536 & 12.46 &&  0.07 & 5.833 & 1.697 & 2.387 & 12.182 & 19.4 \\
			5 & 1.27 & 6.662 & 2.730 & 3.113 & 17.536 & 12.46 &&  0.14 & 5.827 & 1.698 & 2.391 & 12.150 & 19.9 \\
\hline \hline
		\end{tabular}
	\end{table*}

	\begin{table*}[h!]
		\centering
		\caption{Adjusted Luttinger-Kohn parameters and $\text{E}_{\text{p}}$ for \ce{InAs} and \ce{InSb} in function of the adjusted region. The RMSDs are also indicated.}
		\label{}
		\begin{tabular}{SSSSSSSSSSSSSSS}
			
			\toprule
			\hline \\
			
			& \multicolumn{6}{c}{InAs} && \multicolumn{6}{c}{InSb} \\
			\cline{2-7} \cline{9-14} \\
			{{\si{\percent}}}& {RMSD} & {$\gamma_{1}$ }& {$\gamma_{2}$} & {$\gamma_{3}$} &  {{e}} & {$\text{E}_{\text{P}}$} &&
			{RMSD} & {$\gamma_{1}$ }& {$\gamma_{2}$} & {$\gamma_{3}$} &  {{e}} & {$\text{E}_{\text{P}}$} \\
			\hline
			2 &  0.09 & 19.193 & 8.145 & 9.002 & 42.725 & 20.91 &&  0.29 & 40.229 & 18.113 & 19.324 & 81.384 & 28.8 \\
			3 &  0.12 & 19.193 & 8.145 & 9.002 & 42.725 & 20.91 &&  0.90 & 38.150 & 17.096 & 18.267 & 77.318 & 26.9 \\
			4 &  0.25 & 19.193 & 8.145 & 9.002 & 42.725 & 20.91 &&  1.66 & 36.572 & 16.318 & 17.497 & 74.225 & 25.4 \\
			5 &  1.06 & 18.575 & 7.849 & 8.702 & 41.584 & 19.41 &&  2.10 & 36.572 & 16.318 & 17.497 & 74.225 & 25.4 \\
\hline \hline
		\end{tabular}
	\end{table*}

	\newpage
	
	\cleardoublepage
	
	\section{Effective masses and $g$-factors}
	
	\begin{table*}[h!]
		\centering
		\caption{Optimal AlN effective masses and electronic g-factors.
			Heavy ($m_{hh}$) and light ($m_{lh}$) hole effective masses are presented for 
			three different high-symmetry directions ([100], [110] and [111]),
			while spin-orbit splitting hole ($m_{so}$) and electron ($m_{e}$) effective masses are isotropic. RMSD values
			are also presented.}
		\label{}
		
		\begin{tabular}{SSSSSSSSSSSSSSSS}
			
			\toprule
			\hline
			& && \multicolumn{2}{c}{{[}100{]}} && \multicolumn{2}{c}{{[}110{]}} && \multicolumn{2}{c}{{[}111{]}} && &&\\
			\cline{4-5} \cline{7-8} \cline{10-11} 
			{\si{\percent}}       & {RMSD} 	     &&  {$m_{hh}$}   & {$m_{lh}$}    && {$m_{hh}$ }   & {$m_{lh}$}    && {$m_{hh}$}    & {$m_{lh}$}    && {$m_{so}$ }   & {$m_{e}$}     & {$g_{e}$}   \\
			\midrule
			2  & 0.36 && 1.117 & 0.435 && 2.345 & 0.361 && 1.907 & 0.375 && 0.628 & 0.292 & 1.991\\
			3  & 0.38 && 1.190 & 0.434 && 2.820 & 0.359 && 2.193 & 0.372 && 0.638 & 0.293 & 1.991\\
			4  & 0.44 && 1.286 & 0.436 && 3.723 & 0.357 && 2.668 & 0.371 && 0.653 & 0.295 & 1.991\\
			5  & 0.52 && 1.306 & 0.436 && 3.898 & 0.357 && 2.757 & 0.371 && 0.656 & 0.295 & 1.991\\
			6  & 0.62 && 1.306 & 0.436 && 3.898 & 0.357 && 2.757 & 0.371 && 0.656 & 0.295 & 1.991\\
			7  & 0.76 && 1.320 & 0.437 && 4.123 & 0.357 && 2.858 & 0.371 && 0.659 & 0.296 & 1.991\\
\hline \hline
		\end{tabular}
	\end{table*}

	\begin{table*}[h!]
		\centering
		\caption{Optimal AlP effective masses and electronic g-factors.
			Heavy ($m_{hh}$) and light ($m_{lh}$) hole effective masses are presented for 
			three different high-symmetry directions ([100], [110] and [111]),
			while spin-orbit splitting hole ($m_{so}$) and electron ($m_{e}$) effective masses are isotropic. RMSD values
			are also presented.}
		\label{}
		\begin{tabular}{SSSSSSSSSSSSSSSS}
			
			\toprule
			\hline
			& && \multicolumn{2}{c}{{[}100{]}} && \multicolumn{2}{c}{{[}110{]}} && \multicolumn{2}{c}{{[}111{]}} && &&\\
			\cline{4-5} \cline{7-8} \cline{10-11} 
			{\si{\percent}}       & {RMSD} 	     &&  {$m_{hh}$}   & {$m_{lh}$}    && {$m_{hh}$ }   & {$m_{lh}$}    && {$m_{hh}$}    & {$m_{lh}$}    && {$m_{so}$ }   & {$m_{e}$}     & {$g_{e}$}   \\
			\midrule                                                               
			2	& 0.19	 &&    0.458	 &  0.273  && 0.923	&  0.210  && 0.777   & 0.219  && 0.344  &  0.188   & 1.957 \\
			3	& 0.18	 &&    0.458	 &  0.273  && 0.923	&  0.210  && 0.777   & 0.219  && 0.344  &  0.188   & 1.957 \\
			4	& 0.20	 &&    0.533	 &  0.253  && 1.402	&  0.196  && 1.065   & 0.205  && 0.348  &  0.190   & 1.916 \\
			5	& 0.26	 &&    0.533	 &  0.253  && 1.402	&  0.196  && 1.065   & 0.205  && 0.348  &  0.190   & 1.916 \\
			6	& 0.49	 &&    0.546	 &  0.253  && 1.484	&  0.196  && 1.112   & 0.205  && 0.350  &  0.191   & 1.932 \\
			7	& 0.71	 &&    0.546	 &  0.253  && 1.484	&  0.196  && 1.112   & 0.205  && 0.350  &  0.191   & 1.932 \\
\hline \hline
		\end{tabular}
	\end{table*}
	
	\begin{table*}[h!]
		\centering
		\caption{Optimal AlAs effective masses and electronic g-factors.
			Heavy ($m_{hh}$) and light ($m_{lh}$) hole effective masses are presented for 
			three different high-symmetry directions ([100], [110] and [111]),
			while spin-orbit splitting hole ($m_{so}$) and electron ($m_{e}$) effective masses are isotropic. RMSD values
			are also presented.}
		\label{}
		\begin{tabular}{SSSSSSSSSSSSSSSS}
			
			\toprule
			\hline
			& && \multicolumn{2}{c}{{[}100{]}} && \multicolumn{2}{c}{{[}110{]}} && \multicolumn{2}{c}{{[}111{]}} && &&\\
			\cline{4-5} \cline{7-8} \cline{10-11} 
			{\si{\percent}}       & {RMSD} 	     &&  {$m_{hh}$}   & {$m_{lh}$}    && {$m_{hh}$ }   & {$m_{lh}$}    && {$m_{hh}$}    & {$m_{lh}$}    && {$m_{so}$ }   & {$m_{e}$}     & {$g_{e}$}   \\
			\midrule                                                        
			2 &  0.02 &  & 0.426 & 0.170 &  & -0.885 & 0.172 &  & 1.085 & 0.137 &  & 0.272 & 0.130 & 1.114 \\
			3 &  0.05 &  & 0.426 & 0.170 &  & -0.885 & 0.172 &  & 1.085 & 0.137 &  & 0.272 & 0.130 & 1.114 \\
			4 &  0.10 &  & 0.429 & 0.171 &  &  2.220 & 0.164 &  & 1.093 & 0.137 &  & 0.271 & 0.130 & 1.183 \\
			5 &  0.18 &  & 0.431 & 0.171 &  &  0.572 & 0.152 &  & 1.099 & 0.138 &  & 0.270 & 0.130 & 1.247 \\
			6 &  0.51 &  & 0.434 & 0.172 &  &  0.374 & 0.136 &  & 1.116 & 0.139 &  & 0.269 & 0.131 & 1.308 \\
			7 &  0.86 &  & 0.449 & 0.175 &  &  0.291 & 0.101 &  & 1.231 & 0.140 &  & 0.271 & 0.133 & 1.439 \\
\hline \hline
		\end{tabular}
	\end{table*}
	
	\begin{table*}[h!]
		\centering
		\caption{Optimal AlSb effective masses and electronic g-factors.
			Heavy ($m_{hh}$) and light ($m_{lh}$) hole effective masses are presented for 
			three different high-symmetry directions ([100], [110] and [111]),
			while spin-orbit splitting hole ($m_{so}$) and electron ($m_{e}$) effective masses are isotropic. RMSD values
			are also presented.}
		\label{}
		\begin{tabular}{SSSSSSSSSSSSSSSS}
			
			\toprule
			\hline
			& && \multicolumn{2}{c}{{[}100{]}} && \multicolumn{2}{c}{{[}110{]}} && \multicolumn{2}{c}{{[}111{]}} && &&\\
			\cline{4-5} \cline{7-8} \cline{10-11} 
			{\si{\percent}}       & {RMSD} 	     &&  {$m_{hh}$}   & {$m_{lh}$}    && {$m_{hh}$ }   & {$m_{lh}$}    && {$m_{hh}$}    & {$m_{lh}$}    && {$m_{so}$ }   & {$m_{e}$}     & {$g_{e}$}   \\
			\midrule                                                               
			2 &  0.02 &  & 0.306 & 0.125 &  & -1.849 & 0.121 &  & 0.748 & 0.101 &  & 0.237 & 0.110 & -0.827 \\
			3 &  0.04 &  & 0.308 & 0.126 &  & -4.144 & 0.121 &  & 0.772 & 0.101 &  & 0.238 & 0.111 & -0.783 \\
			4 &  0.10 &  & 0.310 & 0.127 &  &  3.624 & 0.120 &  & 0.781 & 0.102 &  & 0.238 & 0.112 & -0.713 \\
			5 &  0.24 &  & 0.312 & 0.128 &  &  1.010 & 0.118 &  & 0.790 & 0.102 &  & 0.238 & 0.113 & -0.618 \\
			6 &  0.43 &  & 0.312 & 0.128 &  &  1.010 & 0.118 &  & 0.790 & 0.102 &  & 0.238 & 0.113 & -0.618 \\
			7 &  1.23 &  & 0.314 & 0.129 &  &  0.542 & 0.114 &  & 0.819 & 0.103 &  & 0.237 & 0.114 & -0.499 \\
\hline \hline
		\end{tabular}
	\end{table*}
	
	\begin{table*}[h!]
		\centering
		\caption{Optimal GaN effective masses and electronic g-factors.
			Heavy ($m_{hh}$) and light ($m_{lh}$) hole effective masses are presented for 
			three different high-symmetry directions ([100], [110] and [111]),
			while spin-orbit splitting hole ($m_{so}$) and electron ($m_{e}$) effective masses are isotropic. RMSD values
			are also presented.}
		\label{}
		\begin{tabular}{SSSSSSSSSSSSSSSS}
			
			\toprule
			\hline
			& && \multicolumn{2}{c}{{[}100{]}} && \multicolumn{2}{c}{{[}110{]}} && \multicolumn{2}{c}{{[}111{]}} && &&\\
			\cline{4-5} \cline{7-8} \cline{10-11} 
			{\si{\percent}}       & {RMSD} 	     &&  {$m_{hh}$}   & {$m_{lh}$}    && {$m_{hh}$ }   & {$m_{lh}$}    && {$m_{hh}$}    & {$m_{lh}$}    && {$m_{so}$ }   & {$m_{e}$}     & {$g_{e}$}   \\
			\midrule                                                            
			2 & 0.02 &  & 0.791 & 0.269 &  & 0.394 & 0.131 &  & 1.771 & 0.226 &  & 0.403 & 0.196 & 1.979 \\
			3 & 0.05 &  & 0.796 & 0.269 &  & 0.415 & 0.156 &  & 1.807 & 0.227 &  & 0.405 & 0.197 & 1.976 \\
			4 & 0.13 &  & 0.796 & 0.269 &  & 0.415 & 0.156 &  & 1.807 & 0.227 &  & 0.405 & 0.197 & 1.976 \\
			5 & 0.31 &  & 0.796 & 0.269 &  & 0.415 & 0.156 &  & 1.807 & 0.227 &  & 0.405 & 0.197 & 1.976 \\
			6 & 0.58 &  & 0.808 & 0.274 &  & 0.440 & 0.157 &  & 1.934 & 0.229 &  & 0.411 & 0.199 & 1.976 \\
			7 & 0.69 &  & 0.805 & 0.275 &  & 0.432 & 0.156 &  & 1.899 & 0.229 &  & 0.412 & 0.200 & 1.976 \\
\hline \hline
		\end{tabular}
	\end{table*}
	
	\begin{table*}[h!]
		\centering
		\caption{Optimal GaP effective masses and electronic g-factors.
			Heavy ($m_{hh}$) and light ($m_{lh}$) hole effective masses are presented for 
			three different high-symmetry directions ([100], [110] and [111]),
			while spin-orbit splitting hole ($m_{so}$) and electron ($m_{e}$) effective masses are isotropic. RMSD values
			are also presented.}
		\label{}
		\begin{tabular}{SSSSSSSSSSSSSSSS}
			
			\toprule
			\hline
			& && \multicolumn{2}{c}{{[}100{]}} && \multicolumn{2}{c}{{[}110{]}} && \multicolumn{2}{c}{{[}111{]}} && &&\\
			\cline{4-5} \cline{7-8} \cline{10-11} 
			{\si{\percent}}       & {RMSD} 	     &&  {$m_{hh}$}   & {$m_{lh}$}    && {$m_{hh}$ }   & {$m_{lh}$}    && {$m_{hh}$}    & {$m_{lh}$}    && {$m_{so}$ }   & {$m_{e}$}     & {$g_{e}$}   \\
			\midrule                                                           
			2 &  0.30 &  & 0.333 & 0.161 &  & 0.186 & 0.075 &  & 0.772 & 0.126 &  & 0.222 & 0.134 & 1.816 \\
			3 &  0.31 &  & 0.333 & 0.161 &  & 0.186 & 0.075 &  & 0.772 & 0.126 &  & 0.222 & 0.134 & 1.816 \\
			4 &  0.32 &  & 0.343 & 0.162 &  & 0.192 & 0.077 &  & 0.800 & 0.128 &  & 0.225 & 0.134 & 1.816 \\
			5 &  0.47 &  & 0.354 & 0.162 &  & 0.203 & 0.079 &  & 0.847 & 0.128 &  & 0.227 & 0.135 & 1.813 \\
			6 &  0.85 &  & 0.354 & 0.162 &  & 0.203 & 0.079 &  & 0.847 & 0.128 &  & 0.227 & 0.135 & 1.813 \\
			7 &  1.35 &  & 0.367 & 0.165 &  & 0.220 & 0.079 &  & 0.945 & 0.129 &  & 0.233 & 0.138 & 1.818 \\
\hline \hline
		\end{tabular}
	\end{table*}
	
	\begin{table*}[h!]
		\centering
		\caption{Optimal GaAs effective masses and electronic g-factors.
			Heavy ($m_{hh}$) and light ($m_{lh}$) hole effective masses are presented for 
			three different high-symmetry directions ([100], [110] and [111]),
			while spin-orbit splitting hole ($m_{so}$) and electron ($m_{e}$) effective masses are isotropic. RMSD values
			are also presented.}
		\label{}
		\begin{tabular}{SSSSSSSSSSSSSSSS}
			
			\toprule
			\hline
			& && \multicolumn{2}{c}{{[}100{]}} && \multicolumn{2}{c}{{[}110{]}} && \multicolumn{2}{c}{{[}111{]}} && &&\\
			\cline{4-5} \cline{7-8} \cline{10-11} 
			{\si{\percent}}       & {RMSD} 	     &&  {$m_{hh}$}   & {$m_{lh}$}    && {$m_{hh}$ }   & {$m_{lh}$}    && {$m_{hh}$}    & {$m_{lh}$}    && {$m_{so}$ }   & {$m_{e}$}     & {$g_{e}$}   \\
			\midrule                                                                
			2  &  0.27 &  & 0.277 & 0.080 &  & 0.329 & 0.098 &  & 0.622 & 0.069 &  & 0.157 & 0.065 & -1.433 \\
			3  &  0.34 &  & 0.296 & 0.081 &  & 0.194 & 0.084 &  & 0.688 & 0.070 &  & 0.158 & 0.065 & -1.042 \\
			4  &  0.44 &  & 0.296 & 0.081 &  & 0.194 & 0.084 &  & 0.688 & 0.070 &  & 0.158 & 0.065 & -1.042 \\
			5  &  0.57 &  & 0.305 & 0.082 &  & 0.179 & 0.074 &  & 0.720 & 0.071 &  & 0.158 & 0.065 & -0.837 \\
			6  &  1.38 &  & 0.310 & 0.082 &  & 0.176 & 0.066 &  & 0.752 & 0.071 &  & 0.157 & 0.065 & -0.683 \\
			7  &  2.00 &  & 0.324 & 0.085 &  & 0.184 & 0.056 &  & 0.847 & 0.073 &  & 0.160 & 0.067 & -0.376 \\
\hline \hline
		\end{tabular}
	\end{table*}
	
	\begin{table*}[h!]
		\centering
		\caption{Optimal GaSb effective masses and electronic g-factors.
			Heavy ($m_{hh}$) and light ($m_{lh}$) hole effective masses are presented for 
			three different high-symmetry directions ([100], [110] and [111]),
			while spin-orbit splitting hole ($m_{so}$) and electron ($m_{e}$) effective masses are isotropic. RMSD values
			are also presented.}
		\label{}
		\begin{tabular}{SSSSSSSSSSSSSSSS}
			
			\toprule
			\hline
			& && \multicolumn{2}{c}{{[}100{]}} && \multicolumn{2}{c}{{[}110{]}} && \multicolumn{2}{c}{{[}111{]}} && &&\\
			\cline{4-5} \cline{7-8} \cline{10-11} 
			{\si{\percent}}       & {RMSD} 	     &&  {$m_{hh}$}   & {$m_{lh}$}    && {$m_{hh}$ }   & {$m_{lh}$}    && {$m_{hh}$}    & {$m_{lh}$}    && {$m_{so}$ }   & {$m_{e}$}     & {$g_{e}$}   \\
			\midrule                                                              
			2 &   0.04 &  & 0.226 & 0.043 &  & 0.133 & 0.053 &  & 0.545 & 0.039 &  & 0.141 & 0.039 & -11.373 \\
			3 &   0.12 &  & 0.229 & 0.044 &  & 0.135 & 0.051 &  & 0.570 & 0.039 &  & 0.141 & 0.039 & -11.038 \\
			4 &   0.32 &  & 0.231 & 0.044 &  & 0.135 & 0.049 &  & 0.580 & 0.040 &  & 0.141 & 0.040 & -10.653 \\
			5 &   0.71 &  & 0.232 & 0.045 &  & 0.134 & 0.046 &  & 0.590 & 0.040 &  & 0.140 & 0.040 & -10.218 \\
			6 &   1.18 &  & 0.232 & 0.045 &  & 0.134 & 0.046 &  & 0.590 & 0.040 &  & 0.140 & 0.040 & -10.218 \\
			7 &   2.52 &  & 0.233 & 0.046 &  & 0.134 & 0.043 &  & 0.606 & 0.041 &  & 0.139 & 0.041 &  -9.741 \\
\hline \hline
		\end{tabular}
	\end{table*}
	
	\begin{table*}[h!]
		\centering
		\caption{Optimal InN effective masses and electronic g-factors.
			Heavy ($m_{hh}$) and light ($m_{lh}$) hole effective masses are presented for 
			three different high-symmetry directions ([100], [110] and [111]),
			while spin-orbit splitting hole ($m_{so}$) and electron ($m_{e}$) effective masses are isotropic. RMSD values
			are also presented.}
		\label{}
		\begin{tabular}{SSSSSSSSSSSSSSSS}
			
			\toprule
			\hline
			& && \multicolumn{2}{c}{{[}100{]}} && \multicolumn{2}{c}{{[}110{]}} && \multicolumn{2}{c}{{[}111{]}} && &&\\
			\cline{4-5} \cline{7-8} \cline{10-11} 
			{\si{\percent}}       & {RMSD} 	     &&  {$m_{hh}$}   & {$m_{lh}$}    && {$m_{hh}$ }   & {$m_{lh}$}    && {$m_{hh}$}    & {$m_{lh}$}    && {$m_{so}$ }   & {$m_{e}$}     & {$g_{e}$}   \\
			\midrule                                                            
			2 & 0.75 && 0.788 & 0.082 && 2.624 & 0.076 && 1.692 & 0.078 && 0.152 & 0.057 & 1.685\\
			3 & 0.86 && 0.832 & 0.082 && 2.295 & 0.078 && 1.617 & 0.079 && 0.154 & 0.057 & 1.694\\
			4 & 0.97 && 0.832 & 0.082 && 2.295 & 0.078 && 1.617 & 0.079 && 0.154 & 0.057 & 1.694\\
			5 & 1.27 && 0.832 & 0.082 && 2.295 & 0.078 && 1.617 & 0.079 && 0.154 & 0.057 & 1.694\\
			6 & 2.92 && 0.834 & 0.086 && 2.669 & 0.080 && 1.755 & 0.081 && 0.159 & 0.059 & 1.716\\
			7 & 3.51 && 0.850 & 0.088 && 2.154 & 0.083 && 1.577 & 0.084 && 0.163 & 0.061 & 1.721\\
\hline \hline
		\end{tabular}
	\end{table*}
	
	\begin{table*}[h!]
		\centering
		\caption{Optimal InP effective masses and electronic g-factors.
			Heavy ($m_{hh}$) and light ($m_{lh}$) hole effective masses are presented for 
			three different high-symmetry directions ([100], [110] and [111]),
			while spin-orbit splitting hole ($m_{so}$) and electron ($m_{e}$) effective masses are isotropic. RMSD values
			are also presented.}
		\label{}
		\begin{tabular}{SSSSSSSSSSSSSSSS}
			
			\toprule
			\hline
			& && \multicolumn{2}{c}{{[}100{]}} && \multicolumn{2}{c}{{[}110{]}} && \multicolumn{2}{c}{{[}111{]}} && &&\\
			\cline{4-5} \cline{7-8} \cline{10-11} 
			{\si{\percent}}       & {RMSD} 	     &&  {$m_{hh}$}   & {$m_{lh}$}    && {$m_{hh}$ }   & {$m_{lh}$}    && {$m_{hh}$}    & {$m_{lh}$}    && {$m_{so}$ }   & {$m_{e}$}     & {$g_{e}$}   \\
			\midrule                                                                
			2 &   0.03 &  & 0.410 & 0.108 &  & 0.210 & 0.071 &  & 0.945 & 0.094 &  & 0.182 & 0.082 & 1.342 \\
			3 &   0.04 &  & 0.410 & 0.108 &  & 0.210 & 0.071 &  & 0.945 & 0.094 &  & 0.182 & 0.082 & 1.342 \\
			4 &   0.07 &  & 0.410 & 0.108 &  & 0.210 & 0.071 &  & 0.945 & 0.094 &  & 0.182 & 0.082 & 1.342 \\
			5 &   0.14 &  & 0.411 & 0.108 &  & 0.214 & 0.074 &  & 0.957 & 0.094 &  & 0.182 & 0.082 & 1.326 \\
			6 &   0.29 &  & 0.413 & 0.109 &  & 0.216 & 0.074 &  & 0.966 & 0.094 &  & 0.183 & 0.083 & 1.328 \\
			7 &   0.85 &  & 0.415 & 0.109 &  & 0.218 & 0.074 &  & 0.974 & 0.095 &  & 0.183 & 0.083 & 1.329 \\
\hline \hline
		\end{tabular}
	\end{table*}
	
	\begin{table*}[h!]
		\centering
		\caption{Optimal InAs effective masses and electronic g-factors.
			Heavy ($m_{hh}$) and light ($m_{lh}$) hole effective masses are presented for 
			three different high-symmetry directions ([100], [110] and [111]),
			while spin-orbit splitting hole ($m_{so}$) and electron ($m_{e}$) effective masses are isotropic. RMSD values
			are also presented.}
		\label{}
		\begin{tabular}{SSSSSSSSSSSSSSSS}
			
			\toprule
			\hline
			& && \multicolumn{2}{c}{{[}100{]}} && \multicolumn{2}{c}{{[}110{]}} && \multicolumn{2}{c}{{[}111{]}} && &&\\
			\cline{4-5} \cline{7-8} \cline{10-11} 
			{\si{\percent}}       & {RMSD} 	     &&  {$m_{hh}$}   & {$m_{lh}$}    && {$m_{hh}$ }   & {$m_{lh}$}    && {$m_{hh}$}    & {$m_{lh}$}    && {$m_{so}$ }   & {$m_{e}$}     & {$g_{e}$}   \\
			\midrule                                                            
			2 & 0.09   && 0.345 & 0.028 && 0.841 & 0.027 && 0.624 & 0.027 && 0.103 & 0.023 & -16.998\\
			3 & 0.12   && 0.345 & 0.028 && 0.841 & 0.027 && 0.624 & 0.027 && 0.103 & 0.023 & -16.998\\
			4 & 0.25   && 0.345 & 0.028 && 0.841 & 0.027 && 0.624 & 0.027 && 0.103 & 0.023 & -16.998\\
			5 & 1.06   && 0.348 & 0.029 && 0.855 & 0.028 && 0.633 & 0.028 && 0.103 & 0.024 & -15.643\\
			6 & 1.20   && 0.348 & 0.029 && 0.855 & 0.028 && 0.633 & 0.028 && 0.103 & 0.024 & -15.643\\
			7 & 1.38   && 0.349 & 0.029 && 0.862 & 0.028 && 0.637 & 0.028 && 0.103 & 0.024 & -15.403\\
\hline \hline
		\end{tabular}
	\end{table*}
	
	\begin{table*}[h!]
		\centering
		\caption{Optimal InSb effective masses and electronic g-factors.
			Heavy ($m_{hh}$) and light ($m_{lh}$) hole effective masses are presented for 
			three different high-symmetry directions ([100], [110] and [111]),
			while spin-orbit splitting hole ($m_{so}$) and electron ($m_{e}$) effective masses are isotropic. RMSD values
			are also presented.}
		\label{}
		\begin{tabular}{SSSSSSSSSSSSSSSS}
			
			\toprule
			\hline
			& && \multicolumn{2}{c}{{[}100{]}} && \multicolumn{2}{c}{{[}110{]}} && \multicolumn{2}{c}{{[}111{]}} && &&\\
			\cline{4-5} \cline{7-8} \cline{10-11} 
			{\si{\percent}}       & {RMSD} 	     &&  {$m_{hh}$}   & {$m_{lh}$}    && {$m_{hh}$ }   & {$m_{lh}$}    && {$m_{hh}$}    & {$m_{lh}$}    && {$m_{so}$ }   & {$m_{e}$}     & {$g_{e}$}   \\
			\midrule                                                           
			2 &   0.29 &  & 0.250 & 0.013 &  & 0.124 & 0.015 &  & 0.632 & 0.013 &  & 0.121 & 0.012 & -61.915 \\
			3 &   0.90 &  & 0.253 & 0.014 &  & 0.123 & 0.015 &  & 0.618 & 0.013 &  & 0.120 & 0.013 & -57.634 \\
			4 &   1.66 &  & 0.254 & 0.014 &  & 0.125 & 0.015 &  & 0.634 & 0.014 &  & 0.119 & 0.013 & -54.372 \\
			5 &   2.10 &  & 0.254 & 0.014 &  & 0.125 & 0.015 &  & 0.634 & 0.014 &  & 0.119 & 0.013 & -54.372 \\
			6 &   2.93 &  & 0.252 & 0.015 &  & 0.126 & 0.015 &  & 0.646 & 0.014 &  & 0.118 & 0.014 & -52.211 \\
			7 &   4.33 &  & 0.253 & 0.015 &  & 0.128 & 0.015 &  & 0.657 & 0.015 &  & 0.118 & 0.014 & -50.230 \\
\hline \hline
		\end{tabular}
	\end{table*}

	\bibliography{references}
	
\end{document}